\documentclass[11pt]{article}
\usepackage{color,amsfonts,amsmath,amssymb,mathrsfs,verbatim,epsf}

\def\inn{\in}

\def\TM{\mbox{\it TM}}

\def\numutoe{
\stackrel{\raisebox{-2pt}[0pt][0pt]{\mbox{{\tiny \:\!({\bf --})}}}}
 {\raisebox{-0.0pt}[0pt][0pt]{\mbox{$\nu$}}}_{\!\!\mu} 
                  \to 
 \stackrel{\raisebox{-2pt}[0pt][0pt]{\mbox{{\tiny \:\!({\bf --})}}}}
 {\raisebox{-0.0pt}[0pt][0pt]{\mbox{$\nu$}}}_{\!\! e}}

\def\rrr{{\mathbb{R}}}
\def\ccc{{\mathbb{C}}}

\def\ooo{{\mathbb{O}}}

\def\b1{\mbox{\boldmath $1$}}

\def\v{\mbox{\boldmath $v$}}

\def\bh{\mbox{\boldmath $h$}}

\def\bv{\mbox{\boldmath $v$}}

\def\hG{\hat{G}}

\def\bB{\mbox{\boldmath $B$}}

\def\bE{\mbox{\boldmath $E$}}

\def\lvtfep{L_8(\v_{248})_{\mathrm{E}_8}=1}

\def\lpvng{L_p(\v_n)_{\hG}=1}

\def\htwc{\mbox{h}_2\ccc}
\def\hthc{\mbox{h}_3\ccc}

\def\htwo{\mbox{h}_2\ooo}
\def\htho{\mbox{h}_3\ooo}

\def\sltc{\mbox{SL}(2,\ccc)}

\def\slthc{\mbox{SL}(3,\ccc)}

\def\sltho{\mbox{SL}(3,\ooo)}

\def\uo{\mbox{U}(1)}

\def\sutw{\mbox{SU}(2)}

\def\suth{\mbox{SU}(3)}

\def\soot{\mbox{SO}^+(1,3)}

\def\sootnm{\mbox{SO}^+(1,n-1)}

\def\gt{\mbox{G}_2}
\def\ff{\mbox{F}_4}
\def\ee{\mbox{E}_8}
\def\eeg{\mbox{E}_{8(-24)}}
\def\ese{\mbox{E}_7}
\def\eseg{\mbox{E}_{7(-25)}}
\def\esi{\mbox{E}_6}
\def\esig{\mbox{E}_{6(-26)}}

\def\SML{\mbox{SU}(3)_c \times \mbox{SU}(2)_L \times \mbox{U}(1)_Y}

\def\ph{\phantom}

\def\mcX{{\mathcal X}}
\def\mcY{{\mathcal Y}}

\def\pal{\partial}

\def\mmin{m_{\mathrm{min}}}
\def\mtot{m_{\mathrm{tot}}}
\def\Lsl{L\!\!\!\!\!\!\;{\mbox{\scriptsize$\boldmath{\diagup}$}}}
\def\Lsl{L\!\!\!\!\!\!\;{\mbox{\scriptsize${\diagup}$}}}

\def\gpath{ }

\def\setb{\setlength{\baselineskip}{0.625\baselineskip}}

\begin{document} 

{\setlength{\baselineskip}{0.625\baselineskip}

\begin{center}

 {\LARGE{\bf Generalised Proper Time as a Unifying Basis  \vspace{6pt} \\ 
              for Models with Two Right-Handed Neutrinos }}

   \vspace{40pt}


  \mbox {{\Large David J. Jackson}} \\ 
  \vspace{10pt}  
  {david.jackson.th@gmail.com}  \\

 \vspace{30pt}  
  
 
 { \large May 27, 2019 }

 \vspace{40pt}

{\bf  Abstract}

 
\end{center}


    Models with two right-handed neutrinos are able to accommodate solar and atmospheric 
  neutrino oscillation observations as well as a mechanism for the baryon asymmetry of the 
universe. While economical in terms of the required new states beyond the Standard Model, given 
that there are three generations of the other leptons and quarks this raises the question 
concerning why \textit{only} two right-handed neutrino states should exist. Here we develop 
from first principles a fundamental unification scheme based upon a direct generalisation and 
analysis of a simple proper time interval with a structure beyond that of local 4-dimensional 
spacetime and further augmenting that of models with extra spatial dimensions.
 This theory leads to properties of matter fields that resemble the Standard Model, with an 
intrinsic left-right asymmetry which is particularly marked for the neutrino sector. It will be 
shown how the theory can provide a foundation for the natural incorporation of two right-handed 
neutrinos and may in principle underlie firm predictions both in the neutrino sector and for 
other new physics beyond the Standard Model. 
 While connecting with contemporary and future experiments the origins of the theory are 
motivated in a similar spirit as for the earliest unified field theories.

{

\vspace{-17pt}




\pagebreak
\tableofcontents

}



\section{Introduction and Reviews}
\label{uni1}

  A salient feature of the history of physics is the progression in 
experimental and theoretical insights delving deeper into the structure of 
matter on ever smaller scales. While the regularity of chemical elements in 
the Periodic Table, as originally organised by Mendeleev in 1869,  
 provided the first indirect hint of an inner structure for atoms,  
 the familiar pattern of elementary particles established in the 1970s in 
the Standard Model encapsulates an order in the structure of matter at a 
further
 submicroscopic level.
 A central characteristic of this structure is the empirically uncovered 
distinctive symmetry properties of the elementary particle multiplets.
  Given the very fundamental level of these observations  
  the modern-day quest to theoretically elucidate the underlying source of 
these symmetries of the Standard Model, in preference to positing them as 
`brute facts' about the world, is particularly pressing.

  This quest is analogous then,
 at a more elementary level, to the pursuit of an explanation
 for the Periodic Table that culminated in the discovery of the central 
atomic nucleus and the quantum mechanical properties of electron orbital 
states around one hundred years ago. While having something in common with 
the `unified field theories' of that era, in this paper we describe a new 
theory based upon a generalisation of proper time which has the potential 
to naturally incorporate not only the Standard Model but also the 
phenomenology of physics beyond, including that of the neutrino sector. The 
core arguments for this analysis are summarised in~\cite{Ufield} and will 
be elaborated along with further observations in this paper as outlined 
below.

  In the first subsection to follow we review the empirically established 
properties of neutrinos and several of the models stimulated by these 
results. We then present an historical survey of early unified field 
theories in subsection~\ref{uni12} which will lead to a probing of the 
motivation for extra spatial dimensions and the application of a further 
generalisation for a proper time interval in section~\ref{uni2}. (This 
paper can be read beginning with section~\ref{uni2}, with 
section~\ref{uni1} for reference). In section~\ref{uni3} the greater 
suitability of this generalisation in directly accounting for features of 
the Standard Model will be explained, further justifying the new approach. 
The manner in which this theory might naturally provide a foundation for 
models with two right-handed neutrinos will be described in 
subsection~\ref{uni41}, with broader connections with new physics beyond 
the Standard Model and further possible tests discussed in 
subsection~\ref{uni42}. We return to the interpretation of theory and its 
relation to special and general relativity and  connection with early 
unified field theories in subsection~\ref{uni51},
 with the underlying simplicity of the theory evaluated further in 
subsection~\ref{uni52}. 
 In section~\ref{uni6} we conclude with emphasis on the opportunities for 
the further development of this fundamental theory. The main goals of this 
paper are summarised here:

\begin{itemize}
 
\item A main theme of the paper is to emphasise the simplicity of this 
theory based on a generalisation of proper time, 
 arguing that a unifying basis for fundamental physics can be encapsulated 
correspondingly in `one simple equation'.    

\item We describe how the theory can be seen as a natural progression from 
special and general relativity, while also being complementary to the 
latter, and is related to the early conceptions of a unified field theory.

\item While reviewing the links with the Standard Model here we focus on 
possible connections with the further esoteric properties of physics 
beyond, in particular establishing a link with models
 incorporating two right-handed neutrinos.  

\end{itemize}

  The theory, originating from considerations of the first bullet point 
above, will be shown to provide a connection between the old and the new, 
from the second and the third bullet points respectively, by demonstrating 
the relevance of the theory for contemporary particle physics and 
cosmology. With this aim in mind we begin by reviewing the current status 
of the neutrino sector.

\subsection{Neutrino Physics}
\label{uni11}

 The striking progress in the empirical understanding of neutrino physics 
in recent decades has centred upon the compelling observations of 
oscillations between the left-handed neutrino states. These can be 
described assuming three-flavour neutrino mixing parametrised by the $3 
\times 3$ complex unitary Pontecorvo-Maki-Nakagawa-Sakata, or PMNS, 
matrix~(see for example~\cite{PDG18} sections 14.1 and 14.2, \cite{Drewes} 
section~2.1) which expresses the non-trivial relation between the three 
left-handed neutrino flavour eigenstates 
 $\nu_e$, $\nu_{\mu}$ and $\nu_{\tau}$ and three mass eigenstates  $\nu_1$, 
$\nu_2$ and $\nu_3$ with respective masses $m_1$, $m_2$ and $m_3$. 

 This structure accommodates both the `solar neutrino' oscillations, which 
are predominantly described by the mixing probability $P(\nu_e \to 
\nu_{\mu})$, as confirmed by nuclear reactor experiments, via a mixing 
angle $\theta_{\mathrm{sol}} = \theta_{12}  \simeq 33.6^{\circ}$  and mass 
difference 
  $\Delta m^2_{\mathrm{sol}} = \Delta m^2_{21} := m_2^2 - m_1^2 \simeq 7.5 
\times 10^{-5}\,\mbox{eV}^2$, as well as the `atmospheric neutrino' 
oscillations which can be assumed to be almost completely due to the mixing 
probability 
$P(\nu_{\mu} \to \nu_{\tau})$, as confirmed by accelerator experiments, via 
a
mixing angle $\theta_{\mathrm{atm}} = \theta_{23} \simeq 40$--$50^{\circ}$  
and mass difference 
  $ \Delta m^2_{\mathrm{atm}} =  \Delta m^2_{32} :=  m_3^2 - m_2^2 \simeq 
2.5 \times 10^{-3}\,\mbox{eV}^2$ (with all data in this subsection obtained 
from~\cite{PDG18} unless stated otherwise).   
  The third mixing angle has also been determined as $\theta_{13} \simeq 
8.4^{\circ}$,
 principally from the disappearance of $\bar{\nu}_e$ produced in nuclear 
fission reactors (\cite{PDG18} section~14.12), 
   while an estimate of the phase of the PMNS matrix $\delta_{C\!P} \simeq 
\frac{3\pi}{2}$ has been obtained from long baseline accelerator $\numutoe$ 
appearance experiments
  (\cite{PDG18} section~14.13).

   The third mass difference is determined from the definitions by the 
other two with
 $\Delta m_{31}^2 = \Delta m_{32}^2 - \Delta m_{21}^2$,  and with 
  $\Delta m_{31}^2 \simeq \Delta m_{32}^2$ since $\Delta m_{21}^2$ is 
somewhat smaller than $\Delta m_{32}^2$. There is however no existing 
constraint on the sign of $\Delta m_{32}^2$, that is whether $m_3$ is 
greater or less than $m_2$ (and $m_1$) -- termed the `normal' or `inverted' 
hierarchy respectively. 
 The neutrino oscillation data also does not determine the absolute mass 
scale.
 
 The electron neutrino mass $m_{\nu_{e}}$ can be defined by
 $m_{\nu_{e}}^2 = \sum_{i=1}^3 \vert U_{ei} \vert^2 m_{i}^2$, where 
$U_{ei}$ are elements of the top row of the PMNS matrix. The measured 
parameters above, given that the lightest neutrino mass state has mass 
$\mmin \ge 0\,\mbox{eV}$, set lower bounds of $m_{\nu_{e}} > 
0.01\,\mbox{eV}$ for the normal hierarchy and $m_{\nu_{e}} > 
0.05\,\mbox{eV}$ for the inverted hierarchy.
A direct upper limit on the electron neutrino mass
  is set by tritium $\beta$-decay
  experiments with $m_{\nu_{e}} 
   < 2.0\,\mbox{eV}$, with future experiments aiming to bring this limit 
down to $m_{\nu_{e}} < 0.2\,\mbox{eV}$~\cite{Katrin}.

  By comparison for $\mmin = 0\,\mbox{eV}$  the simple sum of the three 
neutrino masses $\mtot = \sum_{i=1}^3  m_i$ would be $\mtot = 
0.06\,\mbox{eV}$ or $\mtot = 0.10\,\mbox{eV}$ for the normal or inverted 
mass hierarchy respectively (see also~\cite{Drewes} section~2.1). 
 The constraints from cosmological observations, although dependent upon 
the
 $\Lambda$CDM cosmological model (with $\Lambda$ the cosmological constant 
and CDM cold dark matter), imply $\mtot  
   < 0.12\,\mbox{eV}$ at the 95\% confidence level~\cite{Planck}, already 
putting pressure on the inverted hierarchy.
    Future prospects for the three left-handed neutrinos within the  
$\Lambda$CDM model  are for the lowest possible value of  $\mtot = 
0.06\,\mbox{eV}$ to be detectable at the 3--4$\sigma$ level in the coming 
years~(\cite{PDG18} sections~25.4 and 64).
   
   While the existing data cannot distinguish between whether  neutrinos 
are Dirac or Majorana fermions, any observation of neutrinoless 
double-$\beta$ decay  would indicate the Majorana type (that is, with such 
a neutrino identical to its own antiparticle state). If such experiments 
were to determine a non-zero value for the appropriately defined effective 
Majorana mass $m_{\beta\beta}$ for this process with $m_{\beta\beta} 
\lesssim 0.01\,\mbox{eV}$, an order of magnitude beyond the current 
sensitivity, then the inverted hierarchy could be ruled out~(\cite{Agost}, 
\cite{PDG18} section~62, \cite{Drewes} section~3.3).
 Such a measurement with $m_{\beta\beta} \gtrsim 0.005\,\mbox{eV}$ would 
also indicate that $\mmin > 0\,\mbox{eV}$ (\cite{PDG18} figure~62.1, 
\cite{Drewes} figure~5). 
 Similarly,
  there remains the possibility for a positive observation of
 neutrinoless double-$\beta$ decay with $m_{\beta\beta} \gtrsim 
0.06\,\mbox{eV}$ also implying a value for \mbox{$\mmin > 0\,\mbox{eV}$}, 
although in this case in a range seemingly in tension with the above 
current cosmological bound.

 The observation of neutrino oscillations is generally considered to imply 
the existence of right-handed neutrinos $\nu_R$ which are `sterile', that 
is they transform trivially under the Standard Model gauge group $\SML$, 
accounting for the difficulty of their direct detection
 (unlike the familiar `active' left-handed states $\nu_L$ that undergo weak 
interactions). These right-handed states can be utilised to introduce light 
masses for the active neutrinos by extending the Standard Model Lagrangian 
with further Dirac mass terms. Such terms are similar to those for the 
charged leptons and quarks but with unnaturally small Yukawa couplings to 
the Higgs field, by around a factor of $10^{-12}$ relative to that of the 
top quark and even by a factor of $10^{-6}$ or less relative to that of the 
electron in the same weak doublet as the electron neutrino.
 However, owing to their trivial gauge transformation properties, 
   Majorana mass terms  can also be added for the right-handed neutrinos 
which, if sufficiently heavy, can generate the light active neutrino masses  
 via a `seesaw' mechanism  (\cite{Drewes} section 2 and references 
therein). For either of the above cases, since each $\nu_R$ state can only 
generate one $\nu_{L}$ mass, at least \textit{two} right-handed neutrinos 
are required to account for at least \textit{two} finite active neutrino 
masses as implied by the established measurements of $\Delta m_{21}$ and 
$\Delta m_{32}$, unless there is a different source for the $\nu_L$ masses.

   In the absence of a theoretical argument to the contrary a natural 
expectation might be for all three active neutrinos to be massive,  
   via the introduction of \textit{three} right-handed neutrino states, 
   matching the three generation structure of the other leptons and quarks.
 This is the case for the `Neutrino Minimal Standard Model',
  or $\nu$MSM~(\cite{Asaka1,Asaka2}, \cite{Drewes} section~7), proposed as 
a simple economical extension from the Standard Model (for which all three 
$\nu_L$ states are massless and there are no $\nu_R$ states). In the 
$\nu$MSM there are two $\nu_R$ states with nearly degenerate masses in the 
range from $\sim \!\!1\,\mbox{GeV}$ to the electroweak scale which account 
for active neutrino masses via the seesaw mechanism consistent with the 
well-established neutrino oscillation data.
 At the epoch of the Big Bang \mbox{$C\!P$-violating} oscillations of these 
two sterile neutrinos during their thermal production can also in principle 
generate the baryon asymmetry of the universe.

  Compelling empirical evidence for the mass scale of right-handed 
neutrinos may be even harder to establish than for the active left-handed 
states.
  However the lower part of the preferred mass range of $1\!\! \sim\!\! 
100\,\mbox{GeV}$ for two of the three right-handed neutrinos in the  
$\nu$MSM  could be  accessible to direct searches for $\nu_R$ states 
through mixing with active neutrino states in laboratory experiments 
(\cite{Drewes} section~3.4, \cite{Ship}). Such mixing is required to 
generate the active $\nu_L$ masses via the seesaw mechanism.
 Further, much lighter right-handed neutrinos can also play a significant 
role in cosmology~\cite{Bari}.

   In particular, sterile right-handed neutrinos, which
 apart from the above mixing effects could only be observable through 
gravitational interactions,  
    can be considered a natural candidate for dark matter~\cite{Drewe2}. 
However the required Yukawa couplings for this application are very 
different to those associated with neutrino oscillations.
 In the case of the $\nu$MSM the third $\nu_R$ state, with a mass of a few 
keV, acts as a warm dark matter candidate but with a Yukawa coupling too 
small to make a significant contribution to the active neutrino masses, 
hence leaving the lightest active neutrino practically massless. More 
generally, for scenarios with three  $\nu_R$ states there is no intrinsic 
limit on the mass of the lightest $\nu_L$ mass state.

  Neutrino models are further stretched if required to accommodate the 
empirical hints for anomalous observations in $\numutoe$ oscillations, 
which imply a mass difference of \mbox{$\Delta m^2 \sim\!  
1\,\mbox{eV}^2$}, as well as the anomalies observed in reactor and gallium 
experiments. Further neutrino states or other new features are needed to 
provide a phenomenologically complete description of all neutrino particle 
physics data (see for example~\cite{Doring}).

 On the other hand the compelling neutrino oscillation observations can be 
accounted for by models with only \textit{two} right-handed neutrinos, 
which can also accommodate a source of the baryon 
 asymmetry~(see for example~\cite{Framp,Ibar,Antu,Chang}). 
For these models, which imply that $\mmin = 0\,\mbox{eV}$ for the 
left-handed neutrinos,  the two right-handed neutrinos are typically very 
massive in the range from $\sim \!\!10^{10}\,$GeV up to the GUT scale. In 
this case a possible source of the matter-antimatter asymmetry derives 
through a leptogenesis scenario via heavy right-handed neutrino decays in 
the very early universe~(see also~\cite{Fuku,Bari2}).
 The indications that the PMNS phase $\delta_{C\!P}$ may be relatively 
large suggests that \mbox{$C\!P$-violating} effects in the oscillations of 
light active neutrinos  may be significant, and can be linked via the 
neutrino mass matrices in specific seesaw models with the  
\mbox{$C\!P$-violating} phase in the heavy sterile neutrino sector that 
drives the baryon asymmetry via leptogenesis  (see for 
example~\cite{Framp}, \cite{Bari2} section~8, \cite{Shim}).
 While models with two $\nu_R$ states might hence account for 
 active neutrino oscillations and the baryon asymmetry, similarly as for 
the $\nu$MSM, in lacking a third $\nu_R$ state an alternative candidate for 
dark matter will be required.

  In summary, based only upon well-established observations in the neutrino 
sector the main questions to be addressed concern
the absolute value and spectrum of the active neutrino masses, resolving 
the sign of $\Delta m_{32}^2$,  the number and masses of sterile neutrinos, 
and whether neutrinos are Dirac or Majorana particles. Further questions 
concern the PMNS matrix: including the proximity of the value of 
$\theta_{23}$ to  maximal $45^{\circ}$ mixing, the need for a more precise 
evaluation of $\delta_{C\!P}$, and regarding the relation of the neutrino 
mixing matrix to the very different CKM mixing matrix in the quark sector. 

 While it is always possible in principle to \textit{model} a wide range of 
neutrino phenomena, by extending the Standard Model Lagrangian, a more 
fundamental explanation of the \textit{origin} of these empirical 
observations, providing a deeper understanding of the neutrino flavour 
structure and mass generation mechanism,  would of course be desired.
 Ideally such an explanation would take the form of a theory that also 
accounts for the properties of the Standard Model itself, and which might 
be empirically tested in neutrino experiments, in the particle physics 
laboratory more generally or through
 cosmological observations. 
 In this paper we propose such a fundamental theory.
 The theory may provide in particular a possible basis for 
   models with two right-handed neutrinos in a unified structure 
consistently alongside three generations of the other leptons and quarks,
    as will be explained in subsection~\ref{uni41}.
	 While connecting with these contemporary phenomenological issues, as 
described further in subsection~\ref{uni42},  
	 the underlying motivation for the theory is related to that for the 
earliest unified field theories; we hence next review this background in 
the following subsection.


\subsection{Unified Field Theories}
\label{uni12}

  While the field concept as employed with great success for Maxwell's 
equations~\cite{Maxem} had very much influenced the conception of  
    Einstein's theory of general relativity~\cite{Eingr} half a century 
later, after 1915 it was natural seek a unified field theory that would 
generalise the theory of gravity to incorporate electromagnetism, rather 
than the other way around.  
  The reason for this is perhaps that while the gravitational field 
equations are more complicated than those for electromagnetism the 
underlying motivation envisaged for Einstein's theory can be considered 
somewhat simpler. As a particularly elegant aspect of general relativity 
the \textit{assumption} of an extended globally flat 4-dimensional 
Minkowski spacetime is \textit{dropped}, with special relativity holding 
for all non-gravitational physics strictly only in the limit of 
infinitesimally small inertial reference frames by the 
 equivalence principle (\cite{Pais} chapter~9). Within such a local 
reference frame with local coordinates $\{x^a\}$ an infinitesimal proper 
time interval $\delta s$ can be expressed, in a form invariant under local 
$\soot$ Lorentz transformations between such frames, as: 
\begin{equation}
   (\delta s)^2 =  (\delta x^0)^2 - (\delta x^1)^2 - (\delta x^2)^2 -
    (\delta x^3)^2  \; = \; \eta_{ab}\delta x^a \delta x^b \label{sfourd} 
\end{equation}  
  with the Lorentz metric
  $\eta = \mbox{diag}(+1,-1,-1,-1)$ and $a,b = 0,1,2,3$ (the summation 
convention for repeated indices is assumed throughout this paper).

 In special relativity there exist sets of global coordinates
  $(x^0,x^1,x^2,x^3)$ in which equation~\ref{sfourd} holds for arbitrary 
finite spacetime intervals anywhere in the Minkowski spacetime, leaving the 
corresponding finite proper time interval $\Delta s$ invariant 
  under global Lorentz transformations between such reference frames.
   This framework of special relativity 
   was largely motivated through compatibility with Maxwell's equations, 
with the laws of electrodynamics taking the same form in any of these 
global inertial reference frames~\cite{Einsr}.

 In general relativity there are no such global coordinates and inertial 
frames, except in the flat spacetime limit.
  Instead general coordinates $\{x^{\mu}\}$, with \mbox{$\mu,\nu= 
0,1,2,3$}, must be defined on the global scale, with respect to which a 
local infinitesimal 
 proper time interval can be expressed in a manner invariant under general 
coordinate transformations as:
\begin{equation}
   (\delta s)^2  \; = \; g_{\mu\nu}(x)\delta x^{\mu} \delta x^{\nu}
    \label{gfourd} 
\end{equation}  
 which only locally reduces to equation~\ref{sfourd} in suitable local 
coordinates.
 The force of gravity is not present in such a local inertial reference 
frame but rather can be ascribed globally to the metric field 
$g_{\mu\nu}(x)$ which describes the curved geometry of the extended 
spacetime.
 Through a mutual dynamical interplay the spacetime geometry described by 
$g_{\mu\nu}(x)$ is related to the distribution of matter through Einstein's 
field equation (as will be discussed in subsection~\ref{uni51} for 
equation~\ref{Eineq}), with test particles postulated to propagate through 
spacetime along geodesic trajectories.
  Einstein's  theory of gravity, expressed directly in terms of the more 
general and flexible structure of a curved spacetime, surpassed that of 
Newton in accurately  accounting for gravitational observations such as the 
orbit of the planet Mercury. While Newton's law of universal gravitation 
strictly concerned mathematical relations, and was not based on any 
hypothesis regarding the cause of the force of gravity acting at a distance 
across space, for Einstein the curvature of spacetime also succeeded in 
furnishing an
  \textit{explanation} of gravitational phenomena on \textit{relaxing} the 
assumption of spacetime flatness.

   On the other hand for Maxwell's electromagnetism, initially formulated 
in a Newtonian background of absolute space and time, 3-component electric 
$\bE$ and magnetic $\bB$ fields were first \textit{added}, conceived of as 
mechanical states of an underlying `ethereal medium' filling all of space 
(\cite{Maxem} part~I, see also~\cite{Pais} section~6(a) and \cite{Pais2} 
section~12(a) part~1); with Maxwell's equations constructed on the basis of 
empirical observations, as Newton's law of gravity had been in the same 
background arena. 
  After 1905, with
 the theory of electromagnetism readily compatible with special relativity,
    Maxwell's equations could be expressed more succinctly in a Lorentz 
covariant form in terms of the antisymmetric electromagnetic field strength 
tensor
 $F_{\mu\nu}(\bE,\bB)$, and after 1915 could be \textit{accommodated} 
within the curved spacetime of general relativity via the equivalence 
principle. However, it was natural to enquire whether a further 
generalisation from general relativity might itself provide an 
\textit{explanation} of electromagnetism (which alongside gravity was then 
the only other fundamental force known) on relaxing further assumptions 
regarding the 4-dimensional spacetime metric geometry.

 The first proposal for such a unified field theory was made in 1918 by 
Weyl~(\cite{Weyl1}, \cite{Weyl2} chapter~IV section~35, \cite{ORaif} 
chapters 1--3) on dropping the assumption that the length of a 4-vector, 
determined by the metric $g_{\mu\nu}(x)$ of equation~\ref{gfourd}, should 
be path-independent when `parallel transported' in spacetime, an invariance 
which could be interpreted as a residual of rigid Euclidean geometry still 
remaining in general relativity. Hence, similarly as a 4-vector 
\textit{direction} is propagated in a path-dependent manner through a 
curved spacetime via a linear connection 
$\Gamma^{\rho}_{\ph{\rho}\mu\nu}(x)$, a unique function of the first 
derivatives of 
 $g_{\mu\nu}(x)$ in general relativity,
 Weyl introduced a vector field $A_{\mu}(x)$ to induce path-dependent 
changes to 4-vector \textit{magnitudes} -- with the metric 
$\bar{g}_{\mu\nu}(x)$ employed in forming inner products defined by the 
scaling:
\begin{equation}
 \label{WeylU}
      \bar{g}_{\mu\nu}(x) = \lambda\, g_{\mu\nu}(x)
	     \qquad \mbox{with} \qquad 
	  \lambda = e^{\int_{x_1}^{x_2} A_{\mu}(x)dx^{\mu}}
\end{equation}  

  The length of a 4-vector is path-independent under parallel transport
  between any two spacetime points $x_1$ and $x_2$   only when 
  the scale factor $\lambda$ is integrable. This is the case if the new 
connection field $A_{\mu}(x)$ can be expressed as the gradient of a 
continuous function $\chi(x)$, that is 
 $A_{\mu}(x) = \partial_{\mu}\chi(x)$, which in turn implies that the 
quantity defined by 
  $F_{\mu\nu}(x) = \pal_{\mu}A_{\nu}(x) - \pal_{\nu}A_{\mu}(x)$ vanishes.
    In the general case the field $F_{\mu\nu}(x)$ was identified with the 
electromagnetic field strength tensor and $A_{\mu}(x)$ with the 
corresponding vector potential, within normalisation factors, similarly as 
the Riemann curvature tensor together with  
$\Gamma^{\rho}_{\ph{\rho}\mu\nu}(x)$ and $g_{\mu\nu}(x)$ are associated 
with gravity in general relativity. 
In this manner Weyl inferred that on the 4-dimensional spacetime manifold 
`all physical field-phenomena are expressions of the metrics of the world'
 (\cite{Weyl2} chapter~IV section~35).
The theory hence demonstrated that electromagnetism could in principle be 
accorded such a geometrical significance.
  However, since scaling lengths via $\lambda$ in equation~\ref{WeylU} 
implies scaling time intervals also in equation~\ref{gfourd} Einstein 
immediately noted a fatal flaw of the theory -- the sharp lines of atomic 
spectra observed in the laboratory are not dependent upon the history of 
individual atoms.

  Weyl's theory did however provide the first step towards non-Riemannian 
connections and gauge theories, with the term `gauge' retained from the 
length \textit{scaling} factor in equation~\ref{WeylU}.
 In fact both $F_{\mu\nu}(x)$ and the `Action' defined for the theory are 
   `gauge-invariant' under arbitrary re-calibrations, that is under local 
changes of the adopted metric scale.
 Progress was achieved by 1929 on introducing a factor of $i=\sqrt{-1}$ in 
the exponent in equation~\ref{WeylU} and reinterpreting  $\lambda$ as a 
\textit{phase} factor to be applied instead to a complex wavefunction 
$\Psi(x)$ in the then recently invented quantum mechanics~(\cite{Weyl3}, 
\cite{Weyl4} chapter~II section~12, \cite{ORaif} chapters 4--5).
 Correspondingly 
 in~\cite{Weyl3} Weyl concludes with the assessment that `electromagnetism 
is an 
  accompanying phenomenon of the material wave-field and not of 
gravitation'. 
 Hence with the phase factor taking values in the symmetry group $\uo$
 electromagnetism  could be successfully described as a stand-alone $\uo$ 
gauge theory with a gauge field $A_{\mu}(x)$, rather than as a geometric 
augmentation to general relativity. 
 During the 1950s--1970s
 such gauge theories, with field interactions considered a consequence of 
the gauge-invariance of the equations,
  were developed and generalised beyond the $\uo$ gauge symmetry  of 
electromagnetism to non-Abelian gauge symmetries, ultimately incorporating 
electroweak and strong interactions also within the framework of the `gauge 
principle', essentially \textit{detached} from consideration of the 
geometry of external 4-dimensional spacetime -- which could be taken as the 
flat background of special relativity to a very good approximation in a 
laboratory setting. 

  The properties and representations of gauge symmetry groups are central 
to the modern-day structure of the Standard Model and unification schemes. 
It is well known for example that the branching patterns for the smaller 
non-trivial  representations of Lie groups such as SU(5), SO(10), $\esi$ 
and $\ese$ on extracting the Standard Model subgroup $\SML$ bear some 
resemblance to the gauge multiplet structure of leptons and 
quarks~(\cite{GeoGla}, \cite{FriMin} section~13, \cite{Gur1} and 
\cite{Gur7}  respectively) as the basis for a Grand Unified Theory (GUT).
 While the earlier unified field theories were based on generalisations of 
general relativity, for GUT models the focus is on particle physics with 
gravity being neglected and deferred for later consideration. However, in 
 the case of the Lie group $\ee$
 a symmetry breaking structure can be correlated with a full three 
generations of leptons and quarks incorporating also transformations under 
the external local spacetime Lorentz symmetry alongside the Standard Model 
gauge group~\cite{Lisi}.
 Nevertheless for each of the above Lie group structures the match with the 
symmetry properties of the Standard Model is incomplete and significant 
problems remain.
Further, while the three largest exceptional Lie groups $\esi$, $\ese$ and 
$\ee$ are of particular interest, owing to the high degree of symmetry they 
describe and the uniqueness of these mathematical structures, the nature of 
a  \textit{clear underlying conceptual origin}, whether geometric or 
otherwise, to motivate the application of these groups in particle physics 
remains an open question.

   Despite his rejection of Weyl's theory Einstein himself sought a unified 
field theory for gravity and electromagnetism based on generalisations of 
general relativity. 
 From 1925--1955, throughout the last 30 years of his life, Einstein worked 
on generalisations of 4-dimensional Riemannian geometry based in particular 
on dropping the assumption that the metric tensor ${g}_{\mu\nu}(x)$ and/or 
the linear connection  ${\Gamma}^{\rho}_{\ph{\rho}\mu\nu}(x)$ should be 
symmetric in the $\mu,\nu$ indices (\cite{Pais} section~17(e)).
 The most direct attempt  introduced a nonsymmetric fundamental tensor
 $\tilde{g}_{\mu\nu}(x)$  with a full 16 real components  which was 
proposed to decompose into symmetric $g_{\mu\nu}(x)$ and antisymmetric 
$\acute{g}_{\mu\nu}(x)$ parts as:
\begin{equation}
 \label{EinU}
      \tilde{g}_{\mu\nu}(x) \; = \; g_{\mu\nu}(x) \; 
	       [\mbox{gravitational field}]
	  \; + \;
      \acute{g}_{\mu\nu}(x) \; [\mbox{electromagnetic field}]
\end{equation}   

  For this scheme $g_{\mu\nu}(x)$ was retained as the original 
gravitational metric field while $\acute{g}_{\mu\nu}(x)$ was identified 
with the electromagnetic field strength tensor $F_{\mu\nu}(x)$, within a 
normalisation constant. Other attempts involved associating the 
electromagnetic vector potential $A_{\mu}(x)$ with components of a 
nonsymmetric linear connection.
 (In an independent application the study of linear connections
  with an antisymmetric part
   had been initiated by Cartan in 1922 in the geometric context of
 general relativity with finite torsion, later known as Einstein-Cartan 
theories).
While originally motivated by simplicity Einstein's unified field theory 
attempts  became increasingly elaborate, lacking the conceptual elegance of 
general relativity, and none of them led to the free Maxwell equations even 
in the weak-field approximation, 
  nor was there any prospect of incorporating nuclear forces into these 
schemes.
 During the same period, from around 1925, the mainstream physics community 
was also more focussed upon the developments of quantum theory, with the 
unification proposals of Einstein seemingly attracting more attention from
  \textit{The New York Times}~(\cite{Pais} section~17(e)).
  However, while our understanding of fundamental physics has continued to 
be dominated by quantum theory, the general spirit and motivation for 
Einstein's attempts at a unified field theory remains enlightening when 
transplanted into the context of the present-day quest for   
  unification, as will be discussed in subsection~\ref{uni51}.

Einstein had also been initially enthusiastic about the potential of 
Kaluza-Klein theory as also introduced in the 1920s~(\cite{Kaluza,Klein}, 
\cite{Pais} section~17(c)), upon which he worked intermittently himself 
over a number of years~(\cite{Pais} sections~17(c,e)).
In this approach to a unified field theory the assumption that spacetime 
should be limited to the familiar 4-dimensional arena of general relativity 
was dropped.
 With 4-dimensional spacetime augmented by an extra spatial dimension a 
$5\times 5$ metric $\hat{g}(x)$ could be defined on the extended spacetime
 subsuming the original  $4\times 4$ metric ${g}(x)$ of 
equation~\ref{gfourd}.
 In principle the four components of the electromagnetic vector potential 
$A(x)$ could then be accommodated inside the extended 5-dimensional metric:
\begin{equation}
 \label{KaKlU}
    \left[ \begin{array}{c} \\ \vspace{2pt} \\ \end{array} \;\;\, 
	       \hat{g} \;\;\, 
		   \begin{array}{c} \\ \vspace{2pt} \\ 
		           \end{array}\right]_{5\times 5}
		    = \;   
	\left[\!\!\!  \begin{array}{cc}  
	 \left[ \begin{array}{c}  \vspace{1pt} \\  \end{array} 
	    g \begin{array}{c}  \vspace{1pt} \\  \end{array}   \right] &
	     \!\!\!\!\!\! 
		 \left[ \begin{array}{c}  \vspace{1pt} \\  \end{array} 
	   \!\!\!\!\!\!   A   \!\!\!\!\!\!
		 \begin{array}{c}  \vspace{1pt} \\  \end{array}   \right]   \!\!\!
	  \vspace{-2pt}  \\
	    \left[ \;\;\, A^{\!\mbox{\tiny{\it T}}} \;\;\, \right]  & 
		  \!\!\!\!\!\! \phi
		     \!\!\!   \vspace{-2pt}  \end{array} \right]
\end{equation} 
 where the further new component $\phi(x)$,
  alongside the original metric $g(x)$ of general relativity,
  lacked any clear physical significance.
  Certain components of the 5-dimensional Levi-Civita linear connection  
$\hat{\Gamma}(x)$ could then be identified with the electromagnetic field 
strength \mbox{$F_{\mu\nu}(x) = \pal_{\mu}A_{\nu}(x) - 
\pal_{\nu}A_{\mu}(x)$} as a function of the components $A_{\mu}(x)$ in
 equation~\ref{KaKlU} in
 the appropriate way, on taking the field values to be independent of the 
fifth dimension.
 Maxwell's source-free equations for the electromagnetic field and the 
equation of motion for a charged body in an electromagnetic field could be 
obtained under suitable assumptions for the extraction of 4-dimensional 
physics from the embedding in the 5-dimensional spacetime framework. 
However, while hence providing 
 an element of formal geometric unification with general relativity, no 
predictive power or new phenomena could be determined and the question of 
the very different properties required for the fifth dimension remained.

   In the case of equation~\ref{WeylU} with a geometric scale factor 
$\lambda$ and that of equation~\ref{EinU} with a nonsymmetric metric 
$\tilde{g}_{\mu\nu}$ no further natural generalisation is possible, however 
for the case of equation~\ref{KaKlU} an arbitrary number of further extra 
spatial dimensions could in principle be considered. 
 Indeed, despite the lack of empirical support,
 this third means of augmenting the 4-dimensional spacetime structure has 
led in recent decades to a large number of unification models based upon 
various approaches to extra spatial dimensions, motivated in part by the 
elegance and unity of the Kaluza-Klein idea.

The realisation that the geometry of such augmented spacetimes could be 
adapted to incorporate the internal symmetries of non-Abelian gauge theory 
over 4-dimensional spacetime, with a close relation between gauge and 
coordinate transformations described explicitly on a `fibre bundle' 
manifold,  had 
  revived interest in this approach to unification by the 1970s. (See for 
example~\cite{Cho}, the mathematics of fibre bundles had been developed 
since 1935 for the field of topology in differential geometry \cite{Steen} 
independently of any application in physics).
 This framework `combined gravitation with gauge theory in the context of a 
unified geometric theory in the bundle space' (\cite{Cho} section~9) by 
employing an extended higher-dimensional metric defined on the full space.
 In this manner gauge theory, which had parted company from a geometric 
context in the 1920s as described above following equation~\ref{WeylU}, was 
placed in the setting of a higher-dimensional spacetime arena, and in 
particular \textit{reattached} to the geometry of an external 4-dimensional 
spacetime base manifold, in a unifying physical framework 
   that might in principle reach 
    beyond gravitation and electromagnetism alone.

	While the earliest attempts at a unified field theory may have been 
premature, 
 given the hindsight of the subsequent century of accumulated knowledge in 
particle physics, the quest since the 1970s to accommodate the rich 
properties of the Standard Model, or even a Grand Unified Theory, within 
the unifying framework of geometric structures deriving from extra spatial 
dimensions over \mbox{4-dimensional} spacetime continues, as will be 
discussed in the next subsection. 
  The search now includes the need not only to account for the Standard 
Model but also new physics, such as that of the neutrino sector reviewed in 
the previous subsection. In this paper we shall motivate and build a new 
unified theory from first principles with the potential to accommodate both 
the physics of the Standard Model and that beyond, including the possible 
feature of incorporating two, and only two, right-handed neutrinos 
alongside three generations of the other leptons and quarks. We begin by 
reassessing the motivation for employing extra spatial dimensions in the 
following section.

\section{Generalised Proper Time}
\label{uni2}

\subsection{Extra Spatial Dimensions and the Standard Model}
\label{uni21}

   Rather than considering generalisations of the global metric 
$g_{\mu\nu}(x)$ of equation~\ref{gfourd} on an extended higher-dimensional 
spacetime manifold here we focus upon the \textit{local} metric $\eta_{ab}$ 
of equation~\ref{sfourd} associated with a local inertial reference frame.
   At this most elementary level of purely local structure
   theories with extra \textit{spatial} dimensions extend the metric 
geometry of 4-dimensional spacetime, with local coordinates 
$(x^0,x^1,x^2,x^3)$, 
  augmenting the quadratic expression for the proper time interval 
    $\delta s$ of equation~\ref{sfourd} to the $n$-dimensional form:
\begin{equation}
   (\delta s)^2  = (\delta x^0)^2 - (\delta x^1)^2
       - (\delta x^2)^2 - (\delta x^3)^2
   \, - \, (\delta x^4)^2 \ldots\ldots - (\delta x^{n-1})^2 = 
             \hat{\eta}_{ab}\delta x^a \delta x^b  \;
  \label{snd}
\end{equation} 
  where $(x^4, \ldots, x^{n-1})$ are $(n-4)$ extra dimensions, 
  $\hat{\eta} = \mbox{diag}(+1,-1, \ldots, -1)$ is the extended local 
Lorentz metric and $a,b = 0,\ldots,(n-1)$.
The additional components $(x^4,\ldots,x^{n-1})$ are considered extra 
`spatial' dimensions owing to the quadratic structure and the minus signs 
in equation~\ref{snd}, sharing these properties with the three original 
spatial dimensions given the Lorentz metric signature convention of 
equation~\ref{sfourd}.
 On dividing both sides by $(\delta s)^2$ 
   and defining the components $v^a = \frac{\delta x^a}{\delta s}$ on 
taking the limit 
    $\delta s \to 0$
   the above expression can be written as:
\begin{equation}
  \vert \bv_n \vert^2  := \;\! (v^0)^2 - (v^1)^2 - (v^2)^2 - (v^3)^2
   \, - \, (v^4)^2 \ldots\ldots - (v^{n-1})^2 \;\!  = \, 
             \hat{\eta}_{ab}v^a v^b  \, = \, 1
  \label{vnd}
\end{equation} 
  in terms of the components  of the `$n$-velocity' vector
    $\bv_n = (v^0, \ldots, v^{n-1}) \in \rrr^n$.
	The quantities $\delta s$ and $\vert \bv_n \vert$ in 
equations~\ref{snd} and \ref{vnd} respectively are invariant under 
 $\sootnm$ transformations applied to these $n$-component expressions.

  The simplest and most direct means of constructing a physical theory 
based on this structure is to assume the identification of the four 
components $(x^0,x^1,x^2,x^3)$ in equation~\ref{snd} with a set of local 
coordinates on an external spacetime $M_4$,
 without necessarily specifying a mechanism for distinguishing this 
extended manifold itself in four, and only four, preferred dimensions.
 The first four components $\bv_4 = (v^0,v^1,v^2,v^3)$ of 
equation~\ref{vnd} are correspondingly projected 
 onto the local tangent space of the extended 4-dimensional spacetime, with 
$\bv_4 \in \TM_4$, upon which a preferred local \textit{external} $\soot 
\subset \sootnm$ symmetry acts.
 This breaks the 
  full $n$-dimensional $\sootnm$ Lorentz symmetry of equation~\ref{vnd} and 
on taking the residual components of that equation to form the basis for 
`matter fields' \textit{in} the extended spacetime we directly deduce the 
following symmetry breaking structure:
\begin{eqnarray}
        \sootnm & \to &  \soot \; \times \;  \mbox{SO}(n-4)
  \quad\;\!   \mbox{: external $\times$ internal$\;\;$}  \label{sosb} \\
	    \bv_4 \in \rrr^4  \:\, & : &  \,  \mbox{4-vector} 
		 \qquad\quad \!\!\! 
		     \mbox{invariant}   \qquad\!\!\! 
			  \mbox{: tangent vector in $\TM_4\;\;\;\;$}   \nonumber  \\
 \raisebox{0pt}[0pt][0pt]{ {\raisebox{+1.7ex}{$\!\!\bv_n \in 
    \rrr^n \!\to\! 
  \left\{\! \begin{array}{c}  \\ \vspace{-12pt}  \end{array}  \right. 
   \!\!\!\!\!\! $}} }		 
	 \underline{\bv}_{n-4} \in \rrr^{n-4} \!\!\!\! 
	      & : & \;\:\,  \mbox{scalar} 
	    \qquad\;\!\!\!  
		   (n-4)\mbox{-vector} \quad\hspace{-0.3pt}\!\!\!  
		    \mbox{: matter field over $M_4$}   
	     \label{vnbits}
\end{eqnarray}   
  
  This direct generalisation from the structure of a strictly 4-dimensional 
proper time interval of equation~\ref{sfourd} is depicted in 
figure~\ref{vfdvnd}.

\begin{figure}[htbp]  
\centering
\epsfxsize=14.3cm
\leavevmode
\epsffile[0 0 1631 451]{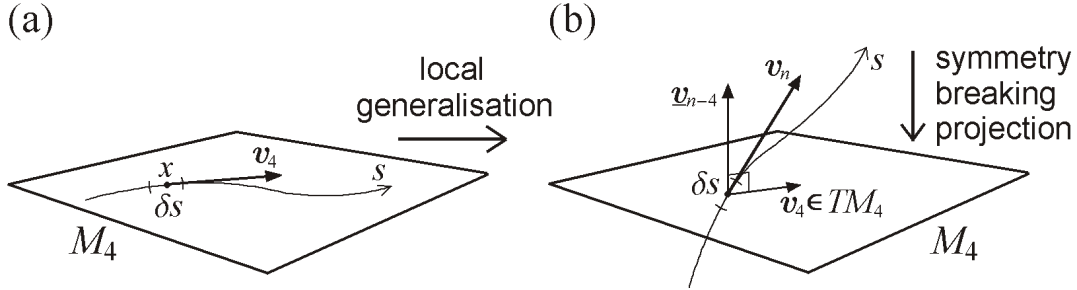}
\caption{\setb  (a) A 4-vector field $\bv_4(x)$ can be 
 constructed from locally Lorentz invariant proper time intervals $\delta 
s$ at each point $x \in M_4$  and (b) augmented for time $s$ propagating 
through a higher-dimensional spacetime, with the corresponding local values 
of $\bv_n(x)$ projected onto the external 4-dimensional spacetime $M_4$ 
here represented by a plane.}
\label{vfdvnd}
\end{figure}  

   We note that although the projection takes place locally on the external 
manifold $M_4$ the action on $\underline{\bv}_{n-4}(x)$ of the full 
\textit{internal} symmetry $G=\mbox{SO}(n-4)$ implies the incorporation of 
this complete gauge group manifold in a `trivial principal fibre bundle' 
structure 
 $P\equiv M_4 \times G$, with `vertical' fibres $G$ attached over each 
point of the `base space' $M_4$ of figure~\ref{vfdvnd}(b).
   (The case for $n=10$ in equation~\ref{sosb}--\ref{vnbits} is depicted in 
  \cite{Novel} figure~1(b) while the relation of this construction to the 
geometry of non-Abelian Kaluza-Klein theories is described 
in~\cite{KKone}). 
  However in this simple picture the matter field 
$\underline{\bv}_{n-4}(x)$ of equation~\ref{sosb}--\ref{vnbits}
 and figure~\ref{vfdvnd}(b) in spacetime $M_4$, as a Lorentz scalar that 
transforms under the $(n-4)$-dimensional vector 
 representation of the residual internal gauge symmetry SO$(n-4)$, does not 
remotely resemble structures of the Standard Model of particle physics. 
Adding further spatial dimensions simply increases $n$ in 
equations~\ref{vnd}--\ref{vnbits} and clearly does not help this situation 
even for models with $n \to \infty$, which might be considered in the 
absence of a natural limit, constituting an extreme case of a `waste of 
space' given the lack of any apparent empirical connection.
  
  A more sophisticated approach is clearly needed for any attempt to 
accommodate the rich properties of the Standard Model within a geometric 
framework deriving from extra spatial dimensions while maintaining a 
reasonably economical level of structure and assumptions. 
 The most direct possibility might be to consider the 14-dimensional 
spacetime case for equation~\ref{sosb}--\ref{vnbits} resulting in a
 $G=\mbox{SO}(10)$ internal symmetry that might in principle be connected 
with a corresponding GUT model incorporating the Standard Model gauge group
 $\SML \subset \mbox{SO}(10)$ (see for example~\cite{Jitt}).
 Adopting a different scheme in
 1981 Witten~\cite{Witt} utilised supergravity in an 11-dimensional 
spacetime, with the seven extra spatial dimensions  `spontaneously 
compactified' over the external 4-dimensional spacetime,
  as a potential
 framework for the unification of the gauge fields of the Standard Model 
with gravity. However a major obstacle is encountered in incorporating the 
appropriate quantum numbers for the leptons and quarks.  
   Particularly challenging more generally is the ambition to incorporate 
the Standard Model in a seemingly natural and unique manner, with the 
search ongoing given the absence of any compelling success.

The question of uniqueness becomes more acute 
for the most technically sophisticated approach via the extra spatial 
dimensions incorporated into string theory.
String theory was primarily motivated in the early 1970s as a candidate for 
a quantum theory of gravity~\cite{Schw}. This provided an independent 
motivation for the introduction of extra spatial dimensions which were 
required in order to obtain a consistent string theory. The original 
bosonic string is only consistent in a 26-dimensional spacetime while 
$n=10$ is the critical dimension for the fermionic string (see 
also~\cite{DuNiPo} section~1.2).
  A major breakthrough came in 1984 with the demonstration that 
  `type I string theory' is finite and free of anomalies for the gauge 
group SO(32), with support then growing for string theory as a promising 
framework for a unification incorporating particle physics as well as 
quantum gravity.
 The observation that the anomaly cancellation property is shared by the 
Lie group $\ee \times \ee$ motivated the construction of `heterotic string 
theory'~\cite{Gross}, combining features of both bosonic and fermionic 
string theory and incorporating an $\ee \times \ee$ gauge group over 
4-dimensional spacetime for the low energy effective theory. This framework 
has been favoured in attempts to connect string theory with the Standard 
Model via a GUT scheme associated with one of the $\ee$ factors. 
 By 1985 superstrings had become a mainstream activity and a total of five 
separate consistent theories (type I, type IIA and IIB, heterotic SO(32) 
and $\ee \times \ee$) had been described; hence also with an element of 
concern over the uniqueness of the theory given these five branches.

  During the above `first superstring revolution' of the mid-1980s Witten 
also became a proponent and played a key role in the `second superstring 
revolution' of the mid-1990s. 
   Marking the latter revolution the five different known types of string 
theory were shown to be interrelated by dualities, or equivalences, and  
subsumed under a single framework with each obtained as a different limit 
of an 11-dimensional `M-theory' (see for example~\cite{Duff}). The 
fundamental objects of M-theory include extended higher-dimensional 
entities called `branes' as well as the original one-dimensional `strings'. 
Combining the five branches of string theory in this way tentatively 
offered some hope  for demonstrating the uniqueness of the theory as a 
unification scheme. 
 However this sophisticated framework provides far from a minimal approach 
to the question of accounting for the Standard Model within a structure of 
extra dimensions.
 Indeed the theory is confronted by the `landscape problem' on attempting 
to deduce a realistic vacuum solution resembling properties of the Standard 
Model of particle physics and observational cosmology out of a vast array 
of possibilities~\cite{Doug,Blum}.
   Resorting to an `anthropic principle' argument to identify our world out 
of a `multiverse' of an estimated $10^{500}$ or more possible string vacua,
  a range of which may be consistent with our world,
  seems barely preferable to positing the properties of the Standard Model 
as `brute facts' as alluded to in the opening of section~\ref{uni1}.

In this paper we describe a more explicit and potentially unique means of 
uncovering familiar features of the Standard Model through a new 
fundamental theory.
 As we have described in subsection~\ref{uni12}
   both gauge theories and extra spatial dimensions have their roots in 
attempts to generalise general relativity dating from a hundred years ago.
 At that time although a relatively modest generalisation was required to 
incorporate  solely electromagnetism alongside gravity a range of possible 
approaches were conceived for example by Weyl, Einstein and Kaluza/Klein as 
reviewed for equations~\ref{WeylU}, \ref{EinU} and \ref{KaKlU} 
respectively.
 Given the fruitfulness and influence of that period in shaping modern-day 
theories of unification, and now with the benefit of hindsight regarding 
both developments in the mathematical description of pertinent symmetry 
structures and the wealth of
empirical data as embodied in the Standard Model, we might reconsider 
whether there is another possibility concerning a broader generalisation 
from equation~\ref{sfourd} or \ref{gfourd} at an elementary level; one that 
might provide more direct access to the structures of particle physics.

\subsection{Motivation for Extra Dimensions Reassessed}
\label{uni22}

  In some models extra spatial dimensions are taken to be infinitely 
extended, with our own 4-dimensional spacetime `brane' world confined to a 
hypersurface in a larger $n$-dimensional spacetime `bulk' (see for 
example~\cite{Ponce,Liu}). More typically the extra spatial dimensions of 
the bulk space are curled up or compactified on a very small scale ranging 
from of order 0.1$\,$mm, if only gravity propagates in the extra 
dimensions, down to 
 the Planck length, accounting explicitly for our inability to observe them 
(see for example~\cite{Csaki} section~2, \cite{Shif}). 

 However, since we neither perceive nor navigate around the extra 
dimensions there is no compelling argument for the additional components to 
be either extended on a \textit{global} scale, as a higher-dimensional 
generalisation of 
 equation~\ref{gfourd}, \textit{or} even to possess the \textit{local} 
structure of the extra components
 $(\delta x^4, \ldots, \delta x^{n-1})$  in equation~\ref{snd} as a 
\textit{quadratic} extension to the local 4-dimensional spacetime form of 
equation~\ref{sfourd}.
  That is, with the minus signs in equation~\ref{snd} adopted from the 
Lorentz metric signature convention, the extra components  have the 
`spatial' property of     
 adding quadratically to form local `lengths' $\delta \Sigma$, with for 
example:
\begin{equation} 
 \label{pythag}
   (\delta\Sigma)^2 \, = \, (\delta x^4)^2 \, + \, (\delta x^{n-1})^2
\end{equation}   
    which via the Pythagorean theorem describes a right-angled triangle 
structure as a basis for a local Euclidean spatial geometry. 
 While this property is \textit{required} for the
  components $(\delta x^1, \delta x^2, \delta x^3)$ of the
  \textit{external} space dimensions  of the world we inhabit  the `extra 
dimensions' are not observed and there seems no essential reason to 
\textit{restrict} the extra components $(\delta x^4, \ldots, \delta 
x^{n-1})$ to also possess this locally Euclidean geometric property.

  This unnecessary restriction seems all the more artificial on considering
  large $n$, or even on taking
   the limit $n \to \infty$, since then \textit{almost all} of the 
components on the right-hand side of equation~\ref{snd} are not required to 
be of a quadratic `spatial' form as the $\{\delta x^a\}$ for all $a>3$ do 
not represent a physical perceived space. 
 However even for this dizzying contemplation of opening up more and more 
extra dimensions and taking $n \to \infty$  the left-hand side of 
equation~\ref{snd} still describes a simple robust interval of proper time 
$\delta s \inn \rrr$, now invariant under 
 SO$^+(1,$`$\infty - 1$') transformations applied to the components on the 
right-hand side. This invariant $\delta s$ is hence pivotal in threading 
together all of the components on the right-hand side and in defining this 
structure, and on shifting our focus to the left-hand side we can in fact 
interpret equation~\ref{snd} as representing a possible arithmetic 
expression 
\textit{for} a real proper time interval $\delta s \in \rrr$. 
 We can then ask what further arithmetic possibilities there may be.

  As an invariant entity 
 proper `time' is in itself something that might be objectively measured, 
as recorded by the readings of a physical clock. Arbitrary intervals of 
time are normally conceived of as an additive linear progression, with 
seconds contained within minutes contained within hours and so on.
 This will be the case for the proper time recorded by the `tick-tock' of a 
pocket watch carried by a pedestrian standing in a street with
 local rest frame  spacetime coordinates $(x^0,x^1,x^2,x^3)$ aligned with 
the local neighbourhood street plan. For the stationary watch an interval 
of proper time $\delta s$, here considered infinitesimal or finite,  can be 
expressed directly as $\delta s = \delta x^0$, preserving the simple linear 
structure. As the pedestrian walks down the street along the $x^1$ 
direction the \textit{same} proper time interval for the watch will be 
expressed in the \textit{quadratic} form $(\delta s)^2 = (\delta x^0)^2 - 
(\delta x^1)^2$ with respect to the local coordinates
(albeit with a walking velocity not too small compared with light speed 
needed for a significant 
 $\delta x^1$  contribution).
 On turning left or right the $\delta x^2$ coordinate will similarly 
augment this expression and upon entering a building and climbing the 
stairs the vertical $\delta x^3$ component will complete the full 
4-dimensional quadratic spacetime expression of equation~\ref{sfourd}.
 The central feature is that the watch itself continues untroubled in 
measuring the invariant `tick-tock' of proper time, with the same 
observation applying hypothetically for the addition of extra spatial 
coordinates in equation~\ref{snd} -- along a trajectory no longer confined 
to 4-dimensional spacetime as represented in figure~\ref{vfdvnd}(b).

  Alternatively, from the original stationary position of the pedestrian,
  recording the linear progression of proper time intervals $\delta s$, for 
 a passive  Lorentz boost in \mbox{4-dimensional} spacetime  to the 
perspective of another local frame in uniform relative motion (which can 
readily approach the speed of light), and with the local coordinates 
$(x^0,x^1,x^2,x^3)$ now assigned to the new frame, the \textit{same} time 
interval $\delta s$ will again be expressed in the \textit{quadratic} form 
of equation~\ref{sfourd} from the new perspective. 
  The four components $\{\delta x^a\}$ for $a = 0,1,2,3$ on the right-hand 
side of equation~\ref{sfourd} are unphysical in the sense that, while 
representing the local external 4-dimensional spacetime geometry, they are 
arbitrary within such local $\soot$ Lorentz transformations.
Similarly all $n$ local components on the right-hand side of 
equation~\ref{snd} are unphysical  in that they depend upon the choice of 
arbitrary $\sootnm$ transformations. These transformations however leave 
the left-hand side invariant.
 We might then consider equation~\ref{snd} to represent a possible 
generalisation of equation~\ref{sfourd} with both interpreted as possible 
expressions for a proper time interval, that is the one objectively 
measurable quantity in these equations, which \textit{can} be 
arithmetically expressed in such a quadratic form, and hence accorded a 
 corresponding geometric spatial interpretation.

 Given then that we can equate proper time with a non-linear quadratic 
structure for the 4-dimensional external spacetime arena that we 
\textit{do} perceive, 
 we might also consider augmentations to more general higher-order 
homogeneous polynomial forms
 that may be utilised by `extra dimensions' that we \textit{do not} observe 
in a geometrical sense.
  This can be achieved by exploiting the basic arithmetic properties of the 
real numbers to obtain expressions for $\delta s \in \rrr$, with this 
infinitesimal proper time interval invariant under a full symmetry group 
$\hG$ that generalises the local Lorentz transformations.
 Indeed expressions can be written down for $(\delta s)$,
  $(\delta s)^2$, $(\delta s)^3,\ldots$, or $(\delta s)^p$ in general for 
any power $p=1,2,3,\ldots$, of which equation~\ref{snd} represents a 
particular case for $p=2$.
  Expressions of quadratic order with $p=2$ are of significance for 
directly identifying components with `spatial' properties, as noted for 
equation~\ref{pythag} and as \textit{needed} for, and \textit{only} for, 
external 4-dimensional spacetime.
 Hence from the perspective of local proper time on the left-hand side, and 
the extra components on the right-hand side, equation~\ref{snd} can be 
generalised to a $p^{\mathrm{th}}$-order homogeneous polynomial expression, 
for $p=1,2,3,\ldots$,
  in $n$ components $\{\delta x^a\}$ with each of $a,b,c,\ldots = 0, 
\ldots, n-1$:
\begin{equation}
 \label{salpha}
  (\delta s)^p  \; = \; \alpha_{abc\ldots}\delta x^a 
                            \delta x^b \delta x^c \ldots
    \quad \mbox{with each} \quad \alpha_{abc\ldots} \inn \{-1,0,1\}
\end{equation}
   \textit{provided} we can extract a specific 4-dimensional quadratic 
substructure
  in four components $(\delta x^0, \delta x^1,\delta x^2, \delta x^3 )$, in 
the form of the right-hand side of equation~\ref{sfourd}, as required
       to represent the local geometric structure of the external spacetime 
$M_4$. 
	   That is, we require that equation~\ref{salpha} can in general be 
written in the form:
 \begin{equation}
 \label{sfourp}
  (\delta s)^p  \; = \; \left[
    \eta_{ab}\delta x^a \delta x^b \right]  
	    (\delta x^4, \ldots ,\delta x^{n-1})^{p-2}
   \; + \;  (\delta x^0, \ldots ,\delta x^{n-1})^{p} 
\end{equation}
  where here in the first term $a,b = 0,1,2,3$ in the first factor and the 
second factor represents a $(p-2)^{\mathrm{th}}$-order polynomial in the 
remaining $(n-4)$ components, while the second term, in all components, 
represents the further $p^{\mathrm{th}}$-order polynomial contributions to 
equation~\ref{salpha}.
This expression clearly generalises the 4-dimensional form for proper time 
in equation~\ref{sfourd} and also reduces to the quadratic form of 
equation~\ref{snd} as a special case, now interpreted as a possible form of 
proper time itself. 

   The sense of a linear `one-dimensional' progression in proper time is 
something we are intimately familiar with.   
  With regards to spatial constructions we can also readily conceive in our 
mind's eye of a one-dimensional straight line. In this case we can picture 
a second dimension adjoined by a right angle to the first, and in turn a 
third spatial dimension adjoined at right angles to each of the first two, 
with each pair forming a basis for the two quadratically added components 
of the Pythagorean theorem.  
 Here the progression ends in terms of our ability to picture such a 
geometric structure with a fourth or more spatial dimension, as does our 
ability to physically perceive or explore such a space given the 
3-dimensional world we inhabit, as described above for the pedestrian 
exploring the neighbourhood streets.

However we can gain some handle on the properties of a fourth dimension of 
space and beyond through a purely mathematical augmentation, by 
incorporating further components into the Pythagorean theorem as for the 
$(\delta x^4)^2$ term and beyond in equations~\ref{snd} and \ref{pythag}. 
This is clearly a mathematical possibility, however since in generalising 
beyond a 3-dimensional space we are \textit{compelled} to employ a 
mathematical extrapolation we should consider what the limits are in a 
purely \textit{algebraic}, rather than geometric, sense. For the case of 
generalising the 4-dimensional spacetime structure of equation~\ref{sfourd} 
this leads beyond the extra spatial dimensions in the quadratic form of 
equation~\ref{snd} to the more general algebraic expression of 
equation~\ref{salpha} which is then open to \textit{mathematical} 
exploration. In this case we can still extract a local 3-dimensional 
spatial structure,  as an integral part of the external 4-dimensional 
spacetime factor in equation~\ref{sfourp}, forming the basis of the locally 
Euclidean world that we \textit{do} physically perceive.

  Essentially we have abstracted the arithmetic composition of 
equation~\ref{sfourd} away from the context of a local inertial reference 
frame and temporarily neglected the Pythagorean
 geometric significance of the quadratic expression on the right-hand side.
  This \textit{initial} arithmetic argument in focussing upon the possible 
mathematical forms for a proper time interval  $\delta s \inn \rrr$ as the 
chief guide is somewhat disorienting in that the prominence of the 
geometric structure of the spacetime background has melted away. From the 
point of view of the flow of time, which is generally conceived of as a 
linear progression, the cubic and higher-order homogeneous polynomial 
expressions for $\delta s$ implied in equation~\ref{salpha}
 are just as mathematically permissible and no stranger than the quadratic 
forms of equations~\ref{sfourd} and \ref{snd}.
 From this more abstract perspective  equation~\ref{sfourd} is considered 
to represent a possible arithmetic composition for an infinitesimal 
interval of proper time $\delta s \in \rrr$ on the left-hand side that 
directly generalises to equation~\ref{salpha}, with  
  $ (\delta s)^p$ invariant under a full symmetry group $\hat{G}$ as a 
generalisation of the Lorentz group $\soot$ acting on the right-hand side 
components.
 We \textit{then} regain our spacetime orientation by extracting out from 
equation~\ref{salpha} a 4-dimensional quadratic substructure, as described 
for equation~\ref{sfourp}, with a $\mbox{Lorentz} \subset \hG$ symmetry as 
a necessary geometric basis for the required external 4-dimensional 
spacetime arena.

 The underlying shift in focus is towards the continuum of proper time as 
the objectively measurable quantity in these expressions. 
 The form of equation~\ref{salpha}, potentially involving cubic or 
higher-order homogeneous compositions, is not problematic for the extra 
dimensional structures provided that we \textit{can} extract the 
4-dimensional quadratic spacetime form of equation~\ref{sfourd}, which 
underlies  the visible external geometry of physical 3-dimensional space.
 However, as described in equation~\ref{sfourp} and as will be explicitly 
demonstrated in the following subsection, we can readily embed the 
quadratic spacetime structure of equation~\ref{sfourd} within specific 
higher-order cases for equation~\ref{salpha}, just as legitimately as we 
can within equation~\ref{snd}. Hence the \textit{assumption} that 
generalisations from the metric structure of equation~\ref{sfourd} should 
be \textit{limited} to quadratic forms can be dropped.

 This generalisation to equation~\ref{salpha}, involving the relaxing of an 
assumption in augmenting the 4-dimensional spacetime metric form, is in 
this sense proposed in a similar spirit as for the earliest unified field 
theories reviewed in subsection~\ref{uni12}.
  In the present case
 the basis is even simpler in that we focus upon generalising
 the expression for a proper time interval $\delta s$ in a local inertial 
reference frame with the
  local metric $\eta_{ab}$ in equation~\ref{sfourd} to that with the 
coefficients $\alpha_{abc\ldots}$ in equation~\ref{salpha},
   and hence begin with a more elementary structure than the global metric 
$g_{\mu\nu}(x)$ of equation~\ref{gfourd} of general relativity in the 
extended 4-dimensional spacetime manifold as incorporated into
 equations~\ref{WeylU}--\ref{KaKlU}.
 While the metric $g_{\mu\nu}(x)$ within equations~\ref{WeylU}--\ref{KaKlU}
 locally reduces to the Lorentz metric $\eta_{ab}$ in appropriate local 
coordinates, the Lorentz metric $\eta_{ab}$ extracted here via 
equation~\ref{sfourp} will be locally equivalent to the metric  
$g_{\mu\nu}(x)$ in such local inertial reference frames in 4-dimensional 
spacetime.
 This contrasting perspective will be discussed further in 
subsection~\ref{uni51} in particular in relation to figure~\ref{grcfme}.

 In order to establish a convenient notation and avoid expressions with 
infinitesimal elements, and similarly as equation~\ref{snd} generalises to 
equation~\ref{salpha}, we can in turn generalise equation~\ref{vnd} by 
again defining an $n$-vector $\bv_n$ with the generally finite components 
$v^a = \frac{\delta x^a}{\delta s} 
          {\big{\vert}}_{\mbox {\tiny $\delta s \! \to \! 0$}}$, 
 and on dividing both sides of equation~\ref{salpha} by $(\delta s)^p$ we 
define:
\begin{equation}
  \label{lpvn}
  L_p(\bv_n)_{\hat{G}} 
  \; := \; 
   \alpha_{abc\ldots} \frac{\delta x^a \delta x^b \delta x^c \ldots} 
                           {\delta s \;\;\delta s \;\;\delta s \,\ldots}
						  \Big\vert_{\delta s \to 0}   
	\;  = \; 	
    \alpha_{abc\ldots}v^a v^b v^c \ldots \; = \; 1
\end{equation}
 with each of $a,b,c,\ldots = 0, \ldots, n-1$ and each $\alpha_{abc\ldots} 
\inn \{-1,0,1\}$,
 while  
 the equality with unity on the right-hand side, via equation~\ref{salpha}, 
is simply from $\frac{(\delta s)^p}{(\delta s)^p} = 1$.  In this equation 
$L_p$ for $p=1,2,3,\ldots$ denotes a $p^{\mathrm{th}}$-order homogeneous 
polynomial expression in the \mbox{$n$ components} of $\bv_n$ with full 
symmetry group $\hat{G}$. (Any of the subscripts in this expression may be 
dropped if their value is implied from the context, see also the discussion 
in~\cite{Novel} between equations~11 and 13 there, although generally this 
notation will be manifestly unambiguous in this paper).
 While the underlying simple conceptual basis for this theory in terms of 
generalised proper time is readily made explicit in equation~\ref{salpha}, 
the equivalent expression in equation~\ref{lpvn} provides a convenient 
notation as a basis for the explicit mathematical analysis and physical 
interpretation of the theory.  
The kernel symbol `$L$' in equation~\ref{lpvn} originates from a 
consideration of
 \mbox{$p^{\mathrm{th}}$-order} multi$L$inear forms that might generalise 
the 
 bilinear metric forms of equations~\ref{sfourd}, \ref{snd} and \ref{vnd}, 
while also having a connection with the role of a conventional $L$agrangian 
in field theory as will be described in the following subsection.

The symmetry breaking identification of the subcomponents 
$(x^0,x^1,x^2,x^3)$ of equation~\ref{sfourp} with a set of local
 coordinates and the local
 geometric structure of the external spacetime $M_4$ now
 corresponds to the projection of 
  the subcomponents \mbox{$\bv_4 = (v^0,v^1,v^2,v^3) \inn \TM_4$} out of 
equation~\ref{lpvn} onto the external tangent space, similarly as described 
for equations~\ref{vnd}--\ref{vnbits}. 
 Indeed equation~\ref{vnd} represents a special case of equation~\ref{lpvn} 
with:
\begin{equation} 
\label{l2vn}
 L_2(\bv_n)_{\mathrm{SO}^+(1,n-1)} \, = \, \vert \bv_n \vert^2 \, = \, 
    \hat{\eta}_{ab}v^a v^b \, = \, 1
\end{equation} 
 while equation~\ref{lpvn} allows generalisation for $p>2$ beyond such 
quadratic spacetime structures. While the case of equation~\ref{l2vn} can 
be `visualised' through a direct lower-dimensional analogy in 
figure~\ref{vfdvnd}(b) (with the projection of the $n$-dimensional vector 
$\bv_n$ over 4-dimensional spacetime $M_4$ for this pseudo-Euclidean case 
depicted as a projection from a 3-dimensional Euclidean space over an 
embedded 2-dimensional plane), the general form of equation~\ref{lpvn} 
cannot be pictured at all in such a geometrical manner. 
 
  In fact 
the necessary extraction of a \textit{quadratic} substructure, to match the 
geometry of the locally Euclidean 3-dimensional spatial arena incorporated 
within the locally
 pseudo-Euclidean 4-dimensional external spacetime background
  against which all physical phenomena are observed, 
  from a \textit{cubic or higher-order}  form for equation~\ref{lpvn} might 
also be interpreted as a central feature of the \textit{mechanism} for the 
symmetry breaking itself, unlike for the uniformly quadratic expression of 
equation~\ref{vnd} or \ref{l2vn}. 
  However the explicit connection with non-Abelian   
 Kaluza-Klein theories for models with extra spatial dimensions,
  as alluded to after figure~\ref{vfdvnd} with reference to~\cite{KKone}, 
remains the same and
  hinges upon the limit of the local structure in which equation~\ref{vnd} 
is a particular case of equation~\ref{lpvn}, as will also be discussed for 
equation~\ref{gchift} in the following subsection.

	 The expression in equation~\ref{lpvn}, equivalent to 
equation~\ref{salpha}, represents the `general form of proper time', as 
distinct from a `spacetime form', emphasising the simple interpretation of 
this theory as deriving directly from the basic arithmetic substructure of 
an infinitesimal interval $\delta s$ of the continuum of proper time alone.
The adjective `proper' essentially refers to the invariance of the time 
interval $\delta s$ under symmetry transformations that can be applied to 
the subcomponents in equation~\ref{salpha} or \ref{lpvn}.
 Via \mbox{equations~\ref{salpha}--\ref{lpvn}} expressions for proper time 
can incorporate the geometric structure of 4-dimensional spacetime as well 
as the physical structures of matter \textit{in} spacetime associated with 
the residual components.
 While this perspective may be unfamiliar
  the new theory has a very simple and conservative interpretation in being 
founded upon the underlying flow of time which we do intimately perceive 
rather than upon the fashionable hypothesis of extra spatial dimensions, 
over and above a 4-dimensional spacetime background, which we do not 
discern at all. 
  For the present theory there are no extra spatial dimensions of a `bulk 
space' to be compactified or otherwise hidden from direct observation, as 
alluded to in the opening of this subsection, rather the properties of the 
additional components in equation~\ref{lpvn} over and above those of
 \mbox{4-dimensional} spacetime are interpreted directly as matter fields.

  Despite this underlying simplicity,
in generalising from equation~\ref{vnd} to equation~\ref{lpvn} on dropping 
the assumption of a local quadratic $p=2$ spatial form for the extra 
components, we now have a seemingly more complicated structure with the 
potential in principle for both $p \to \infty$ and $n \to \infty$, while 
subsuming the $p=2$ and $n=4$ case for equations~\ref{lpvn} and \ref{l2vn} 
for the external 4-dimensional spacetime.
  For the $p=2$ case of equations~\ref{vnd} and \ref{l2vn} any number of 
dimensions through to $n \to \infty$ can be considered, as discussed 
following equation~\ref{pythag}, although particular structures for 
$n$-dimensional spacetime are singled out in \textit{the context of} 
sophisticated theoretical frameworks that employ extra spatial dimensions, 
such as with $n=11$ for supergravity and $n=26$ or $n=10$ for string theory 
as reviewed in the previous subsection. 

   However for $p>2$, as we consider for example possible cubic and quartic 
forms for equation~\ref{lpvn}, particular values for $p$ and $n$ will be 
\textit{intrinsically} preferred as unique mathematical structures which 
possess a high degree of symmetry, while supplanting equation~\ref{sfourd}, 
will be highlighted. In this sense the progression from `spacetime forms' 
to `forms of proper time' is both more general and yet more restrictive, 
and in a manner that will lead to well-known unification symmetry groups as 
we shall describe in section~\ref{uni3}.  
 By comparison with the elementary analysis for the extra spatial 
dimensions in equations~\ref{vnd}--\ref{vnbits}  now applied for the 
generalised form of proper time of equation~\ref{lpvn} 
  the question can then be addressed regarding the form of matter fields 
over 4-dimensional spacetime that can be deduced for this theory in 
practice. In the following subsection we first consider the features and  
consequences of a minimal non-trivial generalisation from the form of 
proper time of equation~\ref{sfourd} in the manner of 
equations~\ref{salpha} and \ref{lpvn}.

\subsection{Minimal Cubic Form of Proper Time}
\label{uni23}

  A source of homogeneous $p^{\mathrm{th}}$-order polynomial forms for
equations~\ref{salpha} and \ref{lpvn} which exhibit a high degree of 
symmetry between the contributions of each component is found in the 
determinant function for $p\times p$ matrices. With the matrix composition 
property $\det(AB) = \det(A)\det(B)$, for any such square matrices $A,B$
 of the same size, these structures are also naturally suited for the 
description of symmetry transformations, 
 via the determinant-preserving multiplication of $B$ by any such $A$ with 
$\det(A)=1$.
 As a means of explicitly embedding the 4-dimensional \textit{quadratic} 
spacetime form of equation~\ref{sfourd} inside a higher-order homogeneous 
polynomial form for the proper time interval $\delta s$ we hence first note 
that there is a standard way of expressing  the norm of a Lorentz 4-vector 
 such as $(\delta x^0, \delta x^1, \delta x^2, \delta x^3) \inn \rrr^4$
 in terms of the determinant of a $2 \times 2$ Hermitian complex matrix:
\begin{equation}
  \label{sqdet}
    (\delta s)^2
	\; = \; \eta_{ab}\delta x^a \delta x^b
	 \; = \;  \det \left(   
	   \begin{array}{cc} \delta x^0 + \delta x^3
	         & \delta x^1 - \delta x^2i \\
		   \delta x^1 + \delta x^2i  
		     & \delta x^0 - \delta x^3  \end{array} \right)   
\end{equation}

   Here $\delta s$ is invariant under the actions of the symmetry group 
$\sltc$ through a $2 \times 2$ matrix composition as the double cover of 
the Lorentz group $\soot$. This structure can be embedded directly within 
the determinant of a $3 \times 3$ Hermitian complex matrix, which we 
interpret as a \textit{cubic} expression in \textit{nine} components 
 for a proper time interval $\delta s$, consistent with 
equation~\ref{salpha} and now with an augmented $\slthc$ symmetry,
  which we can write as:
\begin{eqnarray}
  \label{scdet}
     (\delta s)^3 & = &  \det
	\left( \! \begin{array}{ccc}  
	  \delta x^0  +  \delta x^3  
   &  \delta x^1  -  \delta x^2 i  
   &  \delta x^4  +  \delta x^5 i    \\
	  \delta x^1  +  \delta x^2 i 
   &  \delta x^0  -  \delta x^3 
   &  \delta x^6  +  \delta x^7 i   \\  
      \delta x^4  -  \delta x^5 i 
   &  \delta x^6  -  \delta x^7 i  
   &  \delta x^8     \end{array} \! \right)    \\ 
    &  = &  \left[
    \eta_{ab}\delta x^a \delta x^b \right]  \delta x^8 
   \; + \;  (\delta x^0, \ldots ,\delta x^8)^{3} 
      \begin{array}{c} \\ \vspace{-10pt} \\ \end{array}
    \label{sqinc}
\end{eqnarray}	

  In the construction of this cubic form for proper time in 
equation~\ref{scdet} we emphasise the deviation from the quadratic 
structure of extra spatial dimensions, such as in equation~\ref{snd},   
  while noting that this minimal augmentation from the 4-dimensional 
spacetime form of equations~\ref{sfourd} and \ref{sqdet} maintains a 
balanced contribution from the new components.
  In equation~\ref{sqinc} the same expression of equation~\ref{scdet} is
   written in the form of equation~\ref{sfourp}, where here 
  the first term, with $a,b = 0,1,2,3$, corresponds to part of a standard 
cofactor expansion for a $3 \times 3$ matrix determinant, to complete which 
the second term can be written out explicitly as a cubic function of the 
nine components.
 From the square brackets in the first term  in equation~\ref{sqinc} this 
cubic expression for a proper time interval is seen to directly extend the 
4-dimensional spacetime form of equation~\ref{sfourd}. Indeed 
equations~\ref{scdet} and \ref{sqinc} reduce to equation~\ref{sfourd} on 
setting each of $\delta x^4, \ldots, \delta x^7 = 0$ and $\delta x^8 = 
\delta s$, similarly as equation~\ref{snd} reduces to equation~\ref{sfourd} 
on setting each of $\delta x^4, \ldots, \delta x^{n-1} = 0$. 

 On rearranging equation~\ref{sfourd} in the form of equation~\ref{lpvn} 
the matrix expression in equation~\ref{sqdet} can be written more 
conveniently as:
\begin{equation}
  \label{lqdet}
  L_2(\bv_4)_{\mathrm{SL}(2,\ccc)} \; = \; \eta_{ab}v^a v^b
 \; = \;    \det (\bh)  \; = \;    \det \left( \!\!
	   \begin{array}{cc} v^0 + v^3 & v^1 - v^2i \\
		   v^1 + v^2i  & v^0 - v^3  \end{array}  \!\! \right)
		   \; = \; 1
\end{equation}  
  with the components of $\bv_4 = (v^0,v^1,v^2,v^3) \inn \rrr^4$ embedded 
in the $2 \times 2$ Hermitian complex matrix $\bh \in \htwc$. As indicated 
  this determinant form is again invariant under the actions of the 
symmetry group $\sltc$ as the double cover of $\soot$ 
 (see for example~\cite{KKone} equations~16 and 17). With all four 
components of equation~\ref{lqdet} projected locally onto the external 
spacetime tangent space, with $\bv_4 \in \TM_4$ and no residual structure, 
this effectively represents the `matterless vacuum' case (\cite{KKone} 
subsections~2.1 and 2.2).
 This \mbox{4-dimensional} form can be embedded within a 
  $3 \times 3$ matrix determinant structure, corresponding to 
equation~\ref{scdet} now with the notation of equation~\ref{lpvn}, as: 
\begin{equation}
L_3(\bv_9)_{\mathrm{SL}(3,\ccc)} = 
   \det(\bv_9) =
   \det  \!\!
	\left( \!\! \begin{array}{cc|c}  
	  v^0 \! + \!  v^3  
   &  v^1 \! - \! v^2 i  
   &  v^4 \! + \! v^5 i    \\
	  v^1 \! + \! v^2 i 
   &  v^0 \! - \! v^3 
   &  v^6 \! + \! v^7 i   \\  \hline
      v^4 \! - \! v^5 i 
   &  v^6 \! - \! v^7 i  
   &  v^8     \end{array} \!\! \right) \!  =   \det \!\!
   \left( \! \begin{array}{c|c} 
        \,\,\, \bh \!\!     \begin{array}{cc} 
		   &  \\  &  \end{array} \!\!\!   &
        \:\!  \psi \!\!\!\;  \begin{array}{cc} &  \\  
		   &  \end{array} \!\!\!\!\!\!\!\!\!\! 
				                      \\ \hline
        \,\,\,\,\;\! \psi^{\dagger} \!\!\!\! \begin{array}{cc}  
		        &   \end{array}   &	  
		   n \!      \end{array} \!  \right) \! = 1  \;\:
      \label{lvni}	      
\end{equation}
\begin{equation}
    =     \;  \Big[ \eta_{ab}v^a v^b   \Big] v^8			   
		 \, - \;  2\bh \!\cdot\! (\psi\psi^{\dagger})  \: = \:  1 
		  \qquad\qquad\qquad\qquad\qquad\quad \,
		    \label{lqinc}
\end{equation}
   with $\bv_9 \in \hthc$, $\bh \inn \htwc$, $\psi \inn \ccc^2$ and here 
$n=v^8 \in \rrr$ (consistent with the notation of~\cite{KKone} equation~19) 
while $a,b = 0,1,2,3$. The second term in equation~\ref{lqinc} 
  is the Lorentz inner product
$\bh \!\cdot\! (\psi\psi^{\dagger}) = 
 \frac{1}{2}\mbox{tr}(\bh)\mbox{tr}(\psi\psi^{\dagger}) - 
 \frac{1}{2}\mbox{tr}(\bh \psi\psi^{\dagger})$
  between the Lorentz 4-vectors associated with 
  $\bh, \psi\psi^{\dagger} \in \htwc$ (see for example~\cite{TimeE} 
equations~23 and 70). This cubic expression in
equations~\ref{lvni} and \ref{lqinc} is a specific example of the general 
form of proper time in equation~\ref{lpvn} which, via the first term  in 
equation~\ref{lqinc}, can be seen explicitly as an extension from the 
4-dimensional spacetime form of equation~\ref{lqdet} 
  via a natural minimal symmetric augmentation from a $2 \times 2$ to a
 $3 \times 3$ determinant form.  
    
  As noted in subsection~\ref{uni21} the possibility of embedding the local 
\mbox{4-dimensional} spacetime metric $\eta$ in a higher-dimensional 
spacetime metric $\hat{\eta}$, through the first four components of the 
quadratic form in equation~\ref{snd} or \ref{vnd}, is immediately evident.
 While the case here is a slightly more obscured 
 such a 4-dimensional quadratic metric structure can also be readily 
embedded in a cubic or higher-order expression in a less obvious, but 
nevertheless direct, manner as seen for equation~\ref{lqinc}.  In this form
 equations~\ref{lvni} and \ref{lqinc} reduce to equation~\ref{lqdet} on 
setting each of $v^4, \ldots, v^7 = 0$ and $v^8 = 1$, similarly as 
equations~\ref{vnd} and \ref{l2vn}
 reduce to the form of equation~\ref{lqdet} on setting each of $v^4, 
\ldots, v^{n-1} = 0$. Hence we have no reason to suppose that extra 
components should not be incorporated through the more general form for 
proper time in equation~\ref{lpvn} with the \textit{restriction} to the 
quadratic form of equation~\ref{vnd} being unnecessary. 
In either case in augmenting from the basic matterless vacuum of 
equation~\ref{lqdet} the symmetry of the generalised form will be broken 
through a projection of the local 4-dimensional spacetime substructure.  
 We might then consider the properties of the residual components
  deriving from this symmetry breaking for equations~\ref{lvni} and 
\ref{lqinc},
   interpreted as a basis for matter fields in 4-dimensional spacetime,  
for comparison with  equation~\ref{sosb}--\ref{vnbits} for the restricted 
case of extra spatial dimensions.

  In \textit{beginning} this analysis with the fully $\slthc$-symmetric 
9-dimensional
  cubic form $L_3(\bv_9)_{\mathrm{SL}(3,\ccc)} = \det(\bv_9) = 1$
   there are a number of ways that a 4-dimensional Lorentzian substructure 
could be extracted. However, without loss of generality, from 
equations~\ref{lvni} and \ref{lqinc} we can choose the four components 
originating from equation~\ref{lqdet} that we have effectively extended 
about -- indeed equations~\ref{lvni} and \ref{lqinc} were constructed in 
this way in order to explicitly demonstrate that such an extraction is 
possible.
  These extracted components $\bv_4=(v^0,v^1,v^2,v^3) \in \TM_4$ are then  
aligned with the local coordinates $(x^0,x^1,x^2,x^3)$ of a local inertial 
reference frame of the external \mbox{spacetime $M_4$}. \newline
 In turn a preferred external 
$\sltc \subset \slthc$ symmetry will act upon these subcomponents
  \mbox{$\bv_4 \in \TM_4$}  of
 $\bv_9 \in \hthc$ projected onto the external spacetime from 
equations~\ref{lvni} and \ref{lqinc}, which we can then write as:  
\begin{equation}
  \Lsl_3(\bv_9)_{\mathrm{SL}(2,\ccc)\times
          \mathrm{U}(1)} \, = \, 
		  \det(\bh) n - 
		    2\bh \!\cdot\! (\psi\psi^{\dagger}) \,= \, 1  
\label{lvnib}
\end{equation}  
 The extraction of the necessarily \textit{quadratic} substructure 
  for $\bv_4 \equiv \bh \inn \htwc$ to describe the geometry of the 
external 4-dimensional spacetime results more fully in the broken symmetry
  $\sltc \times \uo \subset \slthc$, with the kernel symbol \mbox{$\Lsl$} 
in equation~\ref{lvnib} denoting the broken form.

 While equation~\ref{lqdet} has been subsumed into equation~\ref{lvnib} the 
latter contains the external $\sltc$-invariant Lorentz 4-vector 
 norm $\vert \bv_4 \vert = \vert \bh \vert$ of the projected  fragment
   $\bv_4 \in \TM_4$ with:
\begin{eqnarray}
 \label{l2v4h}
 \vert\bv_4\vert^2  
 \!\!\! & = & \!\!\!  \eta_{ab}v^av^b = h^2 \\
  \mbox{where} \quad  h \!\!\! & = & \!\!\! \vert \bh \vert 
 = \sqrt{\det(\bh)}\, \in \, \rrr
  \label{hnorm}
\end{eqnarray} 
  which, unlike equation~\ref{lqdet} for the matterless vacuum, is not 
equal to 1 in general. 
   Being central to the symmetry breaking, and now taking a `vacuum value'
  $h = \vert \bh \vert$ in the projection onto $\TM_4$, the four components 
of the vector field $\bv_4(x) \equiv \bh(x) \in \htwc$ of 
equations~\ref{lvnib}--\ref{hnorm} are associated with a non-standard Higgs 
in this theory (also for the further reasons reviewed in \cite{TimeE} after 
figure~4, as also discussed in the following section). 

 In the context of the present theory
 the components $\bh(x) \in \htwc$ play a pivotal role in relating the 
Standard Model of particle physics and the general relativistic theory of
 gravitation by connecting the `origin of mass' in these two frameworks.
 Variations in the value of $h(x)$ 
  in equation~\ref{l2v4h}
 in the projection out of equation~\ref{lpvn}, for the general case, are 
associated directly with a local warping of the external 4-dimensional 
spacetime geometry as can be expressed by the Einstein tensor 
$G^{\mu\nu}(x)$ (see discussion of \cite{Unifi} figure~13.1 and 
equations~13.2--13.4). For the present theory this is proposed to underlie 
the physical property of mass through a contribution to the energy-momentum 
tensor $T^{\mu\nu}(x)$ via Einstein's field equation~\ref{Eineq} (discussed 
in section~\ref{uni5} here), with 
  (\cite{Unifi} equation~13.4) written for $G^{\mu\nu}(x)$ as a function of 
$h(x)$:
\begin{equation}
 \label{gmnconh}
  G^{\mu\nu} \: = \:
  -3 h^{-2} \pal_{\! \rho}h \, \pal^{\rho}\:\!\! h \, g^{\mu\nu} 
  -2 h^{-1} \pal^{\mu} \pal^{\nu} \:\!\! h  
  +2 h^{-1} \square h \, g^{\mu\nu} 
  \: =: \:  -\kappa T^{\mu\nu}
\end{equation} 
  with $\rho,\mu,\nu = 0,1,2,3$ spacetime indices. This direct warping of 
the spacetime geometry and corresponding properties of mass, associated 
with variations $\delta h (x)$, are on a scale set by the vacuum value for 
$h(x)$ in equation~\ref{l2v4h}. 
  
  For the case of equation~\ref{lvnib} the full $\uo$  group manifold is 
incorporated~(\cite{KKone} subsection~2.3 in particular figure~3(b)) in 
place of the group SO$(n-4)$ as described for 
equation~\ref{sosb}--\ref{vnbits} after figure~\ref{vfdvnd}.
 This structure will further generalise for full symmetry groups larger 
than 
 $\hG = \slthc$ in equation~\ref{lpvn}, with a residual internal gauge 
symmetry $G$, in general larger  than $\uo$, related to the geometry of the 
base space $M_4$ in a principal fibre bundle structure analogous to that of 
non-Abelian Kaluza-Klein theories 
 (\cite{KKone} subsection~4.1 in particular points `a)--e)').
  Specifically, the Einstein tensor $G^{\mu\nu}(x)$ can be related to the 
gauge curvature components $F_{\alpha}^{\ph{o}\rho\sigma}(x)$, where 
$\rho,\sigma$ are spacetime indices and $\alpha$ is a Lie algebra index for 
the gauge group $G$, and considered as a further source of energy-momentum 
with (\cite{KKone} subsection~4.2 equation~93, with $\chi$ considered a 
normalisation constant): 
\begin{equation}
 \label{gchift}
  G^{\mu\nu} \: = \: 
  2\chi(  - F^{\alpha \mu}_{\ph{oo} \rho}F_{\alpha}^{\ph{o}\rho\nu}
	             -\frac{1}{4} g^{\mu\nu} \, F^{\alpha}_{\ph{o}\rho
     \sigma}F_{\alpha}^{\ph{o}\rho\sigma})   
	   \: =: \:  -\kappa T^{\mu\nu}   
\end{equation} 
  again explicitly describing a direct warping of the 4-dimensional 
spacetime manifold and a corresponding form of energy-momentum. The 
dynamics of the gauge fields are proposed to be determined in turn by 
geometric constraints such as Bianchi identities (see discussion 
of~\cite{KKone} equations~93 and 94 and the accompanying references). 

 For  equation~\ref{lvnib} 
  with $\sltc$ being the external symmetry the residual internal 
  symmetry $\uo$  
 can be interpreted as a gauge group underlying a theory of 
electromagnetism alongside gravitation (\cite{KKone} subsection~4.2), with 
equation~\ref{gchift} then describing the energy-momentum of the 
electromagnetic field. 
 In this sense this minimal extension from equation~\ref{lqdet} to the 
cubic form of equation~\ref{lvni} is analogous to the early unified field 
theories reviewed here in subsection~\ref{uni12}. 
   There we described how Weyl's original geometric `gauge theory', with 
the scaling factor $\lambda$ for the metric $g_{\mu\nu}(x)$ in 
equation~\ref{WeylU}, was superseded  by a $\uo$ gauge theory for 
electromagnetism independent of the external metric structure. Here we have 
described how a $\uo$ gauge theory \textit{can} be incorporated through an 
augmentation of the local spacetime metric $\eta_{ab}$ via the structure of 
equations~\ref{lvni} and \ref{lqinc}, considered as cubic form for proper 
time.

 In addition to the fragment of equations~\ref{l2v4h} and \ref{hnorm} at 
the elementary local level  the
  broken symmetry reduces the full 9-dimensional  vector space $\hthc$ to 
three parts with 
 the Lorentz $\sltc$ and $\uo$ factors  acting upon these subcomponent 
parts introduced in equation~\ref{lvni} as
 (\cite{KKone} subsection~2.3, \cite{TimeE} subsection~4.1):
\begin{eqnarray}
        \slthc & \to &  \sltc \; \times  \; \uo  
		  \quad\,  \raisebox{-0.75ex}[0pt][0pt]{matter:} \qquad
		  \label{slsb} \\
	    \bh \inn \htwc  \!\!\!\!\! & : &  \,  \, \mbox{vector} \!
		  \qquad\;\;\;\!\hspace{0.1pt} \;\;   0 
		    \quad\; \mbox{: `Higgs-like' role in $\TM_4$} \nonumber  \\
  \raisebox{0pt}[0pt][0pt]{ {\raisebox{+0.0ex}{$\bv_9 \inn \hthc \to 
  \left\{ \begin{array}{c} \\ \\ \vspace{-15pt}  \end{array}  \right. 
   \!\!\!\! $}} }					
	   \psi\inn \ccc^2 \!\!\!  & : & \!\! \, 
	      \mbox{$L$-spinor} \! \qquad\,  \;\;   1 
   \quad \; \hspace{-0.1pt} \mbox{: charged spinor over $M_4$} 
        \nonumber  \\
	    n\inn \rrr    \!   & : &  \; \,
		 \mbox{scalar}\!  \qquad\;\;\,  \;\;   0   
		     \quad\; \mbox{: neutral scalar over $M_4$}   \label{slbits}
\end{eqnarray}  
 with the 2-component Weyl spinor $\psi$ taken to be left-handed by 
convention as denoted by the prefix `$L$-' above.
 A distinct feature of this unification scheme is the direct and natural 
manner in which spinor components such as $\psi$ arise in the local 
symmetry breaking structure, unlike the typical case for non-Abelian 
Kaluza-Klein theories which require a specific additional extension -- for 
example via supersymmetry as alluded to in subsection~\ref{uni21}.
 Hence here not only can the symmetry breaking pattern  be linked with a 
gauge field $A_{\mu}(x)$ for electromagnetism via the internal symmetry 
$\uo$, but this gauge group also acts non-trivially upon the 
spin-$\frac{1}{2}$ field $\psi(x)$ in spacetime, as indicated by the 
normalised unit charge `1' in equation~\ref{slsb}--\ref{slbits}.

 These structures deriving from the residual components and symmetry of 
equation~\ref{lvni} as projected over 4-dimensional spacetime to the broken 
form 
 of equation~\ref{lvnib} then provide a framework for electrodynamics 
incorporating a charged Weyl spinor.
  Given the ambition to ultimately account for properties of the Standard 
Model, with a range of spinor states for the charged leptons and quarks and 
also neutral spinors in the form of neutrinos, there is here then the 
potential  for accommodating such states through further augmentations of 
the form of proper time.
  Hence equation~\ref{slsb}--\ref{slbits} clearly provides a better 
starting point for this goal than the equivalent analysis of 
equation~\ref{sosb}--\ref{vnbits} as applied for the restricted quadratic 
case of extra spatial components in equation~\ref{vnd}. We shall return to 
a possible interpretation for the
 neutral scalar field $n(x) \in \rrr$ of equation~\ref{slsb}--\ref{slbits} 
in the next subsection and in particular in subsection~\ref{uni42}
 
In general
the local symmetry breaking projection of 
 $\bv_4 \in \TM_4$ ($\equiv \bh\in \htwc$) in equations~\ref{l2v4h} and 
\ref{hnorm} out of the full set of components for the $n$-dimensional form 
of equation~\ref{lpvn} partitions the components of $\bv_n \inn \rrr^n$ 
into subsets of subcomponent pieces that transform under irreducible 
representations of the subgroup:
\begin{equation}
 \mbox{Lorentz} \times G \subset \hG
  \label{gbreak}
\end{equation} 
 where the external local Lorentz symmetry group for 4-dimensional 
spacetime can be $\soot$ or its double cover $\sltc$, the group $G$ is the 
internal gauge symmetry and $\hG$ is the original full symmetry, as listed 
for the $\hG = \slthc$ case in equation~\ref{slsb}--\ref{slbits} (and also 
discussed in \cite{KKone} subsection~2.3 for equation~23 there). At the 
same time the corresponding form $\lpvng$ of equation~\ref{lpvn}, which is 
invariant under $\hG$, can be expanded and partitioned into subsets of 
terms with each part invariant under the $\mbox{Lorentz} \times G$ broken 
symmetry of equation~\ref{gbreak} as:
\begin{equation}
 \label{lpvnb}
    \Lsl_p(\bv_n)_{\mathrm{Lorentz}\times G} \, = \, 
		  \sum (\mbox{invariant parts}) \,= \, 1  
\end{equation}

  The individually invariant parts in equation~\ref{lpvnb} which contain a 
factor of $\bh$, or a scalar combination of components such as $\vert \bh 
\vert$ in equation~\ref{hnorm}, are proposed to be associated with `mass 
terms' in an effective Lagrangian deriving from the theory, in part 
motivating the
 kernel symbol `$L$' in equations~\ref{lpvn} and \ref{lpvnb}. For example 
while
 $L_3(\bv_9)_{\mathrm{SL}(3,\ccc)} = \det(\bv_9) =1 $ of
   equation~\ref{lvni} is invariant under the full symmetry 
   \mbox{$\hG = \slthc$}, each of the two terms in equation~\ref{lvnib} is 
invariant under the broken symmetry $\sltc \times \uo$.
 In this case the two terms each contain a factor in the components of 
$\bh(x)$ and might ultimately be interpreted as mass terms for the fields 
$n(x)$ and $\psi(x)$ in spacetime. 
  That is, such terms provide a source of field interactions such as
  $\delta \psi(x) \leftrightarrow \delta \bh(x)$ that can perturb the 
external spacetime geometry in a manner that is proposed to generate the 
property of mass as described for equation~\ref{gmnconh}.

   While the components of $\bh \equiv \bv_4 \in \TM_4$ are composed with 
other fields in the terms of equation~\ref{lvnib} in manner that begins to 
resemble Lagrangian mass terms a closer correspondence will require a more 
complete theory with more components in a higher-order form of proper time 
for equations~\ref{lpvn} and \ref{lpvnb}, as will be discussed further in 
section~\ref{uni4}. Such a construction is possible here for 
equation~\ref{lpvn} unlike the case for the restricted quadratic forms of 
equations~\ref{vnd} and \ref{l2vn}, similarly as spinor states are also now 
readily identified as described above. Indeed for  higher-order forms for 
equation~\ref{lpvnb} there is the potential for spinors to be composed in 
terms incorporating not only an effective Higgs but also further factors 
that might act as a source of Yukawa couplings for possible mass terms, as 
will be described in subsection~\ref{uni42}.

  Equation~\ref{lpvnb} will also act as a constraint on dynamical 
expressions for matter fields in 4-dimensional spacetime, yielding 
equations of motion with explicit interactions between gauge fields and 
spinor fields for example (see discussion of \cite{Unifi}
 equations~5.51 and 11.33).
 As described earlier in this subsection the symmetry breaking projection 
of equation~\ref{lpvn} over 4-dimensional spacetime also has physical 
consequences through the relations of both equation~\ref{gmnconh} and 
\ref{gchift}. Collectively the set of constraints for the full structure of 
the theory will subsume the role of
  effective Lagrangian terms in determining the detailed empirical 
properties of matter at the most elementary level 
  (see discussion of \cite{Unifi} equation~11.29 and table~15.1),
  as will be described further towards the end of subsection~\ref{uni52}.

  For this theory there is no ({\it n}$\,>\,$4)-dimensional extended 
physical or `bulk' space, nor any need for `compactification' since no 
extra spatial dimensions are being considered, in contrast to the models 
described in subsection~\ref{uni21} and in the opening 
 of subsection~\ref{uni22}. 
 Only the components $(x^0,x^1,x^2,x^3)$ underlying $\bv_4 \in \TM_4$ as 
the projected 
   quadratic 4-dimensional part of equation~\ref{lpvn}   are utilized
  as implicit local coordinates in \textit{defining} the local structure of 
an extended spacetime manifold $M_4$. The additional components in an 
expression for a proper time interval, such as equation~\ref{lvni}, are 
\textit{directly} associated with matter fields \textit{in} spacetime, as 
explicitly seen for the $\psi(x) \in \ccc^2$ components of 
equation~\ref{slsb}--\ref{slbits}.
 The identification of such matter fields, which might be described in 
terms of an `associated fibre bundle' related to the principal fibre bundle 
constructed for the internal symmetry $G$ of equation~\ref{gbreak}, follows 
directly from the identification of the distinguished external 
4-dimensional spacetime base \mbox{space $M_4$}.
The only  physical space is this external spacetime $M_4$, upon which 
extended geometric structure and energy-momentum can be defined as for 
example through equations~\ref{gmnconh} and \ref{gchift}.

 The symmetry breaking hence revolves around the necessary choice of a 
 preferred $\mbox{Lorentz} \subset \hG$ subgroup symmetry in 
equation~\ref{gbreak} that acts upon a 
 4-dimensional quadratic substructure of equation~\ref{lpvn} that is 
identified with the local external spacetime geometry.
 This necessary identification and extraction of the geometric structure of 
the spacetime manifold $M_4$ itself, for example via the local 
$(x^0,x^1,x^2,x^3)$ components of equations~\ref{scdet} and \ref{sqinc}, is 
inextricably linked with a 
 complete distinction between the external and internal components that 
hence applies for \textit{all physics} that can be defined in spacetime. 
In turn the full symmetry $\hG$ of equation~\ref{lpvn}, with which we begin 
in the mathematics of the theory as for example with $\hG = \slthc$ in 
equation~\ref{lvni},  is broken \textit{absolutely} to the product of the 
external Lorentz, or $\sltc$, symmetry and internal $G$ symmetry, as for 
$G=\uo$ in equation~\ref{slsb} or for the general case of 
equation~\ref{gbreak}, as a basis for the analysis of physical structures 
in 4-dimensional spacetime. 
Hence while the mathematics of the theory begins with equation~\ref{lpvn} 
the physics begins with equation~\ref{lpvnb}.
There are no surviving symmetries that mix subcomponents of $\bv_n$ 
transforming under different representations of the external Lorentz group.

 The group product structure for the external and internal symmetries in 
equations~\ref{gbreak} and \ref{lpvnb} is consistent with the demands of 
the Coleman-Mandula theorem \cite{ColMan} ultimately for the relativistic 
quantum theory limit (\cite{KKone} subsection~5.3).
 That is, similarly as a physical model that \textit{begins} with 
4-dimensional spacetime $M_4$ and posits the $\sltc \times \uo$ symmetry 
and field structure of equation~\ref{slsb}--\ref{slbits}, without reference 
to $\slthc$, would be compatible with the 
 Coleman-Mandula theorem, the same conclusion applies for the present 
theory in which these structures, as the starting point for physics, 
\textit{derive} from the fundamental origin of the mathematical form of 
proper time in equation~\ref{lvni} through the necessary absolute symmetry 
breaking in the identification of the base space $M_4$ itself.

 These observations apply for the general case of equation~\ref{lpvn} 
resulting in  equation~\ref{gbreak} and also for the restricted quadratic 
case of equation~\ref{l2vn} resulting in the broken symmetry of 
equation~\ref{sosb}, and is hence similar to an
  argument that could be made for some unification schemes based upon extra 
spatial dimensions -- through the necessary extraction of a distinguished 
\mbox{4-dimensional} base space $M_4$ from a more uniform 
higher-dimensional structure,
 as depicted for the case of equation~\ref{l2vn} in figure~\ref{vfdvnd}(b).
 However there are also significant differences, with for example  
 additional assumptions needed to incorporate spinor fields in models with 
extra spatial dimensions as we have noted following 
equation~\ref{slsb}--\ref{slbits}.

   The Higgs, $\uo$ gauge theory and spinor physics alluded to above for 
equations~\ref{l2v4h}--\ref{gmnconh}, \ref{gchift} and 
\ref{slsb}--\ref{slbits} respectively was unknown in 1918 when unified 
field theories based on extending general relativity were first proposed.
 The Higgs mechanism for symmetry breaking was developed much later in the 
1960s (see for example~\cite{Pais2} section~21(e) part~4).
  Even particle spin was not discovered until 1925, with the Pauli matrices 
introduced in 1927 and the relativistic Dirac equation for 4-component 
spinors following in 1928 (\cite{Pais2} chapter~13). As described in 
subsection~\ref{uni12} an understanding of the gauge principle to obtain a 
theory of electromagnetism culminated in 1929~\cite{Weyl3}, incorporating 
an application of 2-component Weyl spinors. Hence
the structures of equation~\ref{slsb}--\ref{slbits} would not have been 
natural to consider as a possible extension from general relativity in the 
years immediately following 1915. 
Also from 1925 the development of the principles of spin and of gauge 
theory were 
 inextricably linked with developments in quantum mechanics 
  led by Heisenberg, Schr\"{o}dinger and others -- the comprehension of 
which itself became the focus for theoretical activity from that time, as 
also noted in
 subsection~\ref{uni12}.

 Although founded at an elementary level the nature of the
  generalisation 
 from the form of the local 4-dimensional spacetime geometry
  for a proper time interval  in equation~\ref{sfourd}
  according to that proposed in equations~\ref{salpha} and \ref{lpvn} might 
also have seemed inappropriate during the period straight after the 
publication of general relativity.
 At that time, around one hundred years ago,  
relatively minimal extensions to general relativity were sought, with only 
electromagnetism to incorporate then alongside gravitation, while the 
general form of equation~\ref{lpvn} allows for much broader and more open 
possibilities.
 Now with the benefit of hindsight, afforded not only by the empirical 
knowledge accumulated in the rich properties of the Standard Model but also 
by the modern-day understanding of particular mathematical symmetry 
structures that exemplify equation~\ref{lpvn} in a manner naturally 
subsuming equation~\ref{lqdet}, direct progress can be made.

We can then explore the possibilities for a generalised proper time 
interval  beyond the initial step of equation~\ref{lvni} to determine and 
assess the nature of further physical structures beyond those of 
equation~\ref{slsb}--\ref{slbits} that might be uncovered for this theory.
 The properties of the new matter fields, again obtained 
 through a symmetry breaking projection of the subcomponents $\bv_4 \inn 
\TM_4$ locally onto the 4-dimensional external spacetime, 
  will be reviewed systematically in the following section.
  Of particular interest will be to observe the extent to which these 
natural mathematical structures dovetail with the empirical features of the 
Standard Model within the context of the conceptual scheme of the theory 
presented here based upon generalised proper time, which might then provide 
a firm foundation for the deduction of new physical phenomena as will be
 explored in section~\ref{uni4}.

\section{Exceptional Lie Groups and the Standard Model}
\label{uni3}

\subsection{Explicit Analysis for $\esi$ and $\ese$}
\label{uni31}

   While the Lie algebras, including the five exceptional cases of $\gt$, 
$\ff$, $\esi$, $\ese$ and $\ee$, were classified by Killing and Cartan in 
the late $19^{\mathrm{th}}$ century (see for example~\cite{Baez1} section~4 
opening) an understanding of explicit expressions for certain 
representations of these algebras
 and their corresponding Lie groups
  developed from the mid-$20^{\mathrm{th}}$ and continues into the 
$21^{\mathrm{st}}$ century. Elements of the space of $3 \times 3$ Hermitian 
octonion matrices $\htho$, which with an octonion incorporating eight real 
components is 27-dimensional over $\rrr$, 
 comprise the `exceptional Jordan algebra' as first described in 1934
 (\cite{Baez1} section~3).
 The smallest non-trivial representation of the Lie algebra $\esi$ is 
27-dimensional and was described in terms of transformations on the
 27-dimensional space $\htho$ in 1950~\cite{Chev}. These actions correspond 
to  $\esig$, one of the non-compact cases of the five real forms for 
$\esi$, and preserve a cubic norm, or determinant, defined on $\htho$. 
  The corresponding explicit $\esig \equiv \sltho$ group action on elements 
of $\htho$ has been described in detail more recently (see for 
example~\cite{Wang2} and references therein, as reviewed in \cite{Unifi} 
chapter~6), making intrinsic use of the octonion algebra composition 
properties.

  In the context of the present theory, while we obtained 
equation~\ref{lvni} from equation~\ref{lqdet} by a natural minimal 
symmetric extension from the $2 \times 2$ to the $3\times 3$ matrix case, 
we can also augment the form for proper time in equation~\ref{lvni} by a 
natural generalisation from the complex numbers $\ccc$ to the octonions 
$\ooo$
 to obtain the cubic 27-dimensional norm:
\begin{eqnarray}
  \label{lvts}
   L_3(\bv_{27})_{\mathrm{E}_6} \!\!\! & = & \!\!\! \det(\bv_{27}) \, 
    = \,   \det \!
   \left( \! \begin{array}{c|c} 
        \,\,\,\,\,\, X \!     \begin{array}{cc} & 
		                \\  &  \end{array} \!\!\!   &
         \:\!  \theta \!\!\!\;  \begin{array}{cc} & 
		               \\  &  \end{array} \!\!\!\!\!\!\!\!\!\! 
				                      \\ \hline
        \,\,\,\,\;\, \theta^{\dagger} \!\!\!\! \begin{array}{cc}  
		        &   \end{array}   &	  
		   n \!      \end{array} \!  \right)  = 1     \\ 
  &  &   \nonumber    \\ 
  & = &   \!\!\! \det(X)n \;		   
		  - \;  2 X \!\cdot\! (\theta\theta^{\dagger})  \; = \;  1 
		    \label{lvts2}
\end{eqnarray}
 with $\bv_{27} \inn \htho$, $X\in \htwo$, $\theta \in \ooo^2$ and again 
here $n\in \rrr$
 (in line with conventions in the main references as for 
equations~\ref{lvni}--\ref{lvnib}, see also~\cite{TimeE} equations~25 and 
72).
  The $\slthc$ symmetry of $L_3(\bv_9)_{\mathrm{SL}(3,\ccc)} = 1$ in 
equation~\ref{lvni} is augmented correspondingly  to $\sltho \equiv \esig$.
 Here equation~\ref{lvts2},
 while of a similar form to equation~\ref{lvnib}, will break down further, 
for the broken form of equation~\ref{lpvnb} at this $\hat{G} = \esi$ level, 
to an augmented set of parts each invariant under the broken symmetry.
 The symmetry breaking now results
 from the extraction of an external Lorentz 4-vector $\bv_4 \equiv \bh \in 
\htwc$ from a set of  subcomponents of $X \inn \htwo$ in 
equations~\ref{lvts} and \ref{lvts2}
  as projected onto the external local 
   tangent space of the 4-dimensional spacetime $M_4$.
 On identifying  an external Lorentz symmetry $\sltc \subset \esig$ acting 
upon the external 4-vector $\bv_4 \inn \TM_4$ 
  a symmetry breaking pattern can also be identified for the components of 
  $\bv_{27} \inn \htho$, as an augmentation from 
equation~\ref{slsb}--\ref{slbits}, as we shall summarise below.

  As a normed division algebra with the octonion norm compatible with 
octonion multiplication, in that $\vert ab \vert = \vert a \vert
 \vert b \vert$ for any $a,b \in \ooo$ (\cite{TimeE} equation~1) similarly 
as for the composition of matrix determinants described in the opening of 
subsection~\ref{uni23}, the octonions naturally describe symmetry 
transformations, for example via the norm-preserving multiplication of $b$ 
by any $a$ 
 with $\vert a \vert = 1$.
 Being non-associative octonion composition can in fact be employed to 
express a high degree of symmetry with relatively few elements 
(\cite{Unifi} section~6.2).
 However since the non-associativity property is counter to the basic 
axioms of group theory, unlike the case for matrix algebras, care is needed 
in applying a standard analysis for symmetry breaking where octonions are 
involved since there may be some deviation from a direct computation via 
the technical tools developed to study Lie groups generally.

  Indeed
  there are subtle differences between the explicit analysis for the real 
form Lie group $\esig$  that here intrinsically involves the octonion 
composition as an instantiation of
 equation~\ref{lpvn} and the parallel Dynkin analysis for the complex Lie 
algebra $\esi$ (as described for~\cite{TimeE} tables~1 and 2 respectively). 
The full explicit analysis is presented in detail in (\cite{Unifi} 
chapters~6 and 8, \cite{Novel} sections~4 and 5, \cite{TimeE} 
subsection~4.2) with the resulting branching properties for the 
subcomponents of equation~\ref{lvts} obtained as: 
\begin{eqnarray}
        \esig & \to &   \sltc \; \times \; \suth_c \; 
		   \times \; \uo_Q   
  \,\;\;\;\:  \raisebox{-1.5ex}[0pt][0pt]{matter:} \quad\qquad\;\!	       
		           \label{esisb} \\
	     X\inn\htwo \!\!\!\!\! & : &  \!\!\!\!\!\!   
    \left\{  \begin{array}{cccl}
       \, \mbox{vector}\,  \quad\;  & \quad\,  \mathbf{1} \quad\, & 
	                       \quad\;\;  0 & \quad\! 
						    \mbox{: $\nu$-lepton$/\bh$}  \\
	   \, \mbox{scalar}\,  \quad\;  & \quad\, \mathbf{3} \quad\, &
	                       \quad\;\;  \frac{2}{3}  & \quad\! 
						    \mbox{: $u$-quark}
	  \end{array}   \right.                  \nonumber  
	                               \\   
\raisebox{0pt}[0pt][0pt]{ {\raisebox{+2.5ex}{$\bv_{27} \inn \htho  \to 
  \left\{ \begin{array}{c} \\ \\ \\ \\ \vspace{-3.0ex} 
       \\ \\   \end{array}  \right. 
   \! $}} }	
	     \theta\inn\ooo^2 \!\!\!\!\! & : &  \!\!\!\!\!\!   
    \left\{ \!\! \begin{array}{cccl}
        L\mbox{-spinor}  \quad\!\!\;\hspace{-0.2pt}  & \quad\,  \mathbf{1} 
\quad\, & 
	                     \quad\;\;  1 & \quad\!   \mbox{: $e$-lepton}  \\
	    L\mbox{-spinor}  \quad\!\!\;\hspace{-0.2pt} 
		 & \quad\, \mathbf{3} \quad\, &
	                       \quad\;\;  \frac{1}{3} 
						    & \quad\! \mbox{: $d$-quark}
	  \end{array}   \right.             
	          \nonumber                     \\
	       n\inn \rrr   \!\!\!\!\! & : &    \quad\!\!
	      \mbox{scalar}  \qquad \quad\, \,\,\hspace{0.2pt}  \mathbf{1} 
		     \quad\: \hspace{-0.2pt} 
	                       \qquad\;\:   0  
						   \quad\;\;\; \mbox{: $n$}
						     \begin{array}{c} \vspace{0pt} \\
						    \end{array}    
						 \label{esibits}           
\end{eqnarray} 

  With the complex numbers $\ccc$ and octonions $\ooo$ being respectively 
2-dimensional and 8-dimensional over $\rrr$ a consequence of this fourfold 
increase is that there are now four spinor states identified in 
equation~\ref{esisb}--\ref{esibits} compared with the single spinor of 
equation~\ref{slsb}--\ref{slbits}. In contrast to the parallel Dynkin 
analysis here as a feature arising from explicit use of the octonion 
algebra structure in the symmetry transformations all four spinors have the 
\textit{same}
 chirality, taken to be \textit{left}-handed by convention, as demonstrated 
explicitly in
  (\cite{Unifi} equations~8.10--8.13).
  
   Through this natural augmentation from equation~\ref{slsb}--\ref{slbits} 
we also find an internal non-Abelian symmetry, which is proposed to 
correspond to the colour gauge group $\suth_c$ of quantum chromodynamics 
(QCD), alongside the original Abelian gauge group of electromagnetism, now 
denoted $\uo_Q$. 
   The three $\suth_c$ singlet $\mathbf{1}$ parts for the $\esig$ symmetry 
breaking in equation~\ref{esisb}--\ref{esibits} are subsumed  from the 
$\slthc$ case of equation~\ref{slsb}--\ref{slbits}  while the $\suth_c$ 
triplet $\mathbf{3}$ parts are new.
The pattern of $\uo_Q$ fractional relative charge magnitudes listed in 
equation~\ref{esisb}--\ref{esibits}, deriving directly from the intrinsic  
mathematical structure of the $\esig$ group action, as aligned with the 
$\suth_c$ singlets and triplets leads to the provisional matter field 
interpretation of the subcomponent decomposition of $\bv_{27} \inn \htho$ 
under the broken symmetry as representing a generation of Standard Model 
leptons and quarks, as listed in the final column of 
equation~\ref{esisb}--\ref{esibits}. 
 The appropriate features are here determined by the algebraic structure 
without needing to be introduced by hand.

    While a full electroweak $\sutw_L \times \uo_Y$ symmetry,
 with $L$ signifying a non-trivial action on left-handed spinors and $Y$ 
denoting hypercharge,	
	 is not identified at this stage there are $\sutw \subset \esig$ 
symmetries that act between the real subcomponents of the three octonion 
elements of $\htho$ in equation~\ref{lvts} in a manner resembling 
properties of the Standard Model $\sutw_L$ doublets $\binom{\nu}{e}_{\! L}$ 
and 
 $\binom{u}{d}_{\! L}$ 
  which, along with the $\suth_c \times \uo_Q$ transformation properties in 
equation~\ref{esisb}--\ref{esibits}, further justifies the association of 
the $\nu$-lepton and $u$-quark states with particular components 
(\cite{Unifi} subsection~8.3.1).
 Given that with respect to the external Lorentz $\sltc$ symmetry the 
$e$-lepton 
 and $d$-quark states in equation~\ref{esisb}--\ref{esibits} transform 
uniformly as a set of four 2-component left-handed Weyl spinors the 
identification of complete
 doublets of an internal $\sutw_L$ symmetry
 would require the $\nu$-lepton and $u$-quark states also to transform as 
left-handed spinors.
  However at this stage for equation~\ref{esisb}--\ref{esibits}
 the \mbox{`$u$-quark'} components transform as Lorentz scalars while the 
neutral components most directly associated with the neutrino, with respect 
to the internal symmetries, are incorporated into a Lorentz 4-vector. 

 Compounding these discrepant features, this natural slot for the 
`$\nu$-lepton' in equation~\ref{esisb}--\ref{esibits} is in fact already 
occupied specifically by the Lorentz 4-vector \mbox{$\bh \inn \htwc$} 
subcomponent of $X \inn \htwo$, which is projected onto the tangent space 
of the external spacetime with $\bh\equiv \bv_4 \inn \TM_4$ and associated 
with the Higgs as discussed after equation~\ref{hnorm} and similarly as 
listed 
 in equation~\ref{slsb}--\ref{slbits}. 
 In fact the partial identification of elements of an 
  $\sutw_L \times \uo_Y$  electroweak symmetry breaking structure as 
alluded to above further motivates the association of the $\bh \inn \htwc$ 
components with the Higgs (\cite{Unifi} section~8.3), as will be discussed 
further in subsection~\ref{uni42}.
 This ambiguity in the assignment of the corresponding components, 
indicated by the `$\nu$-lepton/$\bh$' entry in 
equation~\ref{esisb}--\ref{esibits}, already at this $\esig$ stage suggests 
an intimate link between neutrino and Higgs physics in the context of this 
theory.

  Despite these caveats
 the above overall observations  with the identification now of a set of 
four Weyl spinors and the overall $\suth_c \times \uo_Q$ representation 
pattern in equation~\ref{esisb}--\ref{esibits} are encouraging for this 
direct development from the simple underlying basis of the theory in 
generalised proper time that motivated equations~\ref{salpha} and 
\ref{lpvn}.
 Through this $\esig$ symmetry a connection has also been established with 
the exceptional Lie groups, which are known to be of interest for 
unification as noted in subsection~\ref{uni12}.
 This leads in turn to the potential for further natural extension beyond 
the $\esi$ symmetry as we describe below.

  The smallest non-trivial representation of the next largest exceptional 
Lie group $\ese$, specifically  $\eseg$ of the four real forms, can be 
described by an action on the \mbox{56-dimensional} space of the 
`Freudenthal triple system' $F(\htho)$, investigations of which had begun 
by 1954 (\cite{Freud} section~4.11), as again has been studied in detail 
more recently (see for example~\cite{Rios} and references therein). The 
$\eseg$ action preserves a homogeneous \textit{quartic} norm $q$ defined on 
the space $F(\htho)$ which, while not being expressed as a matrix 
determinant function itself, contains an \mbox{$\esig \subset \eseg$} 
subgroup action on the 27-dimensional \textit{cubic} norm of 
equation~\ref{lvts} and hence can be considered a further natural 
augmentation consistent with equation~\ref{lpvn}. 
  This quartic 56-dimensional form for proper time, identified with the 
norm $q$, can be written explicitly as: 
\begin{equation}
 \label{lvfsq}
    L_4(\bv_{56})_{\mathrm{E}_7} \: = \:
   q(\bv_{56}) \: = \: -2\lbrack \alpha\beta - (\mcX,\mcY)\rbrack^2 \, - \,
       8\lbrack\alpha \det(\mcX) + \beta\det(\mcY) - (\mcX^{\sharp},
	                                          \mcY^{\sharp})\rbrack
    \: = \: 1 \;\;
\end{equation}
 where $\bv_{56} \equiv (\mcX,\mcY,\alpha,\beta) \in F(\htho)$, 
  the bilinear form $(\mcX,\mcY)$ is the trace of the Jordan product of 
$\mcX,\mcY \inn \htho$ and the quadratic adjoint map $\mcX^{\sharp}$ is 
also defined in~\cite{Rios},
 while $\alpha,\beta \inn \rrr$ (see also~\cite{TimeE} equations~30 and 
63). 
 The components of the 27-dimensional representation of $\esig$ in 
equation~\ref{lvts}
  can be associated explicitly with subcomponents in equation~\ref{lvfsq}
   via $\bv_{27} \equiv \mcX \inn \htho$, here with
    $\det (\mcX) \neq 1$ in general,
     while the 27-dimensional complex conjugate representation of $\esig$  
corresponds to the $\mcY \inn \htho$ subcomponents of $F(\htho)$. 
Given this straightforward embedding of the subgroup 
\mbox{$\esig \subset \eseg$} action the main consequences of this further 
augmentation can be inferred directly from 
equation~\ref{esisb}--\ref{esibits}.
  However, there is still only one set of four real  subcomponents to 
project onto the
  external 4-dimensional spacetime with $\bv_4 \inn \TM_4$ in 
equation~\ref{l2v4h} which, without loss of generality given an arbitrary 
choice, we now extract as
  $\bv_4 \equiv \bh\inn \htwc$  from the $\mcY \inn \htho$ subcomponents.
   The resulting breaking pattern for the $\eseg$ symmetry of 
equation~\ref{lvfsq} is then determined (\cite{Unifi} section~9.2, 
\cite{Novel} section~6, \cite{TimeE} subsection~4.3) as summarised here:
\begin{eqnarray}
        \eseg \!\! &  \,\, \to \!\! &  \,\, 
		  \sltc \; \times \; \suth_c \; 
		   \times \; \uo_Q  
		\,\;\;\:  \raisebox{-1.5ex}[0pt][0pt]{matter:} \quad\qquad\;\;\!	   
		    \label{esesb} \\
	     \mcX\inn\htho \!\!\!\!\! & : &  \!\!\!\!\!\!   
    \left\{  \begin{array}{cccc}
       \, \mbox{vector}\,  \quad\;  & \quad\,  \mathbf{1} \quad\, & 
	                   \quad\;\;  0 
					    \qquad\; : & \!\!\!\!  \mbox{`$\nu_L$'}  \\
	   \, \mbox{scalar}\,  \quad\;  & \quad\, \mathbf{3} \quad\, &
	          \quad\;\;  \frac{2}{3} \qquad\;
			    \hspace{-0.5pt} : \hspace{0.5pt} 
				  &  \!\!\!\! \mbox{`$u_L$'}  \\
       \! L\mbox{-spinor} \,  \quad\!\!\;  & \quad\,  \mathbf{1} \quad\, & 
	                       \quad\;\;  1  \qquad\; : &  \!\!\!\!  e_L  \\
	   \! L\mbox{-spinor} \,  \quad\!\!\;  & \quad\, \mathbf{3} \quad\, &
	                       \quad\;\;  \frac{1}{3} \qquad\;
			 \hspace{-0.5pt}   : \hspace{0.5pt}  &  \!\!\!\!  d_L   \\
		\, \mbox{scalar}\,  \quad\;  & \quad\, \mathbf{1} \quad\, &
	                       \quad\;\;  0  \qquad\; : & \!\!\!\!	   n			   
	  \end{array}   \right.          
	          \nonumber                     \\
\raisebox{0pt}[0pt][0pt]{ {\raisebox{+9.3ex}{$\bv_{56}
              \!\inn\! F(\htho) \to 
  \left\{ \begin{array}{c} 
    \\ \\ \\ \\ \\ \\ \\ \\ \\ \\  \vspace{+1.0ex} \\
   \end{array}  \right. 
   \!\!\!\!\!\! $}} }		  
	 \mcY\inn\htho \!\!\!\!\! & : &  \!\!\!\!\!\!   
    \left\{ \hspace{-0.8pt} \begin{array}{cccc}
       \, \mbox{vector}\,  \quad\;  & \quad\,  \mathbf{1} \quad\, & 
	                       \quad\;\;  0 \qquad\; : & \!\!\!\!   \bh  \\
	   \, \mbox{scalar}\,  \quad\;  & \quad\, \mathbf{3} \quad\, &
	                \quad\;\;  \frac{2}{3} \qquad\;
		   \hspace{-0.7pt} : \hspace{0.7pt} & \!\!\!\! \mbox{`$u_R$'}  \\
       \! R\mbox{-spinor} \, \quad\!\!\;  & \quad\,  \mathbf{1} \quad\, & 
	                       \quad\;\;  1 \qquad\; : & \!\!\!\!   e_R  \\
	   \! R\mbox{-spinor} \, \quad\!\!\;  & \quad\, \mathbf{3} \quad\, &
	                       \quad\;\;  \frac{1}{3} \qquad\;
				 \hspace{-0.7pt} : \hspace{0.7pt} & \!\!\!\!  d_R  \\
		\, \mbox{scalar}\,  \quad\;  & \quad\, \mathbf{1} \quad\, &
	                       \quad\;\;  0 \qquad\; : &	\!\!\!\!   N		   
	  \end{array}   \right.       
	              \nonumber                     \\ 
	     \alpha,\beta\inn \rrr   \!\!\!\!\! & : &    \quad\!
	   \hspace{-0.8pt}   \mbox{scalar}  
	       \quad\, \qquad \;\;\:\!  \mathbf{1} \quad\, 
	                       \qquad\;\;   0 
		  \qquad\; \hspace{-0.1pt} :   \:\!   \alpha,\beta  
						   \begin{array}{c} \vspace{0pt} \\ \end{array}  			   
						 \label{esebits}           
\end{eqnarray} 
   where the upper sector for $\mcX\inn\htho$ is essentially a copy of 
equation~\ref{esisb}--\ref{esibits}. 
 The immediate consequence of this augmentation in the symmetry of 
equation~\ref{lpvn}
 through the exceptional Lie groups from $\hG = \esi$ to $\hG = \ese$ is 
the incorporation of \textit{right}-handed spinor states as well as the 
original \textit{left}-handed states, via the inclusion of components 
corresponding to the complex conjugate 27-dimensional representation of 
$\esi$.
 That is, in addition to the four left-handed spinors of 
equation~\ref{esisb}--\ref{esibits} a corresponding set of four 
right-handed spinors is identified in the $\mcY\inn \htho$ subcomponents, 
with
  the $\mcX$ and $\mcY$ components of equation~\ref{esesb}--\ref{esebits} 
hence referred to respectively as the `left-handed' and `right-handed' 
sectors of the theory. With the internal symmetry transformations being the 
same for both sectors, the \mbox{2-component} Weyl spinors for the $e_L$ 
and $d_L$ states in equation~\ref{esisb}--\ref{esibits} have been augmented 
to \mbox{4-component} Dirac spinors $\binom{e_L}{e_R}$ and 
$\binom{d_L}{d_R}$
 in \mbox{equation~\ref{esesb}--\ref{esebits}}. 
 Provisionally anticipated through the study of $\sutw \subset \esig$ 
doublet actions alluded to in the discussion of 
equation~\ref{esisb}--\ref{esibits} 
 corresponding $L$ and $R$ subscripts are also added to the `$\nu$-lepton' 
and `$u$-quark' states in equation~\ref{esesb}--\ref{esebits}, albeit 
within quotation marks since the need to identify an explicit Lorentz 
spinor structure for these states remains and will require yet further 
augmentation beyond this $\ese$ stage.

  Nevertheless  the branching patterns obtained for this elementary 
symmetry breaking for natural augmentations of the form of time 
$L_p(\bv_n)_{\hG} = 1$ of equation~\ref{lpvn}, through 
equations~\ref{lvni}, \ref{lvts} and \ref{lvfsq}, 
 over the local structure of \mbox{4-dimensional} spacetime via the 
projected fragment of equation~\ref{l2v4h}, leading to 
equations~\ref{slsb}--\ref{slbits}, \ref{esisb}--\ref{esibits} and 
\mbox{\ref{esesb}--\ref{esebits}} respectively, achieve far more success in 
terms of the \textit{direct} emergence of Standard Model properties than 
the equivalent case of extra spatial dimensions described for 
equations~\ref{vnd}--\ref{vnbits}. 
 In all cases we are uniformly applying the simplest symmetry breaking 
scheme around the extraction of the external $\bv_4 \in \TM_4$ subcomponent 
part, as was depicted for equations~\ref{vnd}--\ref{vnbits} in
 figure~\ref{vfdvnd}(b). 
 It is particularly striking that these results have been obtained by 
\textit{dropping the unnecessary assumption} that extra components should 
have the `quadratic spatial' form of the Pythagorean theorem at an 
elementary level, rather than by \textit{adding new structures} 
specifically tailored to accommodate Standard Model features, as are 
required for models with extra spatial dimensions
    as discussed in subsection~\ref{uni21}.
	
 In addition, while there is some overlap between the Standard Model 
properties identified in equation~\ref{esesb}--\ref{esebits} and features 
known to arise from the abstract mathematical analysis of the symmetry 
breaking patterns for certain candidate unification groups such as
 $\esi$ and $\ese$ as alluded to in subsection~\ref{uni12}, here we do have 
a \textit{clear underlying conceptual origin} for the significant role 
played by these exceptional Lie groups as symmetries of generalised forms 
of proper time. 
 As a distinct feature, compared with a unification of the internal 
symmetry alone for the GUT models of \cite{Gur1,Gur7} for example,
  here we are generalising from the local structure of 4-dimensional 
spacetime and begin by identifying the external Lorentz symmetry as a 
subgroup of $\esi$ and $\ese$, with residual structures then identified as 
matter fields in spacetime.

  While additional structures beyond those motivated directly by the 
conceptual basis of the theory have not been added by hand, we also find 
that Standard Model features are identified very efficiently in 
equation~\ref{esesb}--\ref{esebits}
 with very little redundancy of components.
 Indeed of the 56 real components in equation~\ref{lvfsq} only four,
 $\{n,N,\alpha,\beta\}$, have not been utilised for the above 
correspondence with Standard Model structures -- with $N$ a subcomponent of 
$\mcY \in \htho$ corresponding to the $n$ subcomponent of 
 \mbox{$\mcX \in \htho$} as incorporated from equation~\ref{lvts}, while 
$\alpha$ and $\beta$ are two further new components. 
Potentially looking beyond the Standard Model we note that at this stage 
these four components, as the scalar invariants $\{n,N,\alpha,\beta\}$ 
listed in equation~\ref{esesb}--\ref{esebits}, provide  candidates for dark 
matter or even a source for `dark energy' phenomena in this theory. These 
augment the original single scalar invariant $n\in \rrr$ of 
equations~\ref{slsb}--\ref{slbits} and \ref{esisb}--\ref{esibits} and will 
be discussed further in subsection~\ref{uni42} where an alternative 
interpretation of such components as Yukawa couplings will also be 
considered.

   Indeed further structure is still needed to describe the complete  
Standard Model itself.  In addition to the required spinor structure for 
the $\nu$-lepton and $u$-quark  states in 
equation~\ref{esesb}--\ref{esebits} the principal Standard Model symmetry 
and particle multiplet features that remain to be accounted for are
 that of a full electroweak theory, with an \mbox{$\sutw_L \times \uo_Y$} 
symmetry that breaks to $\uo_Q$, 
 and a full three generations of leptons and quarks.
 We note however that while a larger unification group beyond $\esi$ and 
$\ese$ may incorporate  appropriate $\sutw$ and $\uo$ internal factors in 
the symmetry breaking we do already possess a natural explanation for a 
left-right asymmetry, which will be needed for the $\sutw_L$ factor of a 
complete electroweak theory.
  The left-right asymmetry is a significant feature of particle physics 
that in general is very difficult to account for in an uncontrived      
manner in model building
 (see for example~\cite{PaSa} as discussed for \cite{TimeE} equation~65).
  The empirical asymmetry between left and right-handed states is 
particularly conspicuous in the neutrino sector, with such properties again 
usually imposed by hand
 in neutrino models as described
 in subsection~\ref{uni11}.

 Here an intrinsic left-right asymmetry is 
 implied in the full symmetry breaking through the necessary choice
 of a preferred set of subcomponents  $\bv_4 \in \TM_4$ projected onto the 
local 4-dimensional external spacetime, upon which a unique 
 external $\sltc\subset\eseg$ subgroup acts.
 The asymmetry  arises as 
  $\bv_4 \in \TM_4$  ($\equiv \bh \in \htwc$) is necessarily
   projected out from \textit{either} the left-handed sector $\mcX \in 
\htho$ \textit{or} the right-handed sector
   $\mcY \in \htho$ components of $\bv_{56} \inn F(\htho)$, as described 
before \mbox{equation~\ref{esesb}--\ref{esebits}}.
  We can observe at this stage for equation~\ref{esesb}--\ref{esebits} that 
this left-right asymmetry is indeed particularly marked for the neutrino 
states since the embedding of the external \mbox{4-vector} 
$\bv_4 \equiv \bh\inn\htwc$, associated with the Higgs, within the $\mcY 
\in \htho$ components \textit{prohibits} the accommodation of a neutrino 
state `$\nu_R$' in the right-handed sector while implying that the 
corresponding slot is now \textit{available} for a neutrino state `$\nu_L$' 
in the left-handed sector 
associated with the
 components of $\mcX \in \htho$, without the conflict described for 
equation~\ref{esisb}--\ref{esibits}.
 The theory is hence in principle able to provide a natural framework for 
left-right  
  asymmetric properties in the neutrino sector as an intrinsic feature of 
the symmetry breaking structure,
  as we explore further in subsection~\ref{uni41}.

 Although the analysis of equation~\ref{esesb}--\ref{esebits} strongly 
suggests that the theory is progressing in a favourable direction,
 since none of this structure has been pragmatically tailored for this 
purpose a perfect fit to Standard Model structures is not necessarily to be 
expected until the full picture has been established. 
 While additional elements of electroweak theory and Higgs physics have 
been partially identified as noted earlier in this subsection (with 
reference to \cite{Unifi} section~8.3, see also \cite{Novel} section~5) and
 the required spinor structures for equation~\ref{esesb}--\ref{esebits} 
might be contrived through the introduction of further components 
 (\cite{Unifi} section~9.1)
   the necessary threefold augmentation to account for two additional 
generations of leptons and quarks clearly requires a substantial extension, 
and in all cases ideally via a further \textit{natural} mathematical 
augmentation. Such
 a further possible extension for
  the generalised form for proper time of
  equation~\ref{lpvn} beyond the $\esi$ and $\ese$ symmetries described in 
this subsection will be considered in that to follow.


\subsection{Predicted Role for $\ee$}
\label{uni32}

  The intrinsic preference for certain values for the polynomial order $p$ 
and the number of components $n$ for the general form of proper time in 
equation~\ref{lpvn}, as suggested at the end of subsection~\ref{uni22}, has 
been explicitly demonstrated in subsections~\ref{uni23} and \ref{uni31}. 
The augmentation from the matterless vacuum case with $p=2,n=4$ in 
equation~\ref{lqdet} with the local Lorentz symmetry of 4-dimensional 
spacetime to the $p=3,n=9$ case of equation~\ref{lvni} with $\slthc$ 
symmetry was obtained by a minimal symmetric extension from a $2\times 2$ 
to a $3 \times 3$ matrix determinant structure. The possibility of a 
corresponding further augmentation to a $4\times 4$ matrix determinant or 
beyond is seemingly ruled out by the implication of a non-compact internal 
symmetry which would be incompatible with the requirements of the quantum 
field theory (QFT) limit that will ultimately need to be considered (as 
discussed for
 \cite{TimeE} equations~86--88).
 
 An augmentation from the $p=3,n=9$ form of proper time to the case of 
$p=3,n=27$ in equation~\ref{lvts} with an $\esi$ symmetry at the $3\times 
3$ matrix determinant level was however achieved by a generalisation from 
the complex numbers $\ccc$ to the octonions $\ooo$, with a compact internal 
symmetry obtained as presented in equation~\ref{esisb}. This means of 
extension in turn terminates here since the octonions are uniquely the 
largest normed division algebra~(\cite{Baez1} sections~1 and 1.1).

 However, while leaving behind matrix and division algebra extensions, from 
the properties of exceptional Lie groups a further natural augmentation 
from \mbox{$p=3,n=27$} to $p=4,n=56$ in equation~\ref{lvfsq} has been 
identified through the embedding of $\esi$ within the $\ese$ symmetry 
action on the components of this quartic norm. 
 In considering the possibilities for further extension a further 
progression from the $\ese$ symmetry to $\ee$  as uniquely the largest 
exceptional Lie group is naturally suggested, potentially terminating the 
series of augmented forms for equation~\ref{lpvn} with a high degree of 
symmetry that might be of most significance for physics.   

 This line of argument hence converges with the latter end of the 
well-known sequence of unification groups
 $\mbox{SU}(5) \to \mbox{SO}(10) \to \esi \to \ese \to \ee$, alluded to in 
subsection~\ref{uni12}, 
 which are linked by a progression of augmented Dynkin diagrams that also 
terminates uniquely in $\ee$ (see for example~\cite{Unifi} 
figures~7.2(c,b,a) and 9.1(a,b,c)). While in most cases these groups are 
employed in Grand Unified Theories, accommodating only the Standard Model 
gauge symmetry $\SML$, for the present theory we initially incorporate the 
external Lorentz or $\sltc$ symmetry and \textit{then} include an internal 
gauge symmetry in this unification. Indeed here a natural connection has 
been made with the unique high-symmetry structures of the exceptional Lie 
groups $\esi$ and $\ese$
  through a generalisation \textit{from} the quadratic form of 
4-dimensional spacetime and extra spatial dimensions
  by exploring higher-order homogeneous polynomial forms of proper time as  
   described in the previous subsection. 

   The case for an extension to an $\ee$ symmetry is further strengthened 
on noting that
  the three largest exceptional Lie groups $\esi$, $\ese$ and $\ee$ are 
also known to describe a sequence of symmetries acting on structures that 
can be interpreted as `generalised spacetimes', in particular based on the 
space $\htho$~(\cite{Gunay2} equations~64, 66 and 67), and with those 
\textit{same} $\esig$ and $\eseg$ actions interpreted here as symmetries of 
`generalised proper time' for the cubic and quartic expressions of 
equations~\ref{lvts} and \ref{lvfsq} respectively. 

  Hence with $\ee$ as uniquely the largest exceptional Lie group the above 
observations  lead to the proposal of a homogeneous polynomial form:
\begin{equation}
 \label{lvto}
 \lvtfep
\end{equation}
as the ultimate instantiation for equation~\ref{lpvn}, as 
originally suggested in (\cite{Unifi} section~9.3, \cite{Novel} section~7)
 and considered in detail
 in~\cite{TimeE}.
 The progression through $\esig$ and $\eseg$ suggests the 
 symmetry action of $\eeg$, one of the three real forms of $\ee$, in
  equation~\ref{lvto}. 
  This provisional  form is potentially of octic order with $p=8$ (see for 
example~\cite{CedP}), and a close connection with the smallest non-trivial 
$\ee$ representation with $n=248$ dimensions is here assumed; although 
other values for $p$ and $n$ might be conceivable.
  The nature of this structure and  
 the plausibility of encompassing the principal remaining Standard Model 
features required (as summarised at the end of the previous subsection) in 
a correlated manner is the main focus of~\cite{TimeE}.

 As a unification symmetry the Lie group $\ee$ itself is comfortably able 
to incorporate a broken symmetry  
 corresponding to a product of the external Lorentz $\sltc$ and internal 
Standard Model gauge groups in the form of equation~\ref{gbreak} with:
\begin{equation}
 \label{eebrk}
    \sltc \, \times \, \suth_c \times \sutw_L \times \uo_Y \: 
	     \subset \: \ee
\end{equation}	
 While the external \textit{and} internal symmetries derive from the same 
single unifying mathematical source in $\ee$ the \textit{absolute} nature 
of the above symmetry breaking \textit{prior} to the 
derivation of any physics in 4-dimensional spacetime is again compatible 
with the requirements of the Coleman-Mandula theorem for the QFT limit,
 as also for the $\esi$ and $\ese$ levels of the previous subsection and
 as discussed towards the end of
  subsection~\ref{uni23}.

 On the other hand as an extension from the $\mathbf{27}$ representation of 
$\esi$ underlying equation~\ref{esisb}--\ref{esibits}, as
 combined with the complex conjugate $\mathbf{\overline{27}}$ for  
equation~\ref{esesb}--\ref{esebits}, 
	    in broad terms a possible factor of three for three generations of 
leptons and quarks is suggested by 
 the factors of 
  $\stackrel{\raisebox{-2.1pt}[0pt][0pt]{\mbox{{\tiny (\!\!
 \raisebox{-1.8pt}[0pt][0pt]{{\large{\bf --}}})}}}} 
{\raisebox{-0.0pt}[0pt][0pt]{$\mathbf{3}$}}$
 in
the subgroup embedding of $\esi \subset \ee$ with the  representation 
branching pattern~(\cite{TimeE} equation~68):
\begin{equation}
  \label{eetoesi}
  \ee \, \supset \, \esi \times \suth : \quad
    \mathbf{248} \, \to \, 
	(\mathbf{27},\mathbf{3}) +
	 (\mathbf{\overline{27}},\mathbf{\overline{3}})
	+ (\mathbf{78},\mathbf{1}) + (\mathbf{1},\mathbf{8}) 
\end{equation}	
 
  However, as also explained in~\cite{TimeE}, unlike the case for the 
direct embedding of the subgroup $\esi \subset \ese$ action described for 
equation~\ref{lvfsq} the embedding of  $\esi$ and $\ese$  in the $\ee$ 
action for the form $\lvtfep$ that we are seeking is expected to be less 
straightforward than that suggested by equation~\ref{eetoesi} if the needed 
spinor structures for \mbox{$\nu$-leptons} and $u$-quarks under $\sltc$ 
 together mutually with a
 complete electroweak symmetry action under \mbox{$\sutw_L \times \uo_Y$} 
are to also be identified compatible with the symmetry breaking pattern of 
equation~\ref{eebrk}.
 It is possible for example that features of the full Standard Model may be 
more closely aligned with a different maximal subgroup embedding such as 
 $\ee \supset \ff \times \gt$, in terms of the two other exceptional Lie 
groups
  (as discussed for~\cite{TimeE} equation~81), or another possible 
algebraic decompositions of $\ee$ (such as reviewed in~\cite{Baez1} 
section~4.6).
  The central ambition is to identify an explicit structure for 
equation~\ref{lvto}
  and supplant \mbox{equation~\ref{esesb}--\ref{esebits}} with a full 
matter field listing  under equation~\ref{eebrk} for the $\ee$ case.

    It is also known that the structure of the 248-dimensional $\ee$ Lie 
algebra itself exhibits some correlation with the full symmetry structure 
of the Standard Model~\cite{Lisi}, albeit with seemingly prohibitive flaws, 
as alluded to in subsection~\ref{uni12}.  While a full three generations of 
`leptons' and `quarks' are identified in the $\ee$ Lie algebra structure 
within that analysis there is an irreconcilable inconsistency in the 
representations under the Lorentz and electroweak symmetry subgroups as 
explained in~\cite{Lisi,DiGa}. That difficulty 
 may be related to a key issue for the present theory regarding the  
  need to identify further Lorentz spinors and a full electroweak theory in 
augmenting from equations~\ref{esisb}--\ref{esibits} 
 and \ref{esesb}--\ref{esebits} for the proposed $\ee$ level.

   However rather than reading off particle states directly from the 
abstract composition of the complex $\ee$ Lie algebra structure and its 
representations, which may be diagrammatically illustrated in an 
aesthetically appealing manner (\cite{Lisi} figures~2--4),  here we seek an 
explicit $\ee$ symmetry action on a homogeneous polynomial form as 
described for equation~\ref{lvto} as a unique expression for 
equation~\ref{lpvn} with a high degree of symmetry and consistent with the 
simple underlying  motivation 
 of this theory, based upon the generalisation of proper time as described 
for equation~\ref{salpha}. In particular as an augmentation from the 
$\esig$ action for equation~\ref{lvts} and the $\eseg$ action for 
equation~\ref{lvfsq} the proposed $\eeg$ symmetry action on the components 
of $\bv_{248} \in \rrr^{248}$ for equation~\ref{lvto} is expected to 
incorporate octonion composition in an essential way. As noted before and 
after equation~\ref{esisb}--\ref{esibits} the application of the octonion 
algebra in this way implies subtle but significant differences compared 
with a standard analysis for the abstract structures of the corresponding 
complex Lie algebra, which hence cannot be fully relied upon for a rigorous 
assessment here.  
 This implies that the present theory, while pointing towards a full 
unification based on $\ee$ or a very closely related structure, is not 
explicitly 
constrained by the prohibitive conclusions of~\cite{DiGa}.

   The octonion composition, exhibiting properties such as `triality', is 
anticipated to be at the heart of the unravelling of the full Standard 
Model spinor structure for a full three generations of leptons and quarks  
(\cite{TimeE} section~5, \cite{Unifi} discussion of equations~9.9--9.12).
 As noted for equations~\ref{slsb}--\ref{slbits}, 
\ref{esisb}--\ref{esibits} and \ref{esesb}--\ref{esebits} spinor states do 
here directly arise for a natural progression in extending the form of 
proper time
 (unlike for the restricted case of extra spatial dimensions in 
equation~\ref{sosb}--\ref{vnbits}).
   The nature of the augmentation from the one spinor identified in 
equation~\ref{slsb}--\ref{slbits} to four spinors with the property of the 
same handedness in equation~\ref{esisb}--\ref{esibits} was already 
intrinsically determined by properties of the octonion algebra. With both 
left-handed and right-handed components then identified in the extension to
 equation~\ref{esesb}--\ref{esebits} the spinor states exhibit 
  an accumulation of properties that increasingly resemble structures of 
the Standard Model, and in particular can be associated with $e$-leptons 
and $d$-quarks. The ideal situation would be to continue this progression 
by identifying a spinor structure also for $\nu$-leptons and $u$-quarks 
through a subsequent  $\ee$ stage without needing to contrive the final 
Standard Model symmetry features in any way.
 That this might in principle be achieved through an underlying octonion 
structure for an $\eeg$ action for a homogeneous form $\lvtfep$ for proper 
time in a natural mathematical manner constitutes a non-trivial prediction 
of the theory~\cite{TimeE}.

 Central to this aim is an understanding of the interconnections between 
various algebraic structures related to $\ee$ (including those reviewed 
in~\cite{TimeE} section~2). For example with the proposed $\eeg$ octic 
invariant in 248 real variables of equation~\ref{lvto} pursued as a 
generalisation from the $\esig$ cubic invariant defined on $\htho$ 
 for equation~\ref{lvts}
and the $\eseg$ quartic invariant on $F(\htho)$ for equation~\ref{lvfsq} a 
close relation to the properties of the exceptional Jordan algebra $\htho$ 
is implied. 
 As well as the central role as a `generalised spacetime'~\cite{Gunay2}, 
alluded to before equation~\ref{lvto}, 
the properties of $\htho$ are also inextricably linked with the $\esi$, 
$\ese$ and $\ee$ entries of the $4 \times 4$ `magic square' of Lie algebras 
(as noted in~\cite{TimeE} towards the end of subsections~2.1 and 5.2, 
citing for example~\cite{Baez1} section~4.3 and \cite{BaSu}, with the 
historical background described in \cite{Freud} section~4.12).
 In turn this magic square can be constructed in terms of `trialities' in a 
manner intimately relating to spinor structures~\cite{Evans1}, and it is 
the corresponding properties of octonion triality in particular that we 
wish to relate to a construction of an octic $\ee$ invariant as proposed 
for equation~\ref{lvto} -- in the pursuit of uncovering three generations 
of Standard Model spinor states for $\sltc \subset \ee$ through a symmetry 
breaking structure under equation~\ref{eebrk}.
 A link between the role of octonion triality in the magic square and that 
for the proposed $\ee$ octic invariant might for example be forged via the 
shared association of the space $\htho$ with these applications.

  With the properties of the octonions, uniquely the largest division 
algebra, central to these connections the mathematical structure required 
for equation~\ref{lvto} in obtaining the appropriate spinor properties for 
$\nu$-lepton and $u$-quark states might also be related to that employed in 
some studies of 
supersymmetry~\cite{Kugo,Evans2,Baez2,Baez3,Durak,Borst,Anast}. 
  These studies also relate explicitly to the above discussion of the magic 
square and triality (see for example~\cite{Durak} chapter~9, 
\cite{Borst,Anast}). 
 Such structures  in mathematical physics might then provide a pivotal 
guide for the further advancement of the present theory.

 Here however we have constructed and developed the theory from the 
elementary first principles described for equations~\ref{salpha} and 
\ref{lpvn}. While significant elements of the Standard Model have already 
been obtained through the $\esi$ level of 
\mbox{equations~\ref{lvts}--\ref{esibits}} and $\ese$ level of 
equations~\ref{lvfsq}--\ref{esebits} a more complete structure is predicted 
to emerge through an explicit analysis of the $\ee$ level based on 
equation~\ref{lvto}, as provisionally denoting this ultimate form for 
proper time. In the meantime, guided by the successes of the $\esi$ and 
$\ese$ levels and general considerations such as equations~\ref{eebrk} and 
\ref{eetoesi}, we can already anticipate possible features of new physics 
that may emerge beyond the Standard Model for the $\ee$ level, as we 
describe in the following section.

\section{Structures Beyond the Standard Model}
\label{uni4}

\subsection{The Case for Two Right-Handed Neutrinos}
\label{uni41}

   In general terms, while keeping in mind the caveats regarding the need 
to incorporate further spinor structures and a full electroweak theory,
given the potential of incorporating three generations of 
 charged leptons and quarks 
 at the $\ee$ level as suggested by equation~\ref{eetoesi} 
we might anticipate some of the implications for neutrino physics  
of an explicit expression for equation~\ref{lvto} with a symmetry breaking 
decomposition under equation~\ref{eebrk}.
 Extrapolating from the ambiguity of neutrino and Higgs components for the
 $\esi$ level in equation~\ref{esisb}--\ref{esibits} and the resolution of 
that ambiguity  at the $\ese$ level in equation~\ref{esesb}--\ref{esebits} 
the assumption of the simplest further augmentation for the neutrino sector 
within a three generation
 symmetry breaking pattern at the $\ee$ level
 points towards the progression:
\begin{eqnarray}
 L_3(\bv_{27})_{\mathrm{E}_6} = 1 \!\!\! & : &  \begin{tabular}{|c|} \hline 
        \raisebox{+0.25ex}{$\nu_L / \bh$} \\ 
                  \hline \end{tabular}
\qquad\qquad\qquad\qquad\qquad\qquad\;\;\;
         \mbox{(equation~\ref{esisb}--\ref{esibits})}	 \nonumber  \\
 L_4(\bv_{56})_{\mathrm{E}_7} = 1 \!\!\! & : &  \begin{tabular}{|c|} \hline 
        \raisebox{+0.25ex}{$\nu_L$}          \\ \hline \end{tabular} 
               \qquad \qquad \;\,\, \mbox{and} \quad
	           \begin{tabular}{|c|} \hline 
	    \raisebox{+0.25ex}{$\:\bh\:$} 	    \\ \hline \end{tabular}                    
   \qquad\qquad\;\;   
     \mbox{(equation~\ref{esesb}--\ref{esebits})}	   \nonumber  \\  
 \lvtfep \!\!\! & : &  \begin{tabular}{|c|} \hline 
        \raisebox{+0.25ex}{$\nu_L \;\;\, \nu_L \;\;\, \nu_L$}  
		                    \\ \hline \end{tabular}
	           \quad \mbox{and} \quad
	           \begin{tabular}{|c|} \hline 
		\raisebox{+0.25ex}{$\:\bh\:\;\;\, \nu_R \;\;\,\nu_R$} 	   
			    \\ \hline \end{tabular} 
			    \label{eebits}		
\end{eqnarray}
 Here the necessary projection of the original external components
  \mbox{$\bh \equiv \bv_4 \inn \TM_4$} is taken, without loss of 
generality, from the right-handed sector of components at this $\ee$ level, 
as for the case of the $\ese$ level in equation~\ref{esesb}--\ref{esebits}.
This schematic augmentation then suggests that the accommodation of a full 
three generations of both left and right-handed  charged leptons and quarks 
at the $\ee$ level  may be accompanied by 
  three generations of left-handed neutrinos but only  \textit{two} 
right-handed
   neutrinos, with the external components of $\bh$ associated with the 
Standard Model Higgs now prohibiting the identification of a third $\nu_R$ 
state (developing the comment made in \cite{TimeE} section~7, second bullet 
point).
 
  In this manner the present theory, developed from
  the first principles underlying equation~\ref{lpvn} upon generalising the 
form of proper time, could provide a unifying theoretical
   basis for phenomenological models based on two right-handed neutrino 
states, including~\cite{Framp,Ibar,Antu,Chang} as reviewed in 
subsection~\ref{uni11}.
 Given the firm conceptual foundation the further mathematical development 
of this theory then offers the potential to greatly narrow down the 
specific features of such models and lead to robust predictions.

 As a further implication of equation~\ref{eebits} the  Higgs sector, 
associated with the components of  $\bh \equiv \bv_4 \in \TM_4$, will be 
intimately connected with the neutrino sector, as had already been 
 suggested by the $\esi$ level of equation~\ref{esisb}--\ref{esibits}, 
  and we can now consider at the $\ee$ level how
  some of the Higgs and neutrino properties may be closely correlated. For 
example the need to identify a spinor structure for both the $\nu_L$ and 
$\nu_R$ states under the $\sltc \subset \ee$ action of equation~\ref{eebrk} 
on the components underlying $\lvtfep$ for the symmetry breaking pattern of 
equation~\ref{eebits} suggests that $\bh$ may also have an underlying 
spinor composition, rather than being directly extracted as $\bh \inn 
\htwc$ vector subcomponents as was the case for 
equations~\ref{slsb}--\ref{slbits}, \ref{esisb}--\ref{esibits} and 
\ref{esesb}--\ref{esebits}.

 Indeed in general spinor components can be combined to form vector objects
  as for $\psi\psi^{\dagger} \in \htwc$ with $\psi\inn \ccc^2$ in 
equations~\ref{lqinc} and \ref{lvnib}, suggesting that $\bh\inn\htwc$ in 
these equations might have a similar decomposition, or a generalisation of 
it, within a higher-dimensional form for proper time such as at the full 
$\ee$ level. On the other hand spinors can also be combined to form 
  scalar objects as is the case for Dirac or Majorana mass terms in a 
Lagrangian and also for $\vert \bh \vert$ in equation~\ref{hnorm} if $\bh$ 
is composed of spinors. This latter case, with the scalar $\vert \bh \vert$ 
associated with the Higgs, might then have some similarities with  
composite models in which the Higgs field is analogous to pion fields for a 
scaled up adaptation of QCD
(see for example~\cite{Azat,Goer} and references therein). While pions are 
composed of quark spinor states 
the possibility of Higgs-like composites constructed specifically 
from right-handed neutrino states is described 
in~\cite{AKLR,Smet,KrHi,Baren}, a connection which is here for 
equation~\ref{eebits} proposed to leave a distinct residual of two free 
$\nu_R$ states.
 Here the components of the would-be third $\nu_R$ state are proposed to be 
fused into an effective Higgs field through the necessary identification of 
the 4-vector projection $\bh \equiv \bv_4 \in \TM_4$
 of equation~\ref{l2v4h}, representing the local external spacetime 
structure.
 
  In the Higgs sector itself the implications of a composite structure can 
be investigated and constrained at the Large Hadron Collider (LHC;
   see for example \cite{Goer} section~4, \cite{Caren,Baner}).
 For composite Higgs models the coupling of the Higgs to fermion pairs can 
deviate from the Standard Model expectation by of order 10\%, sufficient 
for this new physics to also be observable at a 250$\,$GeV $e^+e^-$ linear 
collider  (\cite{Peskilc} section~5).

  Here for a detailed analysis and specific predictions an explicit 
expression for equation~\ref{lvto} and a full symmetry breaking structure 
for the components of $\bv_{248}\in \rrr^{248}$ under equation~\ref{eebrk} 
will be needed. As a progression from the broken $\slthc$ symmetry case of 
equation~\ref{lvnib}, via the $\esi$ and $\ese$ levels, this will also 
involve a full expansion of invariant terms for the broken $\ee$ form:
\begin{equation}
  \Lsl_8(\bv_{248})_{\mathrm{SL}(2,\ccc)\times \mathrm{SU}(3)\times
        \mathrm{SU}(2)\times  \mathrm{U}(1)}   \, = \, 
		  \sum (\mbox{invariant parts}) \,= \, 1  
\label{lvtob}
\end{equation}

 This expansion at the $\ee$ level is proposed to yield an
   explicit form for mass terms in a corresponding effective Lagrangian
    as described for equation~\ref{lpvnb}.
   In particular for
    the neutrino sector such terms will depend upon the nature of the 
embedding of the neutrino spinor structures, under the
  external $\sltc \in \ee$ symmetry  of equations~\ref{eebrk} and 
\ref{lvtob}, and their composition with a factor of the vacuum value for
 a Lorentz invariant scalar combination of the  components of $\bh$ such as
 $h=\vert \bh \vert$ in equation~\ref{hnorm}. These structures relating to 
spinor compositions will be intimately connected with the triality 
properties of the octonion composition that is anticipated to play a 
central role in the construction of the homogeneous polynomial form 
$\lvtfep$ itself, as described towards the end of the previous subsection.

 When partitioned into invariant pieces under the broken symmetry
 in the expansion of equation~\ref{lvtob} Dirac mass terms containing a 
factor of the form $\psi^{\dagger}_{L,R}\psi^{\ph{\dagger}}_{R,L}$ can be 
sought for the $e$-lepton and
 $d$-quark spinor states $\psi^{\ph{\dagger}}_{L,R}$ subsumed from the 
$\ese$ level of
  equation~\ref{esesb}--\ref{esebits}. The `$u$-quark' scalar components in 
equation~\ref{esesb}--\ref{esebits} will also need to be identified as 
spinors in the components of $\bv_{248}$ at the $\ee$ level, through the 
octonion composition triality properties, and similarly incorporated into 
such effective mass terms, and again for a full three generations. 
 For the case of the neutrino sector
  the nature of this spinor embedding at the $\ee$ level may determine both 
the number of $\nu_L$ and $\nu_R$ states as well as  
        the form and combination of Dirac and Majorana, or other, mass 
terms for this sector.

 In general a Majorana mass term contains a factor
 of the form $\psi^T_{L,R} \sigma^2 \psi^{\ph{T}}_{L,R}$, where $T$ is the 
transpose and 
 $\sigma^2 = \binom{0 \,\:\! -\;\!\! i}{i \;\,\; 0}$ is a Pauli matrix, 
which is numerically equal to zero for any spinor values 
$\psi^{\ph{T}}_{L,R} \in \ccc^2$.
 For a Lagrangian field theory with such a term the component 
$\psi^{\ph{T}}_{L,R}(x)$ hence needs to be treated as an anticommuting 
field even at the classical level, anticipating the statistical fermionic 
properties of such a spinor in the corresponding quantum theory
 (see for example~\cite{Pesk} problem~3.4).
 Similarly here the expansion of terms in equation~\ref{lvtob} may also 
require an understanding of
the role of the `spin-statistics theorem' in the QFT limit  in order to 
fully interpret and assess the nature of the 
 particle properties that might be deduced (as alluded to in \cite{KKone} 
subsection~5.3).
 Indeed the connection with QFT will be needed to assess the conception of 
particle states generally for the present theory, as will be discussed for 
equation~\ref{gfromavt} in subsection~\ref{uni51}.

    While the above possibilities are hence provisional one distinctive 
feature concerning the lightest left-handed neutrino can already be 
surmised.  
 With the components of $\bh \equiv \bv_4 \inn \TM_4$ being associated with 
the Higgs and the origin of mass (as discussed for 
equations~\ref{l2v4h}--\ref{gmnconh}) for this preliminary structure a 
clear basis for a mass asymmetry in the neutrino sector is also implied at 
the $\ee$ level in equation~\ref{eebits}, which hints at a form of `seesaw' 
imbalance between the left and right-handed states.  
 As reviewed in subsection~\ref{uni11} in a standard 
  neutrino model seesaw mechanism each $\nu_R$ state generates one $\nu_L$ 
state mass.
  Hence with only two $\nu_R$ states available in the projected structure 
outlined in equation~\ref{eebits}  there is a strong hint that the lightest 
active neutrino mass state may be massless, that is 
$\mmin = 0\,\mbox{eV}$.

   As also discussed in subsection~\ref{uni11},
  while ongoing and future experiments on tritium $\beta$-decay and 
neutrinoless double-$\beta$ decay will improve the corresponding 
constraints on $\mmin$, the most sensitive measurement is currently 
provided by the cosmological observations  limiting  the total mass of the 
active neutrino states, currently with $\mtot  < 0.12\,\mbox{eV}$ implying 
an upper bound of around 
$\mmin < 0.03\,\mbox{eV}$ (for the normal hierarchy)  
  or $\mmin < 0.02\,\mbox{eV}$ (for the inverted hierarchy).
 The prospects for further observations, alluded to in  
subsection~\ref{uni11}, suggest that within the $\Lambda$CDM cosmological 
model
 the consequences of even the most challenging case with $\mmin = 
0\,\mbox{eV}$ in the normal hierarchy (hence with $\mtot =0.06\,\mbox{eV}$) 
   could be detectable within the foreseeable future. If the value of 
$\mtot$ could be determined to be greater than $0.06\,\mbox{eV}$ and 
different from $0.10\,\mbox{eV}$ (allowing for the inverted hierarchy case) 
with statistical significance then the case for $\mmin = 0\,\mbox{eV}$ 
would be disfavoured and is hence testable. 
 As suggested in subsection~\ref{uni11} with reference to 
  (\cite{PDG18} figure~62.1, \cite{Drewes} figure~5) the most stringent 
test of a prediction for $\mmin = 0\,\mbox{eV}$ might be provided in the 
coming years by a combination of cosmological and neutrinoless 
double-$\beta$ decay data.

  The property of a lightest active neutrino mass $\mmin = 0\,\mbox{eV}$ is 
naturally shared with  models that \textit{assume} there are only two 
right-handed neutrinos~\cite{Framp,Ibar,Antu,Chang}, as noted in 
subsection~\ref{uni11}. Such models are `minimal'  in the quantitative 
sense of accounting for the well-established neutrino oscillation data with 
the smallest number of additional states over and above those of the 
Standard Model. 
 On the other hand 
 the $\nu$MSM~\cite{Asaka1,Asaka2}, for which $\mmin$ is non-zero but still 
far too small for the deviation from $\mmin = 0\,\mbox{eV}$ to ever be 
detected by any known means, 
  is `minimal' in the symmetric sense that a  uniform three-generation 
pattern of lepton and quark states is maintained without needing to assume 
an exception for the case of $\nu_R$ states -- although two of the $\nu_R$ 
states in the $\nu$MSM are distinguished in being much heavier than the 
third.
In all cases the above models incorporate two right-handed neutrinos that 
can account for the observed solar and atmospheric oscillations between 
left-handed neutrinos, via an appropriate structure of mass terms, 
 while also providing a mechanism for 
  the baryon asymmetry originating in the early universe, via 
\mbox{$C\!P$-violating} properties of these two heavy $\nu_R$ states.
  The additional state of the $\nu$MSM, the third and lighter $\nu_R$ 
component, provides a dark matter candidate, as also alluded to in 
subsection~\ref{uni11}.
 In the event of confirmation of any
 of the anomalous neutrino observations, discussed in 
subsection~\ref{uni11} in connection with reference~\cite{Doring}, a 
further extension from these models or an alternative approach would be 
required.

  In the absence of decisive observations, or  
 the guide of building up a theory from first principles, it may prove 
difficult to empirically distinguish between the above neutrino models.
 For all models the stark contrast between the empirical properties of
   left and right-handed neutrinos is essentially built in by hand, with 
the
 $\nu_R$ states being typically far heavier, sterile to Standard Model 
gauge interactions, and in some cases fewer in number, compared with the 
$\nu_L$ states.
 The present theory however \textit{does} derive from the first principles 
of a clear underlying  conceptual motivation
 in a generalisation of proper time as described for equations~\ref{salpha} 
and \ref{lpvn}.
 Through equations~\ref{esisb}--\ref{esibits} and 
\ref{esesb}--\ref{esebits} we are led to a proposed symmetry breaking 
pattern for the neutrino sector at the 
$\ee$ level as described for equation~\ref{eebits} in which a stark 
contrast between the left and right-handed states is indeed 
\textit{implied}. Hence  
 we do have a clear origin for a significant asymmetry between the 
properties of the $\nu_L$ and $\nu_R$ states, regardless of any further 
specific features.
 However,
   until further developed  there is an open
  question concerning how much new physics in the neutrino sector might be 
   accommodated, with the need to account for the present and future 
requirements 
   in empirical neutrino physics providing a possible means to test the 
theory.
  More generally the challenge will be to address a range of outstanding 
questions in 
   neutrino physics including those listed in the penultimate paragraph of 
subsection~\ref{uni11}. 
    
 In summary, the provisional structure of
  equation~\ref{eebits} 
 favours the case for the accommodation of only two right-handed neutrino 
states.
  In light of the corresponding neutrino models, if these two $\nu_R$ 
states are found to have the appropriate properties, this
 may be sufficient to account for both the compelling neutrino oscillation 
phenomena observed and in principle the baryon asymmetry of the universe. 
On the other hand in equation~\ref{eebits} there is no room to accommodate 
a third $\nu_R$ state, suggesting that $\mmin=0\,\mbox{eV}$. The third 
$\nu_R$ is not needed to account for dark matter here since such candidates 
may be provided by another sector of the symmetry breaking or through an 
alternative form for proper time itself, as we describe in the following 
subsection.

\subsection{Further New Physics and Potential Tests}
\label{uni42}

  The four scalar components $\{n,N,\alpha,\beta \}$ at the $\ese$ level in 
equation~\ref{esesb}--\ref{esebits}, in being invariant under the internal 
symmetry gauge group, provide a prototype set of dark matter candidates as 
noted in subsection~\ref{uni31}.
 As a progression from the lone scalar invariant $n$ of 
equations~\ref{slsb}--\ref{slbits} and \ref{esisb}--\ref{esibits} these 
components may
   generalise further into a broader range of possible dark matter states 
at the full $\ee$ symmetry level, invariant under the full Standard Model 
gauge group in equation~\ref{eebrk}, and offer the possibility to test this 
new physics and explore the corresponding  cosmological phenomena.
 Such components could in fact relate to an extended `dark sector' more 
generally, incorporating also the
  dark energy impact on the universe at the present epoch  
 and in principle the nature of any `inflationary' period during the very 
early stages of cosmic evolution (as considered in \cite{Unifi} 
chapter~13).

  While in the discussion of equation~\ref{eebits} in  the previous 
subsection  we described a close link between the neutrino and  Higgs 
sectors, here we note that a
    dark sector associated with a set of scalar invariants  
is also anticipated to be closely related to the Higgs.
 With the Higgs  associated with the projected components of
 $\bh \equiv \bv_4 \in \TM_4$ in equations~\ref{l2v4h} and \ref{hnorm} onto 
the external spacetime
   this latter connection is made 
  in particular through several dilation transformations
  identified in the symmetries of equations~\ref{lvni}, \ref{lvts} and 
\ref{lvfsq} 
   as described for (\cite{TimeE} equation~90),
with their possible role in the very early universe considered in more 
detail in  (\cite{Unifi} section~13.2).

  High energy physics experiments might also probe 
the link between the Higgs and scalar invariant components based on a 
generalisation of $\{n,N,\alpha,\beta \}$ through the possibility of 
`invisible Higgs decays'. Limits can be set on the branching ratio
 of the Higgs  to such a dark or `hidden' sector at the LHC via missing 
energy in events with other features observed that might accompany standard  
Higgs production~\cite{CMSinv,ATLASinv}. Similarly
 invisible Higgs decay events might also be detectable via a visible 
recoiling $Z^0$ boson decay at a future $e^+e^-$ collider (\cite{Peskilc} 
section~6).

  The connection between the Higgs and dark matter here may be similar to 
that of a `portal interaction' model, in particular for such a model 
incorporating more than one scalar invariant state (see for 
example~\cite{Casas}). Portal interactions with dark matter 
 can also be utilised in studies of a composite Higgs involving the 
neutrino sector~\cite{KrHi}, hence also connecting with the discussion of
  equation~\ref{eebits} in the previous subsection. 
   The nature of any new physics in the Higgs, dark and neutrino sectors
   would need to be mutually consistent within the context of the present 
theory, as well as with empirical observations in particle physics and 
cosmology more generally, broadening the scope for making predictions that 
might then be tested.

 A pivotal element in analysing the physics will be the role of the 
external components $\bh\equiv \bv_4 \in \TM_4$ as extracted from the full 
form for proper time in equation~\ref{lvto}. 
Through the simple conceptual underpinning for this theory there is only 
one fundamental mass scale as associated with the magnitude of this 
projected 4-vector $h = \vert \bh \vert =\vert \bv_4 \vert$ in 
equations~\ref{l2v4h} and \ref{hnorm}.
  That is, mass terms will be identified in the expansion of 
equation~\ref{lvtob} via the composition of matter fields $\psi(x)$
 with components of $\bh(x)$ such that their mutual
 interaction will imply $\delta h(x)$ variations that generate mass via the 
impact on the local external geometry through equation~\ref{gmnconh}, in 
which the overall scale is set by the vacuum value for $h = \vert \bh 
\vert$. 
  Through the dilation transformations
  a very different value for $h$ might have been attained in the very early 
universe, with for example $h\to 0$ for cosmic time $t\to 0$ considered for
 (\cite{Unifi} figure~13.3), which might correspond to a much higher 
primordial mass scale.  
 However
 this magnitude is presumed to have stabilised in the immediate aftermath 
of the `Big Bang' with a transition to a fixed vacuum value (denoted 
$h=h_0$ in \cite{Unifi} section~13.2). At this early epoch of cosmic 
evolution the properties of the Standard Model emerged, in principle with 
any `hierarchy problem' avoided since only a single stable basic mass scale 
remains (as discussed in~\cite{KKone} towards the end of subsection~5.3).

 While we have considered a unique sequence of mathematical structures for 
equation~\ref{lpvn} in augmenting from the Lorentz symmetry of 
equation~\ref{lqdet} to the proposed  $\ee$ case of $\lvtfep$ in 
equation~\ref{lvto} the possibility remains of an alternative extension 
from
 the local 4-dimensional spacetime form
$L_2(\bv_4)_{\mathrm{Lorentz}}=1$ to
 a full form for proper time that might be denoted:
\begin{equation}
\label{lpvndsh} 
  L_{p'}(\bv_{n'})_{\hG'} \; = \; 1
\end{equation}
  and hence also subsuming equation~\ref{sfourd}.
 Indeed the original `extra spatial dimensions' form of 
equations~\ref{snd}, \ref{vnd} and \ref{l2vn} presents such a possibility.
 Variations $\delta h(x)$ in equation~\ref{hnorm} for the magnitude $\vert 
\bv_4 \vert$ in equation~\ref{l2v4h} can be associated with the projection 
of the common fragment $\bv_4 \in \TM_4$ 
 onto the local external spacetime out of
 \textit{both} 
 $\bv_{n'} \inn \rrr^{n'}$ of equation~\ref{lpvndsh}
  as well as $\bv_{248} \inn \rrr^{248}$ of equation~\ref{lvto},
   impacting the geometry of our universe and exhibiting gravitational 
effects as described for equation~\ref{gmnconh}.
 However the symmetry breaking down to
  $\mbox{Lorentz} \times G' \subset \hG'$, corresponding to 
equation~\ref{gbreak} for this new projection over the local structure of 
$M_4$, would yield internal gauge symmetries $G'$ and $\bv_{n'}$ fragments 
from equation~\ref{lpvndsh} independently  of the symmetry breaking pattern 
of $\hG = \ee$ on the original $\bv_{248}$ components of 
equation~\ref{lvto}.

 Hence the `ordinary matter' deriving from the breaking of $\lvtfep$ would 
not interact with 
   matter fields deriving from the parallel form  
\mbox{$L_{p'}(\bv_{n'})_{\hG'} = 1$} via any gauge forces, and hence the 
latter could also be considered as a potential source of `dark matter'. The 
gravitational link between the two forms of proper time for $\bv_{248}$ and 
$\bv_{n'}$  via the common projection $\bh \equiv \bv_4 \in \TM_4$ is 
perhaps even more reminiscent of a kind of `Higgs portal interaction' as 
alluded to above~\cite{Casas}.
 For the example in which such a parallel dark or hidden sector derives
 from the breaking of equations~\ref{vnd} and \ref{l2vn}, as pictured in 
figure~\ref{vfdvnd}(b),
  the fragments of equation~\ref{sosb}--\ref{vnbits} are obtained, and 
hence the matter field $\underline{\bv}_{n-4}(x)$ associated with `extra 
spatial dimensions' might then yet play a role in the present theory as a 
candidate source of dark matter that can interact gravitationally with 
ordinary matter.
   The nature of this quadratic form, with a potential for $n\to \infty$ 
and a role  for `fractal-like' or `Bott periodicity' properties (see also 
discussion in \cite{Unifi} towards end of section~13.3), 
    and the possibility of several parallel hidden sectors each describing 
a form for equation~\ref{lpvndsh}, might also account for the excess of 
dark matter over ordinary matter by around a factor of 
five~(\cite{PDG18} sections~2 and 26).
While hence identifying a possible role for alternative expressions for 
equation~\ref{lpvn}
 here we focus on the properties of matter fields deriving from the $\ee$ 
form of equation~\ref{lvto}, proposed to be responsible for the properties 
of our visible universe.

  With the physics beyond the Standard Model
  deriving from equation~\ref{lvto}
   potentially involving an extended Higgs sector closely related to 
  the neutrino sector as well as elements of   
   an extended dark sector an explicit branching of the components of 
$\bv_{248}$ in equation~\ref{lvto} under the broken symmetry of 
equation~\ref{eebrk}, as well as a full expansion of the invariant terms 
for equation~\ref{lvtob}, may be
  needed to untangle the various physics components according to their
    transformation properties under the external and internal symmetries.
	As noted in the discussion following equation~\ref{lvtob}  the full 
physical picture will require an understanding of the QFT limit and the 
nature of field quantisation itself, the origin of which for the present 
theory will be reviewed for equation~\ref{gfromavt} in the following 
section. However, as for the analysis described in section~\ref{uni3}, a 
number of features can be directly deduced from the elementary symmetry 
breaking structure for comparison with the Standard Model.

 	In the Standard Model  Yukawa couplings are added to mass terms in the 
Lagrangian by hand alongside the Higgs vacuum value to determine the 
fermion mass matrices (as described for~\cite{Unifi} equations~7.69 and 
7.70).
 Given the potential source of dark matter through alternative forms for 
proper time, as suggested for equation~\ref{lpvndsh} above,
 the specific composition of
 stable vacuum values emerging from the Big Bang for scalar invariants, 
such as $\{n,N,\alpha,\beta\}$ at the $\ese$ level, in the expansion of 
equation~\ref{lpvnb} could play the role of Yukawa couplings and 
 complete the `mass term' interpretation.
 Indeed under  augmentation to
 the full $\ee$ case, with terms in 
  the expansion of equation~\ref{lvtob} being of octic order, a combination 
of several such factors into an effective Yukawa coupling might be needed, 
alongside the Higgs represented by
 a vacuum value scalar combination of components of $\bh\equiv \bv_4 \in 
\TM_4$ and the components of spinor states,  if such terms are to 
closely resemble Lagrangian mass terms and account for the wide range of 
observed fermion masses.

 An understanding of these factors may hence be central in particular in 
ultimately calculating specific masses for the left and right-handed 
neutrinos while specifically incorporating any corresponding `seesaw' 
imbalance, as provisionally suggested by equation~\ref{eebits}, with 
potentially a large Majorana mass sought for the $\nu_R$ states 
 consistent with the models described in subsection~\ref{uni11}.  
 As well as accounting for the hierarchy of neutrino masses and `textures' 
of the neutrino mass matrix
  (see for example~\cite{Ibar,LYZ})
  the aim would be to explain the Standard Model lepton and quark mass 
spectrum more generally.

   From equation~\ref{eebrk} at the $\ee$ level the further breaking of the 
electroweak symmetry down to $\uo_Q \subset \sutw_L \times \uo_Y$ is 
proposed to involve a non-trivial action of a subset of the $\sutw_L \times 
\uo_Y$ transformations upon components of the
  4-vector \mbox{$\bh \equiv \bv_4 \in \TM_4$}, itself projected onto the 
local external spacetime tangent space from subcomponents of the full set 
of $\bv_{248} \in \rrr^{248}$ components in equation~\ref{lvto}. 
 This is considered to account for the masses of the $W^{\pm}$ and $Z^0$ 
gauge bosons (as suggested by the study of $\sutw \times \uo \subset \esi$ 
subgroups in~\cite{Unifi} section~8.3, alluded to here in 
subsection~\ref{uni31}).
  For a `composite' Higgs the set of spinors acted upon by the  $\sutw_L 
\times \uo_Y$ symmetry could include the components of a spinor 
decomposition of  $\bh \equiv \bv_4 \in \TM_4$.
  While such a composition of the 4-vector $\bh \equiv \bv_4 \in \TM_4$ in 
terms of spinor subcomponents under the external $\sltc$ symmetry was 
suggested in the previous subsection, the physical Higgs itself might be 
associated with a more extended set of components in the decomposition of 
$\bv_{248}\in\rrr^{248}$ under equation~\ref{eebrk} and need to be 
disentangled collectively from the neutrino states, elements of the dark 
sector and also Yukawa factors as considered in this subsection.

   The Higgs coupling to the $W^{\pm}$ and $Z^0$ gauge bosons, as well as 
to the two heaviest quarks (top and bottom) and two heaviest leptons (tau 
and muon, with an upper limit in the latter case) have been determined at 
the LHC~\cite{LHChb,CMShb,ATLAShb}. These Higgs couplings are seen to be 
directly proportional to the mass of the particle coupled to (\cite{LHChb} 
figure~5, \cite{CMShb} figure~10, \cite{ATLAShb} figure~15), consistent 
with the Standard Model prediction and allowing constraints to be placed on 
new Higgs physics (see for example~\cite{CMShb} section~9).  For the 
present theory
 the underlying origin of this high degree of uniformity arises from the 
direct connection between the couplings to the identified
  Higgs components in the symmetry breaking of equation~\ref{lpvn}, in 
particular in the form of equation~\ref{lvto}, and the conception of mass 
in general relativity as discussed for 
equations~\ref{l2v4h}--\ref{gmnconh}. 
 Particle masses are here to be determined from terms in 
equation~\ref{lvtob} by a combination of specific scalar invariant `Yukawa' 
coupling factors in composition with Higgs components together with the 
uniform vacuum value $h = \vert \bh \vert$ for the Higgs that underpins the 
overall mass scale. 
 Such properties, and
 deviations that might arise in a full development of the theory, might be 
tested as  measurements improve with more data at the LHC, while further 
tests of Higgs couplings would be possible at a future ILC 
experiment~\cite{Peskilc} as noted in the previous subsection.

 For the present theory as well as the Higgs couplings and particle mass 
spectrum 
  a full understanding of electroweak theory is to be sought at the $\ee$ 
level, in particular with an internal $\sutw_L$ not only impacting upon the 
Higgs subcomponents in an appropriate way but also acting
 on doublets of left-handed leptons and doublets of left-handed quarks.
  In augmenting from the $\ese$ level of 
equation~\ref{esesb}--\ref{esebits} this will require the identification of 
Lorentz spinor properties for both $\nu$-lepton and $u$-quark states, and 
for a full three generations of leptons and quarks.  This ambition is 
expected to depend upon the intrinsic incorporation of the properties of 
octonion triality as discussed in
 subsection~\ref{uni32}.

That all of these required features of the Standard Model in augmenting 
from the $\ese$ level to the $\ee$ level are correlated is further 
emphasised by noting that the $\sutw_L$ doublet actions are anticipated to 
meld with the structure of the fermion mass terms and `textures',
 from the partitioning of equation~\ref{lvto} under the breaking of 
equation~\ref{eebrk} to the invariant terms of equation~\ref{lvtob}, in a 
manner relating to inter-generation mixing. With the symmetry breaking 
pattern, including the internal $\sutw_L$ transformations, oriented around 
the external $\sltc$ action on the external 4-vector
  $\bh \equiv \bv_4 \in \TM_4$ the fermion mass and flavour generation 
mechanisms will be intimately related to this projection.
  In particular with $\bh$ projected out of the `neutrino sector' as 
described for equation~\ref{eebits} the very different PMNS and CKM weak 
mixing matrices for the lepton and quark sectors respectively might in 
principle be accounted for.
This will relate to the nature of $C\!P$-violating effects in both sectors, 
and in particular for $\nu_L$ phenomenology in the leptonic case.
 While in turn a link can be made between \mbox{$C\!P$-violation} for
  $\nu_L$ states and that for $\nu_R$ states in a model-dependent manner, 
as reviewed in subsection~\ref{uni11}, here the aim will be to determine 
this link from first principles to establish whether $C\!P$-violating 
effects
 associated with $\nu_R$ phenomenology  
 in the very early universe might be on an appropriate scale to act as a 
potential source of the baryon asymmetry as proposed by models described in 
subsection~\ref{uni11}.

  Since the form of equation~\ref{lvto} for the $\ee$ level involves a much 
larger space of components in $\bv_{248} \in \rrr^{248}$,  beyond the
 27-dimensional
 $\esi$ and 56-dimensional $\ese$ levels of subsection~\ref{uni31}, in 
addition to accommodating the three generations of known leptons and quarks 
together with two right-handed neutrinos as well as a non-standard Higgs
 with associated Yukawa couplings 
 and potential dark sector candidates further new particles beyond the 
Standard Model might also  in principle be predicted, with the potential 
for further laboratory tests.
  In addition, for the full symmetry group $\hG = \ee$,  
 the complete breaking pattern might also accommodate further internal 
gauge groups beyond that of the Standard Model in equation~\ref{eebrk} (see 
for example~\cite{Unifi} equation~9.51), in principle implying new gauge 
interactions that might also have observable consequences in high energy 
physics experiments. While a range of additional states beyond the Standard 
Model might hence be accommodated a large multiplicity of new states is not 
anticipated, unlike for example the case in general for supersymmetric 
models and in particular for extended supersymmetry. 

  With the further required symmetry and representation features of the 
Standard Model beyond
    equation~\mbox{\ref{esesb}--\ref{esebits}} being mutually closely 
correlated, as noted in the discussion above, it is plausible that they may 
all be uncovered together in one further augmentation from the $\ese$ form 
of equation~\ref{lvfsq} to the proposed $\ee$ form described for 
equation~\ref{lvto} (as considered in detail in~\cite{TimeE}). If these 
required features do emerge at the $\ee$ level this will provide a very 
firm basis for precise predictions of a wealth of physics beyond the 
Standard Model. In this section we have described the nature of the new 
physics, including for the Higgs, neutrino and dark sectors, that can 
already be anticipated.
 In the previous subsection we have considered the particular significance 
of the
  intrinsic 
  left-right asymmetry  for the neutrino sector 
 associated with equation~\ref{eebits}, and emphasised the manner in which 
the theory favours models with two, and only two, right-handed neutrino 
states 
 with a close link to the Higgs sector. In this subsection we have
  described in particular how these interconnections might extend into the
  specific structure of mass terms for particle states in 
equation~\ref{lvtob} and into a 
   `portal' interaction with a
   dark sector of the theory as described for equation~\ref{lpvndsh}, 
augmenting the potential scope for empirical tests in the particle physics 
laboratory or through cosmological observations.

 While the analysis for the $\ee$ level has been of a provisional nature
  we note that here we are not building a \textit{model} pragmatically, for 
example by adding terms to a Lagrangian by hand, but rather the theory has 
been developed from \textit{elementary first principles}, as we discuss 
further in the remaining two sections. 
 Precise empirical predictions 
 will require a full understanding of the structure of the theoretically 
predicted $\eeg$ symmetry action for the full form of time $\lvtfep$ in 
equation~\ref{lvto}
 and a symmetry breaking pattern.
  However given the simple unifying basis underlying equation~\ref{lpvn} in 
terms of a generalisation of proper time an opportunity to uncover the 
ultimate origin of the physics of the Standard Model and beyond at the most 
fundamental level is in principle provided by this theory.
  In the following section we return to consider the elementary foundations 
of the theory from an historical perspective.

\section{Interpretation of the Theory}
\label{uni5}

\subsection{Relativity Further Generalised}
\label{uni51}

  In 1922, seven years after the generalisation from special relativity had 
yielded a theory of gravitation, concerning the ambition for a geometric 
unification with electromagnetism, and in light of the first attempts by 
Weyl and Kaluza, 
 Einstein wrote in a letter to Weyl: `I believe that in order to make real 
progress one must again ferret out some general principle from 
nature'~(\cite{Pais} section~17(b)).
  More specifically, in the early 1930s and having embarked upon his own
  endeavour to find such a unified field theory, Einstein summed up the 
nature of the task faced with the question~\cite{Eincont}:
 \begin{quotation}
    Is there a theory of the continuum in which a new structural element 
appears side by side with the metric such that it forms a single whole 
together with the metric?
 \end{quotation}

   This principle, closely associated with the sentiment expressed by 
Einstein that the quest for unification should be guided by simplicity, 
   applies to the unified field theories of Weyl, Einstein himself and 
Kaluza/Klein, for which the metric $g_{\mu\nu}(x)$ of equation~\ref{gfourd} 
is augmented as reviewed here for equations~\ref{WeylU}, \ref{EinU} and 
\ref{KaKlU} respectively in subsection~\ref{uni12}.
 The aim of each of those early unification schemes was to incorporate a 
theory of electromagnetism alongside the theory of gravity via a minimal 
extension from the latter theory. In each case the corresponding 
generalisation was one that could be  
  interpreted as dropping a further geometric assumption from the framework 
of general relativity, as also described in subsection~\ref{uni12}. 

 For general relativity itself a theory of gravity had been obtained on 
dropping the assumption that spacetime should be globally flat, as noted 
before equation~\ref{sfourd}, with special relativity applying only 
locally. This was a somewhat counter-intuitive starting point since the 
external 3-dimensional space with which we are very familiar appears to 
possess extended Euclidean properties, and it 
 had seemed perfectly natural to assume such properties as a basis for all 
science for centuries. As well as requiring a significantly more 
complicated mathematical description taking away the assumption of a stable 
flat background arena was hence
 somewhat disorienting, even though justified by the explanatory power and 
empirical success in accounting for gravitational phenomena.

  Since, like gravity, electromagnetism is a long range force it was 
reasonable to search for a generalisation from general relativity by 
dropping a further assumption concerning the \textit{global} geometry of 
4-dimensional spacetime. Ideally the aim for such a unified field theory 
was not only to incorporate classical electromagnetism alongside 
gravitation but also to account for particle states in terms of classical 
field solutions and to provide an underlying \textit{explanation} for 
quantum theory  without relying on seemingly ad hoc  \textit{postulates} 
(\cite{Pais} chapters~17 and 26).
 Somewhat like the limitations of Newtonian mechanics, for Einstein the 
apparent incompleteness of quantum mechanics, in particular in terms of its 
probabilistic nature, could not be addressed by incremental internal 
refinements or interpretations, but rather demonstrated the need for an 
explanation from a \textit{new} foundation, such as might be achieved 
through a  generalisation of general relativity.
  Indeed for Einstein this incompleteness of quantum theory was in itself a 
significant motivation for the need of a unified field theory,
  from which quantum mechanics might emerge as a limiting case.
    This initially applied to the `old quantum theory' associated with 
Planck's radiation law of 1900 and the Bohr atom of 1913, and later for the 
`new quantum theory' of Heisenberg and Schr\"{o}dinger from around 1925.

 An early conception proposed that particle-like properties might be 
associated with `energy-knots' of very high classical field 
values~(\cite{Weyl2} chapter~III section~25).
  The original paragraph containing the above quote from Einstein goes on 
to consider whether simple field laws might be obtained to describe the 
properties of both gravitational and electromagnetic fields, 
and continues~\cite{Eincont}:
 \begin{quotation}
    Then there is the further question whether the corpuscles (electrons 
and protons) can be regarded as locations of particularly dense fields, 
whose movements are determined by the field equations.
 \end{quotation}

   A similar hypothesis is  alluded to in the `Concluding Remarks' of 
Klein's
 paper~\cite{Klein} where it is also suggested that the properties of 
quantum phenomena may 
 originate out of a projection from a \mbox{5-dimensional} spacetime.   
  That the indeterminacy of quantum mechanics might arise through an 
inherent ambiguity of 4-dimensional physical laws obtained from an 
embedding within the \mbox{5-dimensional} spacetime of Kaluza and Klein is 
described in `Part~III: Unified Field Theories' of Bergmann's 1942 textbook 
on relativity 
  (as quoted and discussed in \cite{Pais} section~17(c)).
 While that quest for such a unification was ultimately unsuccessful 
  many modern-day models build upon
    the elegant idea of Kaluza-Klein theory, as noted in 
subsection~\ref{uni21}, featuring further extra spatial dimensions in 
aiming to address the wider challenge of accommodating the Standard Model 
of particle physics. However the postulates of quantum theory and the 
quantum particle description are generally adopted, or adapted, as a basis 
for such unification schemes without any further
 underlying explanation.

  Upon reconsidering the motivation for such a unified framework in 
subsection~\ref{uni22}
 we noted that since we do not perceive any `extra dimensions' there seems 
little justification for assuming such components, if they exist at all, to 
possess any form of `spatial properties' even in the purely local sense of 
consistency with the Pythagorean theorem for infinitesimal intervals as 
described for equation~\ref{pythag} and as might be depicted in
 figure~\ref{vfdvnd}(b).
  Dropping this assumption is in some sense more natural than relaxing the 
assumption of a flat external spacetime since we do not even \textit{see} 
the extra dimensions, and hence there is no compelling argument to restrict 
the generalisation of a local
proper time interval from the quadratic expression of equation~\ref{sfourd} 
to that of equation~\ref{snd}, leading to the more general expression of 
equation~\ref{salpha}. 
 Given that the notion of extra spatial dimensions is now very familiar 
loosening \textit{that} preconception and generalising to the local form of
 equation~\ref{salpha} might itself seem disorienting,
  with for example the `visualisation' of figure~\ref{vfdvnd}(b) no longer 
applying. 
 However a mathematical path can be followed 
 and affirmation sought through empirical success, 
 as was the case for the extended non-flat geometry of general relativity.

With reference to the first displayed quote from Einstein in the opening of 
this subsection here
  equation~\ref{salpha}, equivalent to equation~\ref{lpvn}, is proposed as 
a \mbox{`single whole'}, deriving from the continuum of proper time, which 
can incorporate the 4-dimensional local spacetime metric $\eta_{ab}$ of 
equation~\ref{sfourd} as described for equation~\ref{sfourp} and 
exemplified by the cubic form of equations~\ref{lvni} and \ref{lqinc}, 
`side by side' with additional structures that are interpreted as a basis 
for matter fields. 
  For this theory the minimal symmetric extension from the matterless 
vacuum case of equation~\ref{lqdet} to equation~\ref{lvni} also 
incorporates a framework for electromagnetism together with gravitation as 
described for equations~\ref{gchift}--\ref{slbits}, similarly as for the 
ambition of early unified field theories. Here the potential for further 
natural extension described in section~\ref{uni3} 
 leads, via a unique sequence of mathematical structures utilised for 
equations~\ref{lvts}--\ref{esebits}, directly and efficiently to a series 
of explicit structures of the Standard Model, hence vindicating this 
approach, and with the direction of further extrapolation providing insight 
into new physics beyond as described in section~\ref{uni4}.

A fundamental dependence upon the  continuum properties of time and 
infinitesimal intervals can in fact be traced back somewhat further to the 
earliest equations of physics in the 17$^{\mathrm{th}}$ century as 
developed by Newton. The methods of calculus invented by Newton for this 
purpose incorporate at the most elementary level the notion of a `fluxion' 
$\dot{x}$ for the rate of change of a quantity $x$ with respect to 
progression in time~\cite{New2}. The definition of such a fluxion, or 
derivative, involved taking the limit of a ratio of two quantities both 
decreasing without any finite bound. (Such a conception was not without 
controversy for many years, with fluxions famously criticised as a `Ghosts 
of departed Quantities' by Berkeley in `The Analyst' in 1734). The same 
continuum property of time is central to the present theory on taking 
infinitesimal intervals approaching the limit $\delta s \to 0$ in writing 
down the general arithmetic form for proper time in equation~\ref{salpha}. 
We can note also that in forming the `fluxions' $v^a$ defined for 
equation~\ref{lpvn} the quantities $\delta x^a$ and $\delta s$ composing 
this limiting ratio are intimately linked through equation~\ref{salpha}.

  Hence the historical roots concerning the conception of time for the 
present theory date back before relativity to the Newtonian worldview of 
the 18$^{\mathrm{th}}$ and 19$^{\mathrm{th}}$ centuries. Throughout that 
era the purely \textit{linear} progression in absolute time provided a 
fundamental independent parameter for the recording of events in a 
3-dimensional Euclidean arena of absolute space with global coordinates 
$(x^1,x^2,x^3)$. Spatial lengths of arbitrary extent $\Delta \Sigma$ could 
be determined by the simple 
 Pythagorean relationship $(\Delta \Sigma)^2 = (\Delta x^1)^2+(\Delta 
x^2)^2+(\Delta x^3)^2$ invariant under 3-dimensional rotations.  
 In the early 20$^{\mathrm{th}}$ century  for special relativity the time 
coordinate $(x^0)$ was also introduced into this \textit{quadratic} 
structure,  now augmented to a 4-dimensional spacetime form for arbitrary 
`proper time' intervals $\Delta s$ with the form
 $(\Delta s)^2 = (\Delta x^0)^2 - (\Delta \Sigma)^2$
  invariant under global Lorentz transformations between inertial reference 
frames in this Minkowski spacetime.
    The space and time components of the corresponding globally defined
	   coordinates $(x^0,x^1,x^2,x^3)$  are hence connected through the 
Lorentz invariant form of equation~\ref{sfourd} expressed for arbitrary 
finite  proper time and coordinate intervals, 
  as described after   equation~\ref{sfourd}.  
  This form for proper time was determined by 
  the postulates of special relativity, and in particular the constancy of 
the speed of light with $\Delta s = 0$ in any inertial frame, and implies 
that each such rest frame carries its own proper time parameter, abandoning 
universal Newtonian time.

 In general relativity the 
 Lorentz metric $\eta_{ab}$  is necessarily replaced by the general 
 symmetric metric function $g_{\mu\nu}(x)$ of equation~\ref{gfourd} with 
respect to global 
 coordinates on the extended
  scale, in a form for the infinitesimal proper time interval $\delta s$ 
now invariant under general coordinate transformations. Through this more 
general global framework a \textit{unification} is  achieved between the 
geometry of 4-dimensional spacetime itself and gravitation, with the 
structure of the former inextricably 
  accommodating a theory of the latter.
  For this progression from special relativity to the more general geometry 
   of general relativity 
inertial frames are strictly local,
   with Lorentz invariance only holding for \textit{infinitesimal} 
   intervals of proper time as introduced here for $\delta s$ in the 
quadratic form of equation~\ref{sfourd}.

  While incorporating the linear flow of time within such quadratic 
expressions might have originally been considered somewhat eccentric, 
 the present theory 
 represents a further natural progression in the conception of time through 
an arithmetic generalisation of invariant expressions for
 \textit{infinitesimal} proper time intervals to forms of 
\textit{greater-than-quadratic} order through equation~\ref{salpha}.
  As a progression from the central role played by the invariance of proper 
time in special and general relativity, and again noting the quote from 
Einstein in the opening of this subsection~\cite{Eincont}, it is the 
elementary properties of this local \textit{continuum of time} that we 
exploit for the theory proposed here
 in permitting this generalisation of the infinitesimal proper time 
interval $\delta s$ 
 described for equation~\ref{salpha}, which can be written as 
equation~\ref{lpvn}.
 With the residual components, over and above a local 4-dimensional 
spacetime structure, and their properties providing the source of matter 
fields this elementary local generalisation of proper time leads to a 
\textit{unification} of 
 4-dimensional spacetime with structures resembling the Standard Model at 
the elementary particle level of matter as reviewed in section~\ref{uni3}.
 The conceptual comparison and contrast with general relativity
  as depicted in figure~\ref{grcfme}.
  
\begin{figure}[htbp]  
\centering
\epsfxsize=14.4cm
\leavevmode
\epsffile[0 0 1644 733]{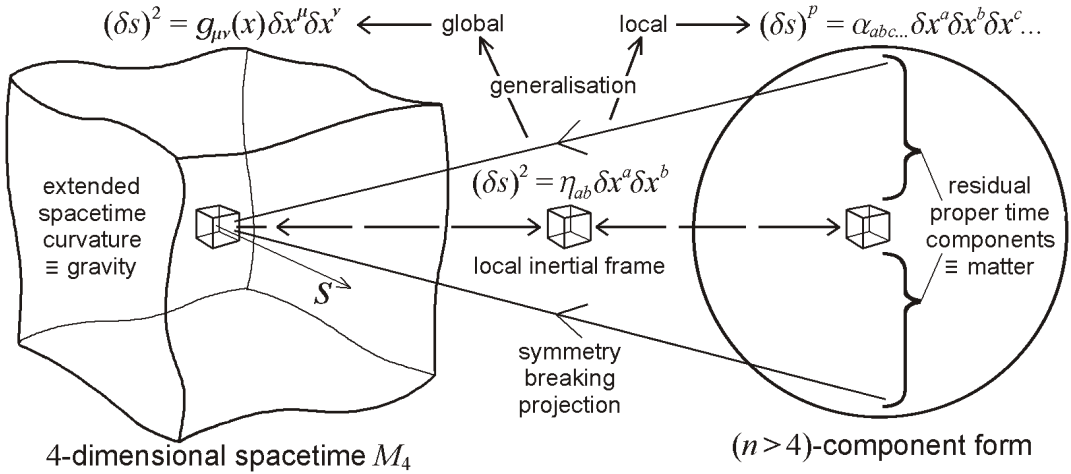}
\caption{\setb  Complementary generalisations of expressions for a  proper 
time interval from the Lorentzian form $(\delta s)^2 = \eta_{ab} \delta x^a 
\delta x^b$: globally to the metric field $g_{\mu\nu}(x)$ underlying a 
theory gravity and
  locally to non-quadratic forms with coefficients $\alpha_{abc\ldots}$ 
providing a basis for matter fields through a local projection over 
4-dimensional spacetime $M_4$.}
\label{grcfme}
\end{figure}

   Essentially figure~\ref{grcfme} describes two complementary 
generalisations from special relativity: globally for general relativity 
and locally for the present theory.
  In general relativity the source of gravitation in the form of spacetime 
curvature, described by the Einstein tensor $G^{\mu\nu}(x)$ which is a 
function of the first and second derivatives of the metric $g_{\mu\nu}(x)$, 
is identified with \textit{posited} forms of matter,
 described by the energy-momentum tensor $T^{\mu\nu}(x)$,
  through Einstein's field equation:
\begin{equation}
 \label{Eineq}
    G^{\mu\nu} = -\kappa T^{\mu\nu}
\end{equation}
 with $\kappa$ a normalisation constant.
  For the present theory
  matter  itself \textit{derives}  from the complementary local 
generalisation of a local proper time interval depicted in 
figure~\ref{grcfme}.
 More precisely the properties of the residual components in the projection 
  of the generalised local form of proper time
 over $M_4$ directly shape the spacetime geometry on the extended scale 
through which the energy-momentum is in turn defined (\cite{Unifi} 
equation~15.1, \cite{TimeE} equation~85): 
\begin{equation}
 \label{gfromavt}
  G^{\mu\nu} \: = \: f(A,\bv_n)   \: =: \:  -\kappa T^{\mu\nu}   
\end{equation}  

 The structure of the spacetime geometry,
 represented by the function $f(A,\bv_n)$,
  includes a contribution from gauge fields $A(x)$ via the corresponding 
field strength tensor $F(x)$, such as the electromagnetic field, in a 
manner similar to modern Kaluza-Klein theories  as reviewed here for 
equation~\ref{gchift}, as well as from subcomponents of $\bv_n(x)$ of the 
full form of proper time for equation~\ref{lpvn} via the impact of the 
projected 4-dimensional spacetime component $\bv_4 \in \TM_4$, as described 
here for equations~\mbox{\ref{l2v4h}--\ref{gmnconh}} and \ref{lpvnb}.    
  The quantum properties of matter are proposed to arise as an intrinsic 
feature through a degeneracy of spacetime solutions for 
equation~\ref{gfromavt}
 (\cite{Unifi} 
section~10.1, chapter~11 and section~15.2),
 with quantum particle states exhibiting the properties of 
equation~\ref{esesb}--\ref{esebits} as generalised further  for the full 
form of time proposed for equation~\ref{lvto}. 
 In the limit of a flat spacetime approximation, with field dynamics 
constrained by equation~\ref{lvtob}, 
  the local interactions of a QFT and quantum particle effects exhibiting 
the properties of the Standard Model are envisaged to arise from the 
symmetry breaking projection of the full form for proper time over the 
local \mbox{4-dimensional} spacetime.
  In this manner the properties of `corpuscles' such as electron states, 
alluding to the second of the above displayed quotes from Einstein, are 
proposed to be identified from this further generalisation of relativity.

 For the Standard Model itself, as usually presented, fields are added to a 
flat 4-dimensional spacetime background and then quantisation rules 
applied, completely detached from any consideration of general relativity. 
 This standard theoretical framework reflects the negligible impact of 
gravity on phenomena recorded in the particle physics laboratory. A more 
technically challenging approach is needed for any attempt to incorporate 
gravitation within a theoretical scheme in which quantisation rules are 
afforded
 precedence, as for the ambition of string theory -- which also aims to 
accommodate the Standard Model as discussed in subsection~\ref{uni21}. Here 
we begin with the very local structure of general relativity and make the 
generalisation for a proper time interval as depicted in 
figure~\ref{grcfme}. Through this unrestricted augmentation, beyond 
quadratic spacetime forms, matter fields are obtained with features both 
resembling  the Standard Model \textit{and} also in principle accommodating 
probabilistic quantum properties through equation~\ref{gfromavt}. Here the 
nature of gravity is inextricably linked with these properties of matter, 
without needing to be subsequently appended.

  Hence the complete theory is not anticipated to involve classical fields 
in spacetime which are \textit{then} `quantised' by applying the postulates 
of a quantum field theory (such as reviewed in \cite{Unifi} chapter~10), 
rather the aim is to account for and \textit{explain} quantum phenomena 
through this generalisation of the local spacetime structure. 
   Matter, together with its quantum properties, is directly correlated 
with the spacetime geometry through equation~\ref{gfromavt},
 which also describes gravity by incorporating Einstein's field 
equation~\ref{Eineq} along with an account of the elementary composition of 
the energy-momentum tensor. The gravitational field $g_{\mu\nu}(x)$ remains 
an essentially classical field on a smooth spacetime manifold, with no 
`quantisation' of the spacetime geometry itself.
This perspective on quantum theory and particle states is then in a similar 
spirit to that of the ambition of the early unified field theories as 
described earlier in this subsection.

  As described in subsection~\ref{uni22} and earlier in this subsection the 
present theory can also be motivated in a similar spirit as for the 
earliest attempts at a unified field theory in that in all cases an 
assumption concerning the 4-dimensional spacetime metric is 
\textit{dropped}, as a possible further progression from the relaxing of 
the flat spacetime assumption of special relativity that underlies general 
relativity.  
   For the Weyl, Einstein and Kaluza-Klein theories of 
equations~\ref{WeylU}, \ref{EinU} and \ref{KaKlU} for the case of 
$A_{\mu}(x) = 0$ and $F_{\mu\nu}(x)=0$, with vanishing electromagnetic 
field, the original metric $g_{\mu\nu}(x)$ still describes the curved 
spacetime of general relativity when related to other matter fields through 
equation~\ref{Eineq}. However for the present theory the trivial case of 
equation~\ref{lqdet} corresponds to the matterless vacuum of a flat 
spacetime,
  with all matter arising in a manner intrinsically related to the 
spacetime curvature under augmentations to
 equation~\ref{lqdet} in the form of equation~\ref{lpvn} as reviewed above 
for equation~\ref{gfromavt}.

  Although in a similar spirit this approach also differs from the early 
unified field theories reviewed in subsection~\ref{uni12}, which were based 
on a generalisation of the geometry of the global metric $g_{\mu\nu}(x)$ of 
equation~\ref{gfourd} on an extended spacetime, in that
  here we focus upon the local metric $\eta_{ab}$ of equation~\ref{sfourd} 
  of a local inertial reference frame
  which is  generalised to the expression for proper time in 
equation~\ref{salpha}. 
The embedding of the general metric $g_{\mu\nu}(x)$ in the generalised 
geometric structures of equations~\ref{WeylU}, \ref{EinU} and \ref{KaKlU} 
is analogous to the embedding of 
 the local  metric $\eta_{ab}$ within the generalised proper time interval 
of equation~\ref{salpha} in the manner of equation~\ref{sfourp} as 
described before equation~\ref{lpvn}.
 With  $\eta_{ab}$ extracted from equation~\ref{salpha} to be locally 
incorporated into the metric
 $g_{\mu\nu}(x)$  of general relativity in the extended external 
4-dimensional spacetime,  matter fields derive from the residual components 
of the generalised form of time of equation~\ref{salpha} in this projection 
over the local spacetime geometry as depicted in figure~\ref{grcfme}. 

  In a sense the present theory is hence based upon a more elementary 
generalisation than that of the early unified field theories.
Compared with the unified field theories that immediately followed general 
relativity 
  the shift in focus here towards the \textit{local} spacetime structure 
also seems reasonable given the aim of accounting for the microscopic 
properties of matter. 
  As figure~\ref{grcfme} implies we can zoom into an infinitesimal local 
inertial reference frame anywhere on the spacetime manifold $M_4$ and 
generalise from the form of proper time in equation~\ref{sfourd} to explore 
the microscopic structure of matter that arises.
 As alluded to in subsection~\ref{uni12} the spirit of Einstein's approach  
to a unified theory, for example via the generalisation of the global 
metric components $g_{\mu\nu} \to \tilde{g}_{\mu\nu}$ in 
equation~\ref{EinU}, remains enlightening as can be seen here in the 
augmentation from the local metric components of equation~\ref{sfourd}
to the coefficients of equation~\ref{salpha}
 with $\eta_{ab} \to \alpha_{abc\ldots}$. 
  This approach, which can hence be considered a further generalisation 
from special and general relativity, has been shown to lead to properties 
of the Standard Model and beyond as presented in the previous sections. In 
the following subsection we further emphasise the underlying simplicity of 
the theory.

\subsection{One Simple Equation}
\label{uni52}

 As implied in figure~\ref{grcfme} the underlying order and structure of 
matter in spacetime essentially arises from the composition of the 
continuous flow of time. 
  Through an elementary and direct analysis, from an abstract mathematical 
perspective, any finite interval of proper time 
  $\Delta s \in \rrr$ can be decomposed down to a limit of infinitesimal 
intervals:
\begin{eqnarray} 
  \Delta s & = & \delta s + \delta s +  \delta s +  \ldots
    \label{sdefin}   \\
    \quad \mbox{with substructure:} \quad
  \delta s & = &  {}^{\substack{ p \vspace{1pt} \\ {} }} \!\!\!\! 
    \sqrt{\alpha_{abc\ldots}\delta x^a \delta x^b \delta x^c \ldots}
	 \qquad \qquad \qquad
		\label{sdecom} 
\end{eqnarray}
 for $\delta s \to 0$ as the $p^{\mathrm{th}}$-root of a homogeneous 
polynomial of $p^{\mathrm{th}}$-order in $n$ infinitesimal components 
$\{\delta x^a\} \in \rrr^n$
 labelled by $a,b,c,\ldots$ with each coefficient $\alpha_{abc\ldots} \inn 
\{-1,0,1\}$, maintaining a consistent order of infinitesimals. Here we are 
simply exploiting the basic arithmetic properties of the continuum of real 
numbers, representing the continuum of time with $\delta s \in \rrr$, which 
together with addition include the operations of multiplication and 
extracting roots. Equation~\ref{sdecom} is equivalent to 
equation~\ref{salpha}, which as a direct generalisation from 
equation~\ref{sfourd} is the starting point for the whole theory.  

There is a simplicity in the foundation of this theory in \textit{dropping} 
the assumption that a proper time interval should be expressed through a 
quadratic form, as for equations~\ref{sfourd} and \ref{snd}, and hence 
generalised to higher-order homogeneous polynomial structures, as described 
for equation~\ref{salpha}. Here we see that this simplification is further 
emphasised by observing that rather than \textit{adding} extra spatial 
dimensions that we do not see, with the components 
 $(\delta x^4, \ldots, \delta x^{n-1})$ in equation~\ref{snd}, here 
equation~\ref{salpha} can be interpreted as expressing a basic general 
arithmetic form \textit{inherent} in an infinitesimal interval $\delta s$ 
of the continuum of proper time  as described for equations~\ref{sdefin} 
and \ref{sdecom}.
 Hence the theory is 
 essentially grounded conservatively in this `one dimension' of time 
\textit{alone}, the passage of which we are intimately familiar with
  (prompting the title of~\cite{Unifi} to emphasise the contrast with 
theories based on extra spatial dimensions).
  In equations~\ref{salpha} and \ref{sdefin}--\ref{sdecom} we are not 
\textit{adding} anything to time, nor \textit{replacing} time with 
anything, but simply expressing an intrinsic arithmetic substructure that 
is carried \textit{simultaneously} with time. 

  Generally in the equations of physics a lot of attention is naturally 
paid to the objects on the left-hand side and the right-hand side but the 
`equals sign' in the middle often carries a significant meaning or 
interpretation in itself. For example in general relativity the equals sign 
in Einstein's field equation~\ref{Eineq} is often interpreted to imply that 
matter (energy-momentum $T^{\mu\nu}$ on the right-hand side) `bends' 
spacetime (the geometry $G^{\mu\nu}$ on the left-hand side). This 
description conforms with the historical development since the notion of 
matter very much preceded the conception of spacetime curvature. However 
for the present theory the form of the spacetime solution 
\mbox{$G^{\mu\nu} \equiv f(A,\bv_n)$} in equation~\ref{gfromavt} takes 
priority, with $T^{\mu\nu}$ being \textit{defined} in terms of this 
structure in a manner proposed to incorporate the particle and quantum 
properties of matter consistently with the field equation of general 
relativity as reviewed in the previous subsection.

  Similarly, given the historical background to equation~\ref{sfourd} the 
initial priority might be given to the components of space 
 $(\delta x^1, \delta x^2, \delta x^3)$ and the interval of coordinate time 
$(\delta x^0)$ on the right-hand side, since these are effectively 
assimilated from the earlier Newtonian worldview as alluded to in the 
previous subsection. In  relativity these components are collectively 
combined in a 4-dimensional spacetime form which can \textit{then} be 
identified with the square of the Lorentz invariant proper time interval
 $\delta s$ on the left-hand side of equation~\ref{sfourd}. For the present 
theory, however, we place all the initial emphasis on the left-hand side 
interval $\delta s$ itself, which \textit{can} be arithmetically expressed 
in the quadratic form on the right-hand side of equation~\ref{sfourd} and 
\textit{interpreted} in geometric terms as a basis for 4-dimensional 
spacetime.

 Consideration of possible arithmetic expressions for the proper time 
interval $\delta s$ then leads to the general form of 
equation~\ref{salpha}, which is equivalent to the expression for $\delta s$ 
in equation~\ref{sdecom}, taken to subsume the 4-dimensional form of 
equation~\ref{sfourd} as described for equation~\ref{sfourp}. Physical 
structures arise through the breaking of the symmetry of the full form of 
time when projected over the \mbox{4-dimensional} spacetime substructure as 
depicted in figure~\ref{grcfme}, leading to properties of matter fields 
that resemble the Standard Model of particle physics as described in 
section~\ref{uni3}.
 Hence the pivotal role of the equals sign in equation~\ref{sfourd} is 
central to the interpretation of the theory as deriving entirely 
\textit{from} generalising and analysing the single simple entity of proper 
time.

  Attempts to account for all physical phenomena through a single simple 
entity date back to the pre-Socratic philosopher Thales of Miletus (circa 
600$\:${\small B.C.}) on identifying water as the basis for all matter 
(\cite{Russ} book 1 chapter 2). As arguably the first `unified theory' this 
proposal was perhaps motivated on observing that water could be transformed 
into three known forms, as a solid, liquid or vapour, from which further 
extrapolation might account for all types of material substances. In light 
of the apparent empirical implausibility of that theory by the time of 
Aristotle (circa 350$\:${\small B.C.}) the basic elements had grown to four 
in number: earth, water, air and fire with a further augmentation by a 
fifth element, or `ether', originally associated with celestial phenomena. 
The properties of each element could in turn be ascribed to tiny 
indestructible atoms of matter, according to the theory of Democritus  
(circa 400$\:${\small B.C.}), which might be attributed to the unique 
geometric structures of the five Platonic solids.

  In 1808 John Dalton proposed the modern atomic theory of matter with 
chemical combinations  of originally around twenty elements, including 
hydrogen, carbon and oxygen, in simple numerical ratios accounting for the 
wide variety of compound substances.
 By 1869 the list had grown to over sixty basic elements which when ordered 
according to their atomic weights were seen to exhibit recurring  physical 
properties at a regular pattern of intervals in Mendeleev's Periodic Table 
of chemical elements.
 By that time the `ether' had been reinvented as the `luminiferous ether' 
proposed to permeate all space, with elastic mechanical properties 
permitting the transmission of light similarly as vibrations in the air 
allow the propagation of sound.  
 However such a   
hypothetical ethereal medium  substratum for the electromagnetic field in 
Maxwell's theory~\cite{Maxem}, discussed here in subsection~\ref{uni12}, or 
an `ether drift',  was never detected and that concept along with the 
notion
 of a Newtonian absolute space with which to associate such an ether 
was ultimately discarded as superfluous  with the advent 
  special relativity in 1905 (\cite{Einsr} introduction).
  
  The regular pattern of Mendeleev's Periodic Table of chemical elements 
  hinted at an inner structure for atoms that foreshadowed the discovery of 
their composition and quantum properties, as alluded to in the opening of 
section~\ref{uni1}. The 1911 Rutherford model of subatomic structure with a 
central 
 massive nucleus surrounded by a spherical cloud of electrons was followed 
by the 1913 Bohr atom with the electrons confined to circular orbits in 
discrete steps of angular momentum in multiples of Planck's constant as an 
application of the `old quantum theory' (\cite{Pais2} chapters~9 and 10), 
subsequently supplanted by the atomic picture of the `new quantum theory' 
in the mid-1920s
  (\cite{Pais2} chapter~12). By 1925 in addition to the two established 
fundamental forces of electromagnetism and gravitation only the electron, 
proton (identified with the hydrogen nucleus) and photon were believed to 
be needed to account for elementary material phenomena, albeit with a third 
force seemingly required to hold the atomic nucleus together (\cite{Pais2} 
section~12(a)).   The empirical study of the properties of those and 
further components discovered over the following half a century led to the
 Standard Model of particle physics as established by 1975. With the 
Standard Model incorporating a collection of regular symmetry patterns 
associated with  multiplets of elementary particles the focus then turned 
to the empirical and theoretical investigation into the underlying source 
of these structures. 
 
 In the late 19$^{\mathrm{th}}$ and early 20$^{\mathrm{th}}$ century gaps 
were filled in and
 the Periodic Table grew in size with the discovery of further chemical 
elements, before an underlying explanation in terms of atomic structure was 
uncovered. Similarly since 1975 the Standard Model has been both confirmed 
with new empirical discoveries and augmented with an extended neutrino 
sector and other models of physics beyond, while the search continues for 
an explanation of the origin of these properties.

   Rather than conceiving of the composition of the world as an indefinite 
sequence of `onion layers' of comparable complexity  
  the intuitive idea is often expressed that the underlying structure of 
matter is expected to become simpler as deeper layers are explored. 
 On the macroscopic scale a huge variety of elaborate material structures
  can be observed. While a great deal of complexity remains for structures 
that can be observed with the most powerful optical microscopes all such 
matter is composed from a wide range of molecules and compounds in turn 
constructed from combinations of less than a hundred chemical elements. 
 As arranged in the Periodic Table these elements
 are associated with a series of basic atomic structures built from a 
common set of a small number of component types. Analysis of these 
components led to the discovery of a range of elementary particles, 
somewhat smaller in number than the variety of atoms, as arranged in the 
Standard Model.
 The thread of ever greater simplicity is proposed here to culminate in 
equations~\ref{sdefin} and \ref{sdecom}, with the basic arithmetic 
decomposition of time itself providing the template for the elementary 
particle multiplet structures.
  The distinctive patterns of the Standard Model arise from the symmetry of 
the generalised multicomponent form of time as this symmetry is broken in 
the projection over the local geometric form of 4-dimensional spacetime as 
described for figure~\ref{grcfme}.

  Atoms listed in the Periodic Table can be broken down into constituent 
pieces while in other experiments the particle states of the Standard Model 
can mutate into each other through interactions. Seemingly no such 
experiment can be performed for `time' as the basic element of the present 
theory. However time can be `broken down' and `mutated' through the simple 
mathematical analysis and identities of equations~\ref{sdefin} and 
\ref{sdecom}. Indeed it is precisely through this analysis that 
substructures of time can be identified as a basis for both the geometry of 
4-dimensional spacetime and the matter content within this arena as studied 
in the laboratory as described explicitly in subsection~\ref{uni31}. In 
this sense particle physics experiments could be considered as an 
investigation into the elementary substructure of time itself. 
 Since it is difficult to conceive of a simpler basis for a theory than 
time alone, and since the `fragmentation' of this entity in 
equations~\ref{sdefin} and \ref{sdecom} is already employed at the heart of 
this theory, there is a sense of reaching the ultimate `bedrock' in 
accounting for the properties of matter, with no further `onion layers' to 
be sought at a yet deeper level. 

On the other hand given
 equation~\ref{sdecom},
which is equivalent to equation~\ref{salpha} and can be conveniently  
written in the form of equation~\ref{lpvn},
  specific mathematical possibilities for this generalisation of proper 
time 
 and their empirical consequences are very much open to further 
exploration. This has been described through to the proposed level of an 
$\ee$ symmetry in subsection~\ref{uni32} with tentative consequences for 
physics beyond the Standard Model considered in section~\ref{uni4}, 
demonstrating the potential of this theory for making testable predictions. 
 The possibility of a yet higher-order fragmentation for 
equation~\ref{sdecom}, with potentially fractal-like properties, or a role 
for alternative multicomponent forms for proper time as suggested
 for equation~\ref{lpvndsh} as a candidate source of a dark sector in 
subsection~\ref{uni42}, can also be considered.   

  The flow of time pervades the entire spacetime manifold in 
figure~\ref{grcfme}, sharing this property with the hypothetical ether.
 However, while the ether was abandoned with relativity theory  
 the notion of proper time, and its local invariance, is central to the 
progression through special and general relativity to the theory presented 
here, as described in the previous subsection. For Maxwell the luminiferous 
ether, while never detected, was a postulated material substratum  
 \textit{underlying} the observed phenomena of the electromagnetic field. 
Here 
on the other hand there is no question of performing an experiment to 
search for `time', rather
we are directly and intimately familiar with the flow of time as an 
irreducible element infusing \textit{all} experiments and  observations,
 including in the high energy physics laboratory,
 with all particle and material phenomena proposed to arise 
\textit{through} the substructure of time.

  Indeed the usual basis for most physical models and theories is to begin 
by positing a basic entity or entities \textit{in} space and time, whether 
for example water, an ether, atoms, particles or fields. Here the left-hand 
sketch in figure~\ref{grcfme} is \textit{not} ultimately to be similarly 
interpreted with the basic entity time $s$ flowing through a 
\textit{pre-existing} space and time. Rather the geometric spacetime 
structure \textit{itself}, as well as the matter within it, is extracted 
\textit{through} the composition of the continuum of time as the sole basic 
entity.  
 While conceptually very different to earlier unification schemes,
   the simplicity of this perspective then provides a further motivation 
for the theory.
The historical and philosophical aspects associated with this simple 
interpretation of the theory as deriving from the intrinsic arithmetic 
substructure of proper time alone are elaborated in~\cite{Struct}. In 
particular this change in perspective with time promoted to the prior role, 
via the substructure of equations~\ref{sdefin} and \ref{sdecom}, as the 
progenitor for both spacetime structure and matter in spacetime, subject to 
laws of physics  
  determined by the constraints implied by this underlying simplicity, is 
further described for (\cite{Struct} figure~1).

The properties of elementary particles are uncovered at the most elementary 
level of the theory through a simple symmetry breaking analysis for 
equation~\ref{sdecom}, 
 written as equation~\ref{lpvn},
  deriving directly from the extraction of an external 
\mbox{$\mbox{Lorentz} \subset \hG$}  symmetry acting on the projected 
$\bv_4 \in \TM_4$ subcomponents of $\bv_n\inn \rrr^n$ that underlie the 
basis of the local external 4-dimensional spacetime itself, without adding 
anything else to the theory.
 This symmetry breaking is implied through the necessity of perceiving a 
physical world in space as well as through time, as also discussed 
in~(\cite{Struct} section~4), with the properties of matter entirely 
determined by the residual components in the form for proper time of 
equation~\ref{lpvn} over the 4-dimensional spacetime  manifold $M_4$.

 As noted above this contrasts with the usual approach of formulating a 
theory through the introduction of entities with particular properties into 
a pre-existing space and time. Such is the case for example in Maxwell's 
theory with the introduction of electric and magnetic fields, whether or 
not supported by an ethereal substratum,  together with a set of equations 
to describe empirical observations. In many modern-day theoretical 
frameworks particle properties are typically built in by hand in a similar 
spirit as for Maxwell's theory
 by introducing the corresponding fields and interactions into the proposed 
Lagrangian for the theory, as is the case for the `Standard Model' itself 
as well as typically for many models beyond. 
 While masses and charges are assigned to physical bodies in the equations 
of gravitation and electromagnetism of Newton and Maxwell, similarly in the 
Lagrangian of the Standard Model masses, charges and couplings generally 
for elementary particles are typically assigned according to empirical 
observations.  

  There are a number of constraints on the type of
  fields and terms that can be included in a Lagrangian. This is  
particularly the case in the context of a quantum field theory (the 
postulates of which are also generally imposed pragmatically)
  for consistency with unitarity and causality.
 Typically the Lagrangian will be a real function consisting of a series of 
terms in the fields and their first or second order derivatives in a 
Lorentz and gauge invariant form, with the renormalisability of the QFT 
placing further restrictions such as avoiding coupling constants with 
negative mass dimension. Nevertheless considerable arbitrariness is still 
permitted, particularly for the construction of models beyond the Standard 
Model.  
  Indeed on occasions when a provisional hint of new physics is seen in the 
high energy physics laboratory, in data otherwise implying a significant 
statistical fluctuation above the expected background,
 a large number of new models may be prompted as there will generally be 
many ways to accommodate the apparent observations through augmenting the 
Standard Model Lagrangian. 
   Even for the Standard Model, with a minimal extension to incorporate the 
current phenomenology of left-handed neutrinos, at least 25 free parameters 
also need to be introduced and determined or constrained from the data 
(eighteen for the Standard Model together with seven neutrino mass and 
mixing parameters). 

In the light of these observations the sentiment is sometimes expressed 
that a key ambition for a unification scheme would be to replace the 
lengthy and complicated Lagrangian of the Standard Model by `one simple 
equation' from which all of the properties of particle physics might be 
derived.
  A number of questions could be raised about any candidate for such a 
primordial equation, with key issues regarding the nature of the basic 
entity described, whether for example a particle, field, string etc., 
concerning why it should exist itself at all, why it should \textit{be} the 
basic entity and why it should be subject to the `one simple equation'.

  For the present theory `time' is the basic entity. As well as being 
fundamental to empirical and theoretical physics, as discussed in this 
section, time is also a necessary, intrinsic and inherent element of any 
subjective experience we can have, unlike the case for water, an ether, 
atoms, particles, fields or other proposed basic physical entities. 
 These fundamental objective and subjective features of time make it an 
appropriate basic element for any theory. For the present theory time is 
the \textit{sole} basic entity. Further, here we are \textit{not} proposing 
or positing an equation to be \textit{imposed} on time, rather we are 
simply utilising a direct arithmetic property innate in the concept of the 
continuum of time as expressed for equations~\ref{sdefin} and \ref{sdecom}.

  The continuum properties of time have been recognised as central to the 
development of the equations of physics in describing the behaviour of 
matter since the days of Newton, as noted in leading up to 
figure~\ref{grcfme} in the previous subsection. Here however we do not 
introduce any material or other entity with a particular `time dependence', 
rather material phenomena derive from the `subcomponents of time' itself 
through equation~\ref{sdecom}, which also accommodates the basis for the 
local Lorentz metric structure of the background spacetime arena via the 
substructure of equation~\ref{sfourp}. 
 In this manner all of physics is proposed to follow from the `one simple 
equation' for the invariant proper time interval $\delta s$ in 
equation~\ref{sdecom}, which is equivalent to equation~\ref{salpha} and can 
be written as   equation~\ref{lpvn} as described in subsection~\ref{uni22}, 
as the basis of the theory. 
 As labelled in equation~\ref{lpvn} the symmetry transformations applied to 
the subcomponents of time belong to a group $\hG$ that generalises the 
Lorentz group of  relativity theory.

  A connection with the Lagrangian approach is anticipated to arise through 
terms in the breaking of the full symmetry $\hG$ of equation~\ref{lpvn} in 
the projection over the local substructure of 4-dimensional spacetime 
$M_4$.
  Matter fields deriving from the broken fragments of $\bv_n \inn \rrr^n$  
will feature in `mass terms' in the resulting expression of 
equation~\ref{lpvnb}  if composed  with a factor derived from the projected 
subcomponent \mbox{$\bh \equiv \bv_4 \in \TM_4$}, which is here central to 
the `origin of mass' itself and associated with the Higgs  as described for 
equations~\ref{l2v4h}--\ref{gmnconh}, with Yukawa coupling factors 
to be identified as proposed in subsection~\ref{uni42}.
 The broken symmetry expression in equation~\ref{lpvnb} also places 
constraints on the dynamics of the matter fields, coupling them with gauge 
fields, as discussed in subsection~\ref{uni23}. 
 The role of Lagrangian kinetic terms for each gauge field $A(x)$ 
associated with the internal symmetry is proposed to be appropriated by the 
quadratic terms in the internal gauge curvature $F(x)$ in the relation with 
the external spacetime geometry of equation~\ref{gchift}, which is similar 
to that in many Kaluza-Klein theories, with the dynamics of the gauge 
fields  constrained by geometric identities as also reviewed in 
subsection~\ref{uni23}.
 Further, rather than \textit{applying} the postulates of a quantum theory,
 a local degeneracy of field solutions for describing the external 
spacetime geometry is proposed to \textit{underlie} the `quantisation' of 
the matter and gauge fields and the corresponding particle phenomena as 
reviewed for equation~\ref{gfromavt}.

 That these particle phenomena will exhibit properties closely resembling 
the Standard Model has been demonstrated for the analysis through to the 
$\hG = \ese$ level of equation~\ref{esesb}--\ref{esebits} in 
subsection~\ref{uni31}. The full picture is predicted to emerge for the  
 proposed $\hG=\ee$ level of equation~\ref{lvto} as discussed in 
subsection~\ref{uni32}, for which physics beyond the Standard Model in the 
neutrino, Higgs and dark sectors can be anticipated as described in 
section~\ref{uni4}.
  It is striking to observe how properties of the Standard Model, and 
contemporary physics beyond, can be uncovered in this manner from such a 
simple underlying basis. 
 In particular the above analysis illustrates how a range of relatively 
complex phenomena, matching empirical observations, can be obtained through  
the `one simple equation' for an infinitesimal interval of time in 
equation~\ref{sdecom}, or the equivalent expressions of 
equations~\ref{salpha} and \ref{lpvn}.
 The ultimate ambition would be to determine not only the elementary 
particle multiplet structure but also the masses, charges and couplings of 
elementary particle states as far as possible from the intrinsic  
constraints of the theory.
 In deriving from the natural generalisation for a proper time interval in 
equation~\ref{sdecom} as a simple elementary basis for the theory, there is 
then the potential for a genuine understanding of the underlying origin of 
the Standard Model and physics beyond, all accounted for by a fundamental 
unified theory.


\section{Discussion and Conclusions}
\label{uni6}

  One of the main aims of experiments in particle physics, observations in 
cosmology and the construction of phenomenological models, typically via a 
proposed Lagrangian function, is to point the way to a unified theory 
incorporating a more elementary understanding of the workings of the 
universe. Given a set of empirical data and an array of associated models
  an intuitive leap, 
 rather than a series of small iterative steps, is likely to be needed to 
uncover the fundamental theory. Having posited such a theory, based largely 
on an internal motivation relating to its conception, the aim would then be 
to work forwards developing the theory and seeking correspondence with 
models
 that have been devised on the basis of empirical observations.
 In this paper we are not building a model pragmatically, for example by 
adding terms to a Lagrangian by hand, but rather 
 we have posited such a fundamental theory and developed the consequences 
from underlying elementary first principles.

  In place of directly adding matter fields independently on top of 
4-dimensional spacetime to match empirical observations, as for the models 
alluded to above, the early approaches to a unified field theory reviewed 
in subsection~\ref{uni12} aimed to account for the empirical properties of 
matter in spacetime within a single unified framework by further 
generalising the structure of spacetime itself.
This principal  goal was present from the beginning, as alluded to in the 
title of Kaluza's paper~\cite{Kaluza}, and continues to motivate modern-day 
unification schemes with extra spatial dimensions aiming to account for the 
Standard Model through an augmented spacetime structure.
 Via such a conception the aim has been to achieve a unification in which 
the field and particle states we observe in experiments reside within the 
internal `mathematical space' of extra spatial dimensions, which is  more 
akin to the external `geometrical space' within which macroscopic 
structures such as ourselves reside and through which we perceive the 
physical world.

  Extra spatial dimensions have then provided the potential to connect the 
physics of the external world of 4-dimensional spacetime and the internal 
world of material phenomena within a more unified structure (see for 
example \cite{Cho} section~9 first two of final three paragraphs). 
 However even the most sophisticated attempts to account for the Standard 
Model via the structures of extra spatial dimensions, as discussed in 
subsection~\ref{uni21}, have if anything been less successful than the 
original 5-dimensional spacetime theory of Kaluza and Klein in 
accommodating electromagnetism, as reviewed for equation~\ref{KaKlU}.

 From a local perspective the addition of extra spatial dimensions augments 
a proper time interval $\delta s$ from the form of equation~\ref{sfourd} to 
that of equations~\ref{snd} and \ref{vnd}, as depicted by the 
generalisation from figure~\ref{vfdvnd}(a) to \ref{vfdvnd}(b).  
 An elementary analysis for the implied matter content in 4-dimensional 
spacetime is described for equation~\mbox{\ref{sosb}--\ref{vnbits}}.
 The present theory is
based on a direct further generalisation of a proper time interval over and 
above  the 4-dimensional form of equation~\ref{sfourd} for the local 
external spacetime  
    to the form of equation~\ref{salpha}.
	Compared with augmentations restricted to extra spatial dimensions with 
the quadratic form of equation~\ref{snd} the more general homogeneous 
polynomial form of equation~\ref{salpha}
	 demands \textit{fewer} assumptions, as explained in 
subsection~\ref{uni22}.
  \mbox{As described} explicitly for equation~\ref{sfourp} this more 
generalised form for proper time can still readily accommodate a 
substructure matching 
 the local external spacetime metric geometry.
 
Adopting the generalisation of equation~\ref{salpha}, which can be 
expressed as equation~\ref{lpvn},
 matter fields are then associated with the properties of the residual 
components resulting from the symmetry breaking of the generalised form of 
proper time when projected over
 the local 4-dimensional geometric spacetime substructure, 
  as depicted in figure~\ref{grcfme}. 
 The elementary physical structures identified in spacetime for a minimal 
case are described in subsection~\ref{uni23}.
  Leading to the structure of equation~\ref{esesb}--\ref{esebits}
 the residual components are found to exhibit a significant correlation 
with the Standard Model of particle physics as reviewed in 
section~\ref{uni3}.
 This direct analysis for
  the general form of proper time 
 then provides a far better template for connecting with the Standard 
Model, and with features beyond as described in section~\ref{uni4}, than 
the equivalent analysis applied for the restricted case of extra quadratic 
spatial components as described for equation~\ref{sosb}--\ref{vnbits}, and 
in this respect also compares favourably with the more sophisticated 
approaches to extra spatial dimensions discussed in subsection~\ref{uni21}.

    Similarly as for extra spatial dimensions this theory based on 
generalised proper time can be seen as a natural further development from 
special and general relativity as described in subsection~\ref{uni51}.
  Compared with the employment of extra spatial dimensions the present 
theory is however more \textit{unifying} in reducing the fundamental entity 
of theory to the more elementary starting point of the single parameter of 
proper time alone. The greater \textit{simplicity} of the theory is 
manifested in its expression through `one simple equation' for 
infinitesimal intervals of time  
 as discussed in subsection~\ref{uni52} for equation~\ref{sdecom}, which is 
equivalent to equations~\ref{salpha} and \ref{lpvn}. The theory is also 
more \textit{conservative} in that the flow of time is something that we 
are intimately familiar with while, essentially by definition, we do not 
perceive any `extra' spatial dimensions. Further the theory is more 
\textit{unique} in terms of having an unambiguous starting point in proper 
time as parametrised by $s\in \rrr$ with a trivial topology, rather than 
having a potentially arbitrary number of extra spatial dimensions with a 
large range of possible topological properties. 
 This uniqueness is further seen in the mathematical development of the 
theory in leading to equations~\ref{lvfsq}--\ref{esebits}, as summarised in 
the opening of subsection~\ref{uni32}.
 Each of the above points applies for the present theory both in comparison 
with the direct approach to extra spatial dimensions of 
equations~\ref{snd}--\ref{vnbits} and figure~\ref{vfdvnd} as well as in 
comparison with the most technically sophisticated approach via string and 
M-theory which, for example with regards to uniqueness, is faced with the 
landscape problem as also discussed in subsection~\ref{uni21}.

 The internal simplicity and uniqueness of the present  theory, which 
nevertheless yields a
  far more direct and efficient connection with empirical structures of the 
Standard Model at the most elementary level of matter
 compared with the models based on extra spatial dimensions,
 strongly suggests that equation~\ref{lpvn} rather than equation~\ref{vnd} 
provides a more appropriate core basis for a unified theory.
 The extraction of a necessarily \textit{quadratic} substructure to 
represent the local external spacetime also underlies a \textit{mechanism} 
for the symmetry breaking itself 
for equation~\ref{lpvn} but not for equation~\ref{vnd}, as described after  
  equation~\ref{l2vn}.
 While models with extra spatial dimensions essentially propose a 
`materialisation of space' the present theory goes further by describing a 
`materialisation and spatialisation of time'.
 The nature of our necessary perception of the world in space as well as 
through time is central to the symmetry breaking as also alluded to in 
subsection~\ref{uni52} with reference to \cite{Struct} where further 
historical background to this conceptual picture is also discussed.
 Here we briefly summarise the historical developments in mathematics and 
physics underlying this theory.

 The first particle of the Standard Model to be discovered was the electron 
at the end of the $19^{\mathrm{th}}$ century, around the same time that the 
exceptional Lie algebras were first classified as noted in the opening of 
subsection~\ref{uni31}.
The  Standard Model itself was fully established in the 1970s, 
 representing an amalgamation of the inferences drawn from many experiments 
over a number of decades, and has been tested subsequently 
 and largely confirmed 
by further decades of empirical data through to the discovery and 
 analysis of the Higgs in recent years.
Mathematical developments from the mid-$20^{\mathrm{th}}$ to early  
$21^{\mathrm{st}}$ century, including~\cite{Chev,Wang2,Freud,Rios} as cited 
in subsection~\ref{uni31},
 have enabled the development of the present theory, as expressed for 
equation~\ref{lpvn}, through the 9, 27 and 56-dimensional forms of 
equations~\ref{lvni}, \ref{lvts} and \ref{lvfsq}. These have led 
respectively to the identification of the symmetry breaking structures of 
equations~\ref{slsb}--\ref{slbits}, \ref{esisb}--\ref{esibits} and 
\ref{esesb}--\ref{esebits}, closely resembling the features of a generation 
of Standard Model leptons and quarks -- including a set of spinor states, 
colour $\suth_c$ singlets and triplets with the appropriate electromagnetic 
$\uo_Q$ fractional charges and an intrinsic left-right asymmetry.

 This progression then represents an intimate coming together of 
mathematical and physical structures which, while contemporary with each
other and studied in parallel, developed largely independently over the 
past hundred or more years, and with very little redundancy as seen for 
equation~\ref{esesb}--\ref{esebits}, drawn together through natural 
mathematical expressions for equation~\ref{lpvn} as the central equation of 
this unified theory.
While the theory is rooted in the firm conceptual basis of generalised 
proper time the subsequent employment of mathematical structures relating 
to the exceptional Lie groups has advanced the theory to the point of 
making connections with empirical phenomena.
 Mutually, the application of these mathematical structures, which have 
been developed in a largely abstract manner, within the context of the 
present theory, presents an explicit means of relating the exceptional Lie 
groups to the physical world. 
 That is, the expressions in equations~\ref{lvts} and \ref{lvfsq} provide 
their respective symmetry groups $\esi$ and $\ese$ with a clear and simple 
conceptual basis within this theory,
  as symmetries of generalised proper time, 
  without needing to introduce them in an 
  ad hoc fashion as might be the case for model building
  (dating back to \cite{Gur1,Gur7} as alluded to in 
subsection~\ref{uni12}),
   here making their relevance for the Standard Model explicit through 
equations~\ref{esisb}--\ref{esibits} and \ref{esesb}--\ref{esebits}. 
Such an interplay of mathematical development and physical application is 
analogous to the incorporation of fibre bundle structures in the framework 
of gauge theories and non-Abelian Kaluza-Klein theories as discussed in 
subsection~\ref{uni12} with reference to~\cite{Cho,Steen}.

  While properties associated with the Standard Model emerge much more 
readily for the present theory, compared to models with extra spatial 
dimensions, a complete correspondence is not to be expected until a 
mathematically complete understanding of the full form of proper time, and 
its corresponding symmetry breaking pattern, has been established. The 
unique development of the theory through to the $\ese$ quartic form of 
equation~\ref{lvfsq} with the implications of 
equation~\ref{esesb}--\ref{esebits} together with properties of the 
exceptional Lie groups
 naturally lead to the prediction of a further unique progression to an 
$\ee$ symmetry of a full form $\lvtfep$, as described for 
equation~\ref{lvto} in subsection~\ref{uni32}, proposed to result in the 
uncovering of the 
full particle multiplet structure of the Standard Model and beyond.
 While elements of the Standard Model have been accounted for through a 
rigorous analysis of the $\esi$ and $\ese$ levels underlying 
equations~\ref{esisb}--\ref{esibits} and \ref{esesb}--\ref{esebits}, the 
investigation at the $\ee$ level, including that for equation~\ref{eebits}, 
has to date been of a more provisional nature. An explicit description of a 
real $\ee$ action constructed on the, currently hypothetical, homogeneous 
form of equation~\ref{lvto} with a symmetry breaking pattern of 
equation~\ref{eebrk}
 applied to the subcomponents of $\bv_{248} \inn \rrr^{248}$, subsuming 
equation~\ref{esesb}--\ref{esebits}, will be needed to attempt a full 
reconstruction of the Standard Model structure and to make firm predictions 
beyond.

  The need to unfold the full Standard Model structure, including three 
generations of leptons and quarks with a complete set of spinor states
 also for the \mbox{$\nu$-leptons} and $u$-quarks
 and with a full electroweak symmetry,  implies that there are significant 
requirements on the desired structure of the homogeneous polynomial form 
$\lvtfep$ which may also be utilised as a
 guide in identifying the $\ee$ symmetry structure and breaking pattern 
itself.
 However the main goal for the predicted $\ee$ structure, as achieved for 
the intermediate $\esi$ and $\ese$ levels, is principally to deduce the 
properties of the Standard Model, and beyond,  \textit{as constrained by} 
the natural mathematical expressions for equation~\ref{lpvn}.
 That it is not obvious that this can be achieved for $\ee$, or a closely 
related symmetry structure consistent with equation~\ref{lpvn}, provides a 
non-trivial test for the theory.
The technical mathematical details of the connections with the Standard 
Model made through to the $\ese$ level and support for the predicted role 
for $\ee$ are described extensively in~\cite{TimeE} and the references 
therein.
 Here
 a significant direction for further progress on the mathematical side, 
through investigating a possible role for octonion triality in the 
construction of an octic $\ee$ invariant in 248 components, has been
  described in subsection~\ref{uni32} -- in particular with the guide of 
the references cited in the last three paragraphs of that subsection.

  The attempt to advance the status of the theory beyond the conceptual 
basis and corresponding mathematical formalism by making connections with 
empirical phenomenology has also been extended beyond the Standard Model. 
 In this paper we have focussed upon the opportunity for a mutual 
development of this theory along with an understanding of new physics, in 
particular in the neutrino sector. Building upon the progression through 
the $\esi$ and $\ese$ levels
  general consideration of the possibility of a significant role for $\ee$ 
leads, via the 
schematic extrapolation from the relevant parts of 
equations~\ref{esisb}--\ref{esibits} and \ref{esesb}--\ref{esebits} to 
equation~\ref{eebits}, to the preliminary prediction for only \textit{two} 
right-handed neutrino states to be accommodated 
 in a manner compatible with a structure of \textit{three} generations for 
the other 
  lepton and quark states.
  As also noted in  subsection~\ref{uni41} this suggests that  
  one of the active left-handed neutrino states will be correspondingly 
massless,
   with forthcoming improving limits in the laboratory and from cosmology 
able to test this tentative but specific empirical feature of the theory.
  By making a connection with models featuring  two right-handed neutrinos
   (for example~\cite{Framp,Ibar,Antu,Chang})
    in this paper the  focus  has been to demonstrate a specific bridge 
from the forefront in developing this theory, based on a very simple 
underlying origin, to a key area in recent developments in particle 
physics.

 If this theory had reached the present stage of development in the 1980s, 
with the Standard Model established but still with relatively limited 
empirical input in neutrino physics, then the argument for two right-handed 
neutrinos would \text{not} have been a compelling prediction of the theory 
at this point. Rather we have aimed to demonstrate how the theory might in 
principle accommodate two right-handed neutrinos 
 \textit{naturally} within a structure of three generations of Standard 
Model fermions, consistent with a series of recent neutrino models that 
have been motivated by contemporary neutrino phenomenology. 
 While at this stage lacking in more explicit detail for the neutrinos,  
since the present theory is based on elementary first principles rather 
than being explicitly tailored  to fit the data  as a pragmatic model, 
through further development there is here an opportunity for a deeper 
understanding of the nature of neutrino physics generally, as discussed in 
subsection~\ref{uni41}.

While the mathematical pursuit of the full symmetry action of $\ee$ for the 
form $\lvtfep$ and the resulting breaking pattern may elucidate  the origin 
of the esoteric properties of the neutrino sector, building upon the basic 
generation structure of equation~\ref{eebits}, existing empirical knowledge 
of neutrino phenomenology as established in recent decades, as expressed by 
models with two right-handed neutrinos (or even other models such as for 
example the $\nu$MSM in which two right-handed neutrinos have distinct 
properties and a distinct role from a third $\nu_R$ 
state~\cite{Asaka1,Asaka2} as also reviewed in subsection~\ref{uni11}),
might itself provide a pertinent clue, on top of the Standard Model itself, 
towards deciphering the detailed structure of the specific mathematical 
application for $\ee$ that is predicted.
 In particular the packing of the neutrino spinor structure, under an 
$\sltc \subset \ee$ action of equation~\ref{eebrk} on the subcomponents of 
$\bv_{248}$ in equation~\ref{lvto}
    utilising octonion triality as alluded to above, and
 their Dirac or Majorana nature in mass terms identified in the expansion 
of equation~\ref{lvtob}, will be central to this ambition.
 A connection might then be made between Lorentz and gauge invariant terms 
in the expansion
  of equation~\ref{lvtob} and Lagrangian mass terms in models with two 
$\nu_R$ states.
   More generally there is an open 
 opportunity  to explore the mutual development and close interplay of the 
underlying mathematical structure of the theory together with an 
understanding of Standard Model physics and empirical consequences  beyond.

   The main approach of this theory is `deductive' in starting with the 
hypothesis of the central role for proper time in equations~\ref{salpha} 
and \ref{lpvn} and then developing the mathematics, for example through to 
equations~\ref{esesb}--\ref{esebits} and \ref{eebits}, to then examine the 
empirical consequences in a manner ideally leading to testable predictions.
  As alluded to in the opening of this section this contracts with the 
`inductive' approach employed in the construction of a model, as inferred 
from patterns in the data, for example by adding fields and terms to a 
Lagrangian.
 While that data hence provides evidence for such a model the inductive 
argument is not in general unique and typically leads to a range of 
plausible models, as is generally the case for neutrino phenomenology. 
Here, while an interplay with inductive elements might be utilised to aid 
in the mathematical construction of the $\ee$ action and the explicit form 
of equation~\ref{lvto} sought, the ultimate aim is for a clear, unique and 
deductive argument unambiguously covering a wide range of physical 
phenomena, and for example homing in on the features of a particular model 
in the neutrino sector.

 A degree of confidence in any predictions of the theory can be gained 
through the intrinsic simplicity of the underlying basis of the theory, as 
described in the previous section, and the significant foothold in features 
of the Standard Model that has already been efficiently attained as 
summarised here for equations~\ref{slsb}--\ref{slbits}, 
\ref{esisb}--\ref{esibits} and \ref{esesb}--\ref{esebits}. 
 Indeed most of the subcomponents of $\bv_{56}$ in 
equation~\ref{esesb}--\ref{esebits} are associated with elements of the 
Standard Model with the only apparent redundancy being the set of four 
scalar invariants $\{n,N,\alpha, \beta \}$, which provisionally form 
candidates for dark matter as described in the opening of 
  subsection~\ref{uni42}.
The prospects for insights into new physics include the nature of the Higgs 
and  electroweak symmetry breaking, for which empirical predictions might  
be sought  for the laboratory.
 The close connection here between the Higgs sector, associated with the 
$\bh\equiv\bv_4 \inn \TM_4$ projection of equations~\ref{l2v4h} and 
\ref{hnorm}, and  neutrino physics, as implied by equation~\ref{eebits}, 
  suggests the Higgs may have a composite structure as described in
  subsection~\ref{uni41}. This feature, together with the possible 
connections between this Higgs and 
a dark sector as explained in subsection~\ref{uni42}, may impact  precision 
electroweak measurements in a manner accessible at the LHC or at a future 
$e^+e^-$ linear collider as also discussed in section~\ref{uni4}.

   In subsection~\ref{uni42} we  described for equation~\ref{lpvndsh} how 
the impact of possible alternative forms for proper time on the 
4-dimensional spacetime geometry might itself provide a quintessential 
source for the dark sector, in only being linked with the visible sector 
deriving from the $\ee$ form of equation~\ref{lvto} through purely 
gravitational means.  
Correspondingly the generalisation from $\{n,N,\alpha, \beta \}$ at the 
$\ese$ level
 of equation~\mbox{\ref{esesb}--\ref{esebits}} could yield a broader set of 
scalar invariants under the full $\ee$ broken symmetry which, on taking 
vacuum values and being composed in terms of the expansion of 
equation~\ref{lvtob}, 
 may form a range of Yukawa couplings and hence in principle augment the 
predictive power of the theory, as also noted in subsection~\ref{uni42}.
More generally analysis of the properties of the full set of subcomponents 
of  $\bv_{248} \in \rrr^{248}$ from the breakdown of equation~\ref{lvto} 
offers the opportunity to explore further
  new physics, as does the possibility to fit a further gauge group 
alongside that of the Standard Model in the $\ee$ symmetry breaking pattern 
of equation~\ref{eebrk}. 
   
   While previous papers (including~\cite{Novel,TimeE}) have described in 
detail the connections made with the Standard Model (summarised here in 
section~\ref{uni3}), in this paper  we have emphasised the potential 
connections with new physics (as assessed in section~\ref{uni4}),
 while always keeping in sight the simple origins of the theory based on an 
elementary generalisation of a proper time interval as described for 
equations~\ref{salpha} and \ref{lpvn}.
 With reference to subsection~\ref{uni12}, 
  in section~\ref{uni2} and subsection~\ref{uni51} we have described how 
the basis for this theory connects with the early unification schemes from 
the immediate post-general relativity era based on generalisations of the 
4-dimensional spacetime geometry.
  Hence in developing the present theory we connect the original conception 
of a unified field theory dating from a hundred years ago with the current 
quest to understand the empirical data from recent and ongoing  laboratory 
experiments and cosmological observations. In particular the 
 modern-day empirical status of neutrino physics, as reviewed in 
subsection~\ref{uni11}, is hence connected
 with the 
historical proposals of subsection~\ref{uni12} through this theory, 
ultimately making contact with models featuring two right-handed neutrinos  
as explained in subsection~\ref{uni41} (with the central argument 
summarised in~\cite{Ufield}).

 As described for equations~\ref{sdefin} and \ref{sdecom} it is difficult 
to conceive of a more elementary basis for a unified theory
 than this simple direct generalisation of proper time.
 On the other hand the natural mathematical possibilities for 
 the invariant infinitesimal proper time interval in   
  equation~\ref{sdecom}, which is equivalent to the `one simple equation' 
of equation~\ref{salpha} and also equation~\ref{lpvn}, are open to further 
exploration. Also open to investigation are the wider implications of the 
theory in relation to the geometric structures of Kaluza-Klein 
models~\cite{KKone},  quantisation and the nature of physical particle 
states as studied in laboratory experiments (\cite{Unifi} chapters~10, 11 
and section~15.2, as also discussed for \cite{Struct} equations~9 and 10) 
and the cosmological questions regarding the dark sector and early universe 
(\cite{Unifi} chapters~12 and 13) as well as the connections with the 
Standard Model itself (\cite{Unifi} chapters~6--9, \cite{Novel,TimeE}).
 With the aim of making a range of testable predictions there is an 
opportunity for mutual advances in these theoretical and related 
experimental investigations.
 The manner in which this theory has been able to reproduce a series of key 
features of the Standard Model, and has already yielded provisional 
connections with neutrino and other new physics beyond, demonstrates this 
open opportunity to further explore this  theoretical structure with the 
ambition of advancing our understanding of the elementary constitution of 
the physical world.


{\setlength{\baselineskip}{0.956\baselineskip}

\par}


\par}

\end{document}